# Thermoelectric films and periodic structures and spin Seebeck effect systems: Facets of performance optimization

Nagaraj Nandihalli[a,✉]

*a,* Department of Chemistry and Waterloo Institute for Nanotechnology, University of Waterloo, Waterloo, Ontario, N2L 3G1, Canada

✉: Nagaraj Nandihalli | nnandiha@uwaterloo.ca

## *Abstract*

The growing market for sensors, internet of things, and wearable devices is fueling the development of low-cost energy-harvesting materials and systems. Film based thermoelectric (TE) devices offer the ability to address the energy requirements by using ubiquitously available waste-heat. This review narrates recent advancements in fabricating high-performance TE films and superlattice structures, from the aspects of microstructure control, doping, defects, composition, surface roughness, substrate effect, interface control, nanocompositing, and crystal preferred orientation realized by regulating various deposition parameters and subsequent heat treatment. The review begins with a brief account of heat conduction mechanism, quantum confinement effect in periodic layers, film deposition processes, thin film configurations and design consideration for TE in-plane devices, and characterization techniques. It then proceeds to alayzing the latest findingd on the TE properties of $Bi_2(Te,Se)_3$ and $(Bi,Sb)_2Te_3$, PbTe, GeTe, SnSe, SnTe, $Cu_{2-x}Se$, and skutterudite films, including superlattices and the performance of TE generators, sensors, and cooling devices. Thickness dependent microstructure evolution and TE characteristics of films in relation to temperature are also analyzed. In the context of spin Seebeck effect (SSE) based systems, SSE mechanism analysis, developments in enhancing the spin Seebeck signal since its first observation, and recent developments are covered from the facets of new system design, signal collection, magnetic manipulation, interface conditions, thickness-dependent longitudinal spin Seebeck signal, and length scale of phonon and magnon transport in longitudinal SSE (LSSE) in different bi-layer systems. At the end, possible strategies for further enhancing $zT$ of TE films and spin Seebeck signals of many systems are addressed.

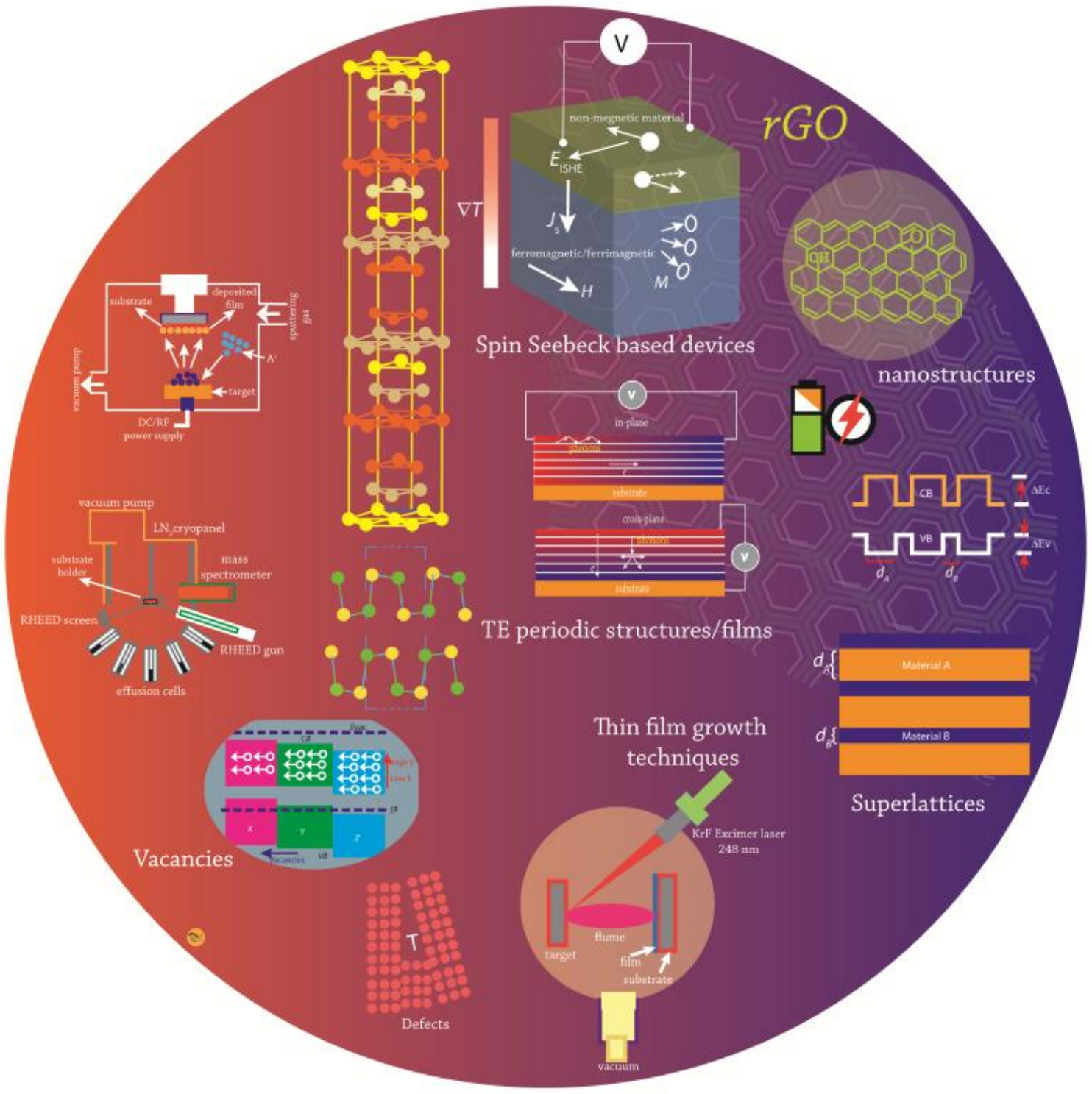

## 1. Introduction

Thermoelectric (TE) energy conversion technology is in the spotlight for its outstanding features. TE devices without any moving parts find applications in power generation, cooling, and wearable devices.[1-5] The performance of TE materials is primarily judged by the dimensionless *figure-of-merit*, $zT = S^2\sigma T/\kappa$, where $S$ is the Seebeck coefficient, $\sigma$ is the electrical conductivity, $T$ is the absolute temperature, and $\kappa$ is the total thermal conductivity. To have a good thermoelectric performance, the material should possess a large $S$, high $\sigma$, and poor $\kappa$. The power factor ($PF = S^2\sigma$) describes the electron energy conversion capability. The major challenge in designing high $zT$ TE materials comes from the strong correlation of $S$, $\sigma$, and $\kappa$ through carrier concentration $n$, as depicted in the following relation[6]

$$S = \frac{8\pi^2 {k_B}^2}{3eh^2} m^* T \left(\frac{\pi}{3n}\right)^{2/3} \qquad \text{Equation 1}$$

where $k_B$ is the Boltzmann constant, $e$ the electron charge, $h$ the Planck constant, and $m^*$ the density of states effective mass of carriers. Therefore, low $n$ materials such as semiconductors or insulators usually exhibit high $S$ and with increasing $n$, $S$ drops rapidly as in metals. On the other hand, in order to ensure large $\sigma$, a large $n$ is required, according to the following relation $\sigma = ne\mu$, where $\mu$ is the carrier mobility. Only within the optimized $n$ range (~$10^{19}$–$10^{20}$ $cm^{-3}$) the $PF$s can be maximized by balancing $S$ and $\sigma$. The total thermal conductivity, $\kappa$ consists of two major parts: $\kappa_l$, the lattice or phonon contribution, and $\kappa_e$, the contribution from heat carrying charge carriers. $\kappa_e$ is related to $\sigma$ according to the Wiedemann–Franz law,[7] $\kappa_e = L_0 T\sigma$, where $L_0$ is the Lorenz number ($2.45 \times 10^{-8}$ $V^2K^{-2}$ for metals and degenerate semiconductors).[8] This trade-off between high $\sigma$ and low $\kappa_e$ makes it even more difficult to maximize $zT$.

The conventional processing of advanced semiconducting materials consists of two steps: (A) the synthesis of materials that can be characterized by (a) a certain chemical composition; (b) a physical state, such as crystal structure or microstructure; and (3) specific properties such as electrical, thermal, optical, magnetic, etc.; and (B) materials fabrication: shape forming and shape-fixing by firing/sintering, pyrolysis, melting, or casting. On the other hand, processing routes such as: chemical vapor deposition (CVD), physical vapor deposition (PVD), metalorganic chemical vapor deposition (MOCVD), co-sputtering,[9] co-evaporation,[10] pulsed laser deposition (PLD),[11] molecular-beam epitaxy (MBE),[12-13] electrochemical deposition,[14] atomic layer deposition (ALD),[15] and other methods can produce shaped materials (films, periodic structures) with desired properties in a reduced number of steps via well controlled various deposition parameters. Thin films produced by such methods are small, light-weight and very competitive in the applications of micro-machines, device miniaturization, sensors/devices in science, engineering, technology, and health.

In microelectronics and opto-electronics, with decreasing ICs size and increasing processor frequency, the amount of heat dissipated by components locally (hot spots) is huge, adversely affecting their reliability and life-time unless the dissipated heat is eliminated. The conventional cooling techniques are getting outdated as the compactness of devices is increasing as per Moore’s law, making thermal management very challenging. Miniature thermoelectric cooling (TEC) devices fabricated using TE films are the ideal alternative considering their high cooling efficiency, flexible geometry, and ease of integration with the ICs. With a proper closed loop circuit, the TE cooling modules can control the temperature precisely.

Thermoelectric generator (TEG) can be used to harness the body heat to generate electrical current to power the wearable electronics and health sensors.[16] Figure 1 shows the amount of energy/power consumption of various portable and wearable devices. The power requirement of many healthcare devices and sensors is not very high. The Internet of Things (IoT), which consists of a network of low-power-consuming microsensors/transmitters deployed in a range of settings, might benefit from miniaturized TEGs. The TE thin films can be used to design small chip-sized devices for micro energy harvesting. Under a small temperature gradient, a μ-TEG can power sensor/IoTs enough to transmit periodic or continuous data. Such μ-TEGs have the potential for multifarious applications in healthcare. Similarly, in the industrial environment, TEGs can harvest the ubiquitously available waste heat to power process-monitoring sensors/IoTs to streamline industrial processes remotely. Solar TEG (STEG) have the potential to convert solar energy to electricity.[17] Figure 1b shows the fabrication and testing of TE film, TEC or TEG device assembly. For testing the film with multiple devices, $\Delta T$ is established between the film's top and bottom surface. Film dimensions, TE properties, and operating temperatures are all linked. It is quite challenging to obtaining high output power and cooling heat flux via optimizing various parameters. Figure 1c-e shows some of the film-based TE devices. The reported characteristics of TEGs, TECs, and thermal sensing elements devised using films of different materials are reviewed in the relevant sections and enumerated in Table 2 and Table 3.

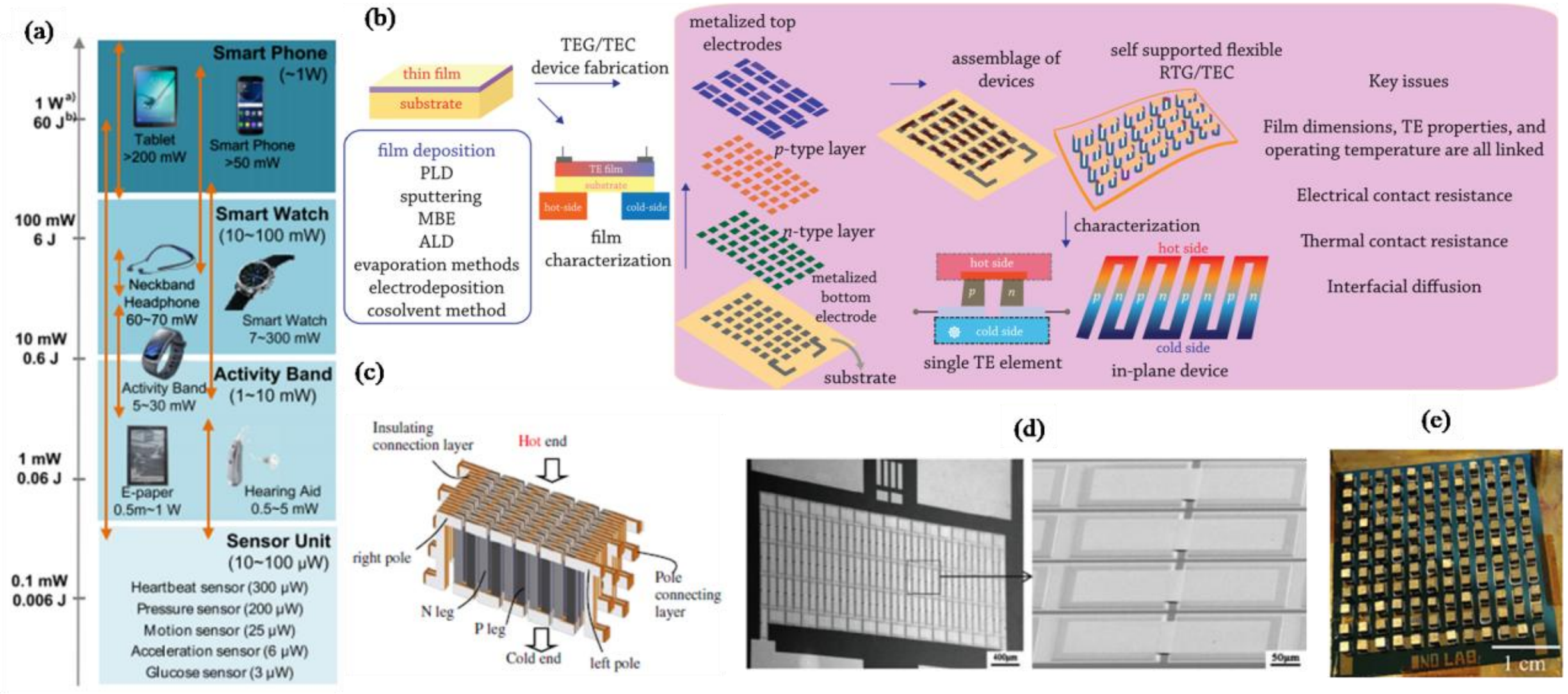

Figure 1: (a) Energy consumption of various portable and wearable devices. The arrows indicate the energy consumption range in different modes for each device.[18] (b) fabrication and testing of TE film, TEC or TEG device assembly; (c) schematic view of the thermoelectric aggregation.[19] (d) SEM micrograph of the TEG with 127 TE pairs and enlarged SEM view of the device.[20](e) *n* and *p* elements obtained via electrochemical deposition method.[21]

TE periodic structures provide freedom to tailor electrical and thermal transport properties. In these structures, the effect of interaction of phonons with interfaces and grain boundaries significantly alters phonon transport. Further, thin-film TE material can change the DOS by inducing a quantum effect, elevating the *S*. Similarly, the magnitude of spin Seebeck effect (SSE) and the dynamics of SSE voltage depends on the interface conditions in spin Seebeck material/heavy metal systems. Concise but key details: heat conduction mechanism and quantum confinement in TE periodic structures; spin Seebeck voltage generation, collection mechanism, and critical elements for longitudinal SSE (LSSE) and transverse SSE (TSSE) configurations are provided at the start of respective sections.

This review has two major components: (a) TE films and periodic structures: Section 2; and (b) spin Seebeck-based materials and devices: Section 4. Section 3 briefly discusses the transverse TE effect, recent reports, and applications. The review specifically focuses on different strategies adapted to obtain high performing TE films of different classes, high transverse Seebeck effect, and high spin Seebeck signals in different material systems. A detailed understanding of the variation of $n$, $\mu$, $S$, $\sigma$, and $\kappa$ with respect to film thickness and temperature is very important as this information ensures the selection of type of material, film thickness, fabrication methods, and fabrication conditions to design and fabricate RTG/TEC. Henceforth, Section 2 will examine recent reports on similar studies and conclude with a summary of the findings. Although a brief account of the intrinsic TE properties of relevant materials are given in advance, readers are urged to refer to some comprehensive articles on $(Bi,Sb)_2(Te,Se)_3$,[22-26] SnSe,[27-29] $Cu_2Se$,[30] and SnTe,[31-32] PbTe,[33-34] GeTe,[35-37] and skutterudites.[38-39] The beginning of the Section 4 covers SSE generation and its usage from the perspective of TE applications and the remainder of that section presents the subtleties of SSE enhancement by examining recent results, albeit not strictly adhering to TE specific features.

## 2. TE periodic films and superlattice structures

### 2.1 Heat conduction mechanism in periodic layers

The following section discusses the phenomenon of thermal phonon conduction in periodic structures by reviewing pertinent theoretical and experimental results. While many of these investigated material systems are not considered favorable for thermoelectrics, experimental work on these systems is the primary source of data for thermal conductivities in periodic structures, which is central to understanding the heat conduction in TE film structures in addition to design principles.

The most important mechanism affecting thermal transport at the nanoscale is the interaction of phonons with interfaces and grain boundaries. Structures wherein these interactions significantly alter the phonon transport are layered/periodic structures such as superlattices (SL) and Quantum well (QW) structures. SLs and QW structures consist of periodic stacking of very thin alternating layers of two different semiconductors that are either lattice matched or lattice mismatched with individual layer thickness ranging from nm to tens of nm (Figure 2c).[40] TE monolayer films and SLs were initially proposed and developed as these structures provide freedom to tailor electrical and thermal transport properties. The process of phonon conduction in periodic structure is dependent on different key length scales: the total size/thickness of the periodic structure *s*; the periodicity of the structure *a*; and interface surface properties (roughness $\eta$ and correlation length $L_c$) (Figure 2a). Apart from these, the phonon length variables such as MFP and wavelength $\lambda$ are important to the conduction phenomenon.[41] A subtle interaction between the effects arising from these various length scales and a wide spectrum of phonons decides the overall heat transport in nanostructures. In periodic structures, $\kappa$ can be largely reduced by phonon scattering at the interfaces (film boundary and grain boundary) or by phonon spectrum modification.[42] Phonons with relatively short wavelengths are scattered diffusely by the SL interfaces and they contribute negligibly to the in-plane $\kappa$ after scattering (Figure 2b) and phonons with large wavelengths are scattered specularly and allowed to transmit and reflect, contributing to heat flux after scattering.[43] SL films are known to exhibit anisotropic TE properties of *S*, $\sigma$, and $\kappa$. Figure 2c shows the $Bi_2Te_3/MoTe_2$ SL structure with a clear line of demarcation between the two layers.

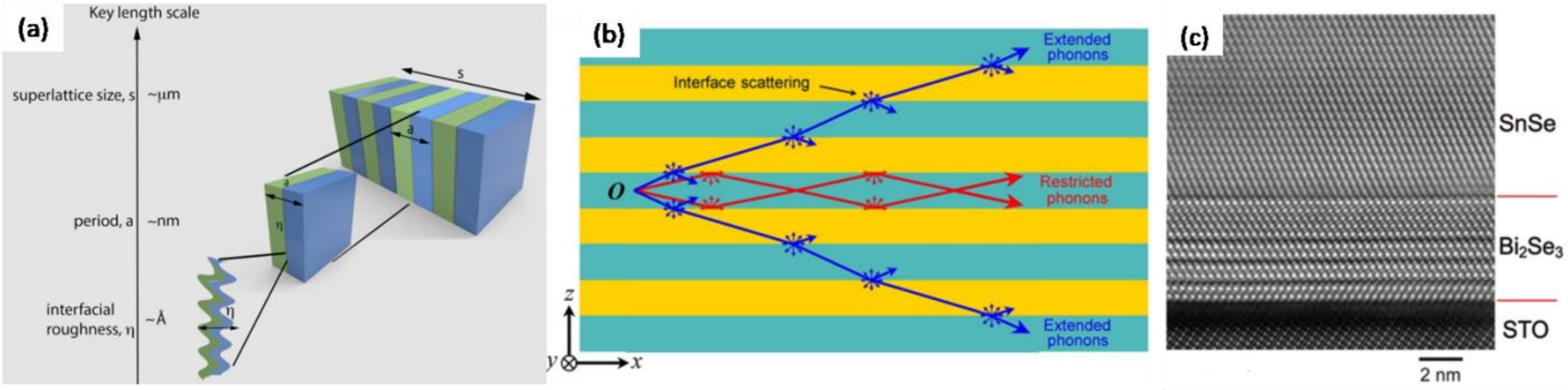

Figure 2: (a) Schematic showing key multiple length scales that control thermal conduction in SL structures. Thermal phonons scatter at rough interfaces (~Å) between each layer within a unit cell of period *a* (~nm) which combine to create a periodic SL nanostructure of total size *s* (~µm). Reproduced with permission from ref.[41] (b) phonons originating at point *O* within the SL can travel within a single layered, i.e. layer-restricted (red) or they can cross multiple interfaces and propagate in different layers, i.e. extended (blue). Reproduced with permission from ref.[44] (c) cross-section HRTEM image of a 16 nm SnSe film deposited on 5 QL $Bi_2Se_3$ film. $SrTiO_3$ (STO) is the substrate. Reproduced with permission from ref.[45]

In SL systems, the periodicity of the SL is known to affect the $\kappa$ in two ways: i) it creates well-defined internal boundaries that increase phonon boundary scattering (i.e., incoherent) or ii) the phonons see the periodicity of the SL as a new material system, resulting in a modified phonon dispersion (i.e., coherent). Theoretical methods to study thermal transport in SLs can be divided into two groups: atomistic (molecular dynamics (MD) and density-functional-theory (DFT)) and continuum (e.g. Boltzmann transport). Using MD, Daly et al. investigated thermal transport in GaAs/AlAs SLs and found that for SLs having smooth interfaces, the cross-plane $\kappa$ shows a minimum as a function of the SL period.[46] In contrast, in Si–Ge SLs with periods of 30 < $L$ < 70 Å, the $\kappa$ decreases with decreasing $L$.[47] Chen et al. reported that in SLs, a minimum in $\kappa$ occurs if the mean free path (MFP) of phonons is close to or > than the period and the lattice constants are similar.[48] According to MD simulations, a minimum of $\kappa$ exists if the SL interfaces are perfect.[46, 48-53] Using ab initio calculations, Chen et al. found that surface segregation and intermixing of atoms may further reduce cross-plane $\kappa$ in Si/Ge SLs.[54] Garg et al. investigated the $\kappa$ of Si/Ge SLs with smooth surfaces and found that, with increasing period, the in-plane $\kappa$ increases monotonically while a minimum was observed for the cross-plane configuration.[55] Liu et al. have evaluated the cross-plane $\kappa$ of $Si/Si_{1-x}Ge_x$ SL thin films by theoretical Boltzmann transport equation (BTE) analysis and experimental study ($3\omega$). The significant reduction in the overall effective $\kappa$ was due to the interfaces scattering.[56] Aksamija et al. have presented a simplified BTE model to calculate the $\kappa$ of $Si_xGe_{1-x}/Si_yGe_{1-y}$ SLs.[57] It has shown that the lifetime of phonons due to scattering with the rough interfaces depends not only on the thickness of each layer and the roughness of the interfaces, but also on the strength of the competing internal scattering mechanisms. This competition between internal and interface scattering leads to a variation in the interface scattering rate between the situation where each layer is a single crystal and the case when the structure comprises alloys. Based on this model, it was proposed that minimal $\kappa$ can be achieved by introducing alloying into both layers in the SL and ensuring that the difference in the alloy concentration of the two layers is large.

Phonon-interface scattering theory considers phonon transport in low-dimensional materials as the combined effects of scattering and transmission at the interfaces.[58-59] When the transmission coefficient is 1 for all the interfaces, phonons can completely transmit the interface without scattering/attenuation, and the $\kappa$ is equal to that of the bulks. However, the transmission coefficient of phonons at the interface is always < 1. The smaller the coefficient, the stronger the interruption of the phonons by the layers. Further, the interfaces are not perfectly specular, which leads to diffusive scattering. Recently, Kothari et al. studied thermal phonon transport in Si/Ge SLs using a Boltzmann transport model considering the factors: phonon coupling between layers, SL periodicity, interface surface conditions, dependence of surface specularity on phonon wavelength, angle of incidence, and roughness.[44]

On the experimental front, there is no dearth of reports on the measurements of $\kappa$ of SLs with different results. Lee et al. showed that the cross-plane $\kappa$ of Si-Ge SLs increased with increasing period length.[47] Capinski et al. measured the cross-plane $\kappa$ of GaAs/AlAs SLs and found that it increases with increasing period length.[49] The cross-plane $\kappa$ of four $Si/Si_{0.7}Ge_{0.3}$ SLs (period thicknesses of 300, 150, 75, and 45 Å) along with three $Si_{0.84}Ge_{0.16}/Si_{0.76}Ge_{0.24}$ SLs (period thicknesses of 100, 150, and 200 Å) were measured over a temperature range of 50 to 320 K. For the $Si/Si_{0.7}Ge_{0.3}$ SLs, the $\kappa$ was found to decrease with a decrease in period thickness and, at a period thickness of 45 Å, it approached the alloy limit. For the $Si_{0.84}Ge_{0.16}/Si_{0.76}Ge_{0.24}$ samples, no dependence on period thickness was observed. The interfacial acoustic impedance mismatch, which is substantially greater for $Si/Si_{0.7}Ge_{0.3}$ than for $Si_{0.84}Ge_{0.16}/Si_{0.76}Ge_{0.24}$, was the factor for the differences in behavior between the two SLs. The $\kappa$ increased slightly up to ~200 K, but was relatively independent of temperature from 200 to 320 K.[60]

Koga et al. proposed a concept called carrier pocket engineering to improve the TE performance of SLs.[61] They designed a dual quantum well structure by collectively tuning the lattice period, thickness of potential barrier and/or potential well, and crystallographic orientation. This dual quantum well structure enabled the two different types of carriers to be selectively confined in the quantum well layer and quantum barrier layer, respectively and allowed the holes and electrons to participate together in the electronic transportation, resulting in enhancement of TE performance. This concept was applied to fabricating GaAs/AlAs[62] and Si/Ge[63] SLs. A phonon-blocking/electron-transmitting structure in multiple-quantum-well $Bi_2Te_3/Sb_2Te_3$ SL was fabricated and the highest $zT$ value was 2.4.[64] The cross-plane $\kappa_l$ measurement (using 3 ω method)[65] on both $Bi_2Te_3$ (1 nm)/$Sb_2Te_3$ (5 nm) and $Bi_2Te_3$ (3 nm)/$Sb_2Te_3$ (3 nm) SLs yielded a $\kappa_l$ of 0.25 W/m-K. A reduction of $\kappa_l$ to 0.3 W/m-K (< 0.75 W/m-K for bulk $Bi_2Te_3$) was also estimated from thermoreflectance measurements[66] for $Bi_2Te_3/Sb_2Te_3$ SLs, with an appreciable reduction for period, $d$ < 6 nm.

Figure 3 exhibits the change in $\kappa_l$ along the cross-plane direction and the MFP of phonons with different periods of $Bi_2Te_3/Sb_2Te_3$ SLs.[65] When the SL period is ~5 nm (50 Å), the $\kappa$ (cross-plane) reaches the minimum value of 0.22 W/m-K (half of the bulk $Bi_2Te_3$-$Sb_2Te_3$ alloys).

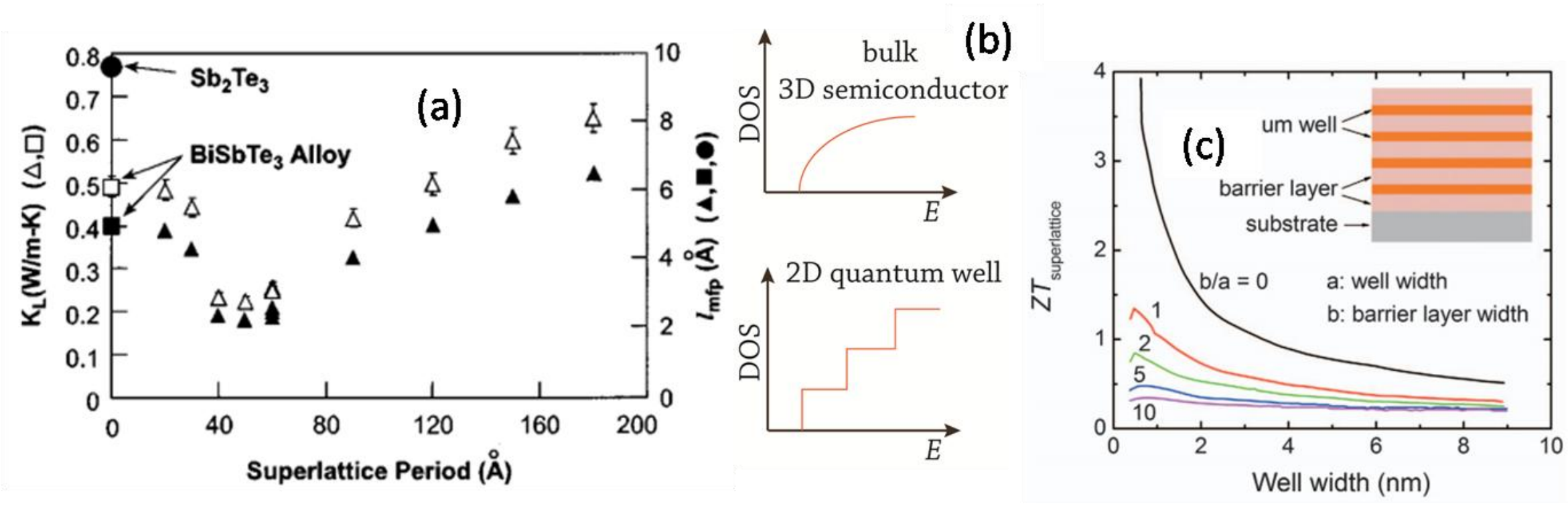

Figure 3: (a) Experimental $\kappa_l$ and calculated avg. phonon MFP as a function of the period in $Bi_2Te_3$ /$Sb_2Te_3$ SLs and other reference materials. Note: There are three data points, almost on top of each other, at the 60 Å period, corresponding to 30 Å/30 Å, 10 Å/50 Å, 20 Å/40 Å structures. Reproduced with permission from ref.[65] (b) when the dimension of the materials decreases and reaches nm scale, distinct changes in the DOS occur; (c) $zT$ of SL composed of QWs of $Bi_2Te_3$ and barrier layers of $Pb_{0.75}Sn_{0.25}Te$ as a function of ratio $b/a$, where $a$ is the thickness of QW and $b$ is the thickness of barrier layer. Adapted with permission from ref.[67]

Further reducing the SL period brings an unanticipated increase of $\kappa$ approaching to the $\kappa_l$ of the bulk $Bi_2Te_3$-$Sb_2Te_3$ material. This behavior can be explained by the coexistence of phonon coherent conduction and interfacial diffuse reflection effects.[43, 55, 68] When the MFP of phonon is < the SL period, the phonon coherence effect will be reduced and the phonon transport obeys the BTE. Then the thermal conductance is decided by the interfacial diffusion/reflection of phonon as well as its relevant impedance of thermal boundary. Therefore, within a certain range, reducing SL period will increase the interface ratio and the phonon collision frequency, resulting in a reduced $\kappa$. However, when the SL period is reduced further to a critical value (< than the MFP of phonons), an increase in the tunneling of low frequency phonons can also be anticipated. The tunneled low-frequency phonons contribute the thermal conductance due to their high heat-carrying ability. When the SL period is reduced even further, the phonons with higher frequencies start to participate in the coherence heat conduction, resulting in a further increase of $\kappa$.

While fabricating the SL films, two key issues should be addressed. First, to keep the crystallographic compatibility between the two layers (matched lattice constants) to prevent the formation of dislocation around the interfaces; second, to keep the chemical compatibility between the layers to prevent interfacial diffusion or reaction.[69]

## 2.2 Quantum confinement effect

When the dimension of materials is in nanometer length scales, the band structure tends to be more flattened as a result of quantum confinement effect and produces sharp density of states (DOS) peak, which would generate enhanced $\sigma$ and $S$ concurrently. Thin-film TE material can change the DOS by inducing quantum effect (Figure 3b).[70-71] Calculations show that for 2D $Bi_2Te_3$ materials, when the size of a QW is < 40 Å, the $zT$ values in all the directions are higher than the bulk. When the size is reduced to 10 Å, $zT$ of the 2D films can approach 1.5 or higher ~2.3. When the size is below the MFP of phonons, the expected $zT$ is even higher (Figure 3c).[71] According to the Mott expression (Equation 2), $S$ depends on the energy derivative of energy-dependent $\sigma$, $\sigma = n(E)q\mu(E)$ close to the Fermi energy $E_F$,[72] with $n(E) = g(E)f(E)$, the carrier density at the energy level $E$ considered; $g(E)$ is the DOS/unit volume and per unit energy; $f(E)$ is the Fermi function; $q$ the carrier charge, $\mu(E)$ the mobility, and the rest have their usual meanings.

$$\boldsymbol{S = \frac{\pi^2 k_B^2 T}{3q}\left\{\frac{d[\ln(\sigma(E))]}{dE}\right\}_{E=E_F} = \frac{\pi^2 k_B^2 T}{3q}\left\{\frac{1}{n}\frac{dn(E)}{d(E)} + \frac{d\mu(E)}{\mu dE}\right\}_{E=E_F}} \quad \text{Equation 2}$$

$$S = \frac{8\pi^2 k_B^2 T}{3qh^2} m_d^* \left(\frac{\pi}{3n}\right)^{2/3} \quad \text{Equation 3}$$

$$\boldsymbol{m_d^* = \left(\frac{g(E)\,\hbar^3 \pi^2}{\sqrt{2E}}\right)^{3/2}} \quad \text{Equation 4}$$

Equation **2** shows there are two pathways to increase $S$ by implementing strategies where, (a) an increased energy-dependence of $\mu(E)$, for example a scattering mechanism that strongly depends on the energy of charge carriers, or (b) an increased energy-dependence of $n(E)$, for example by a local increase in $g(E)$. This concept can also be expressed in terms of density of state effective mass $m_d$* (Equation 3 and Equation 4) as reported for degenerate semiconductors.[72] These strategies have been employed in many nanostructured periodic layers. PbSe nanocrystal SL structure with different nanocrystal sizes exhibited a significant enhancement in $S$ compared with the bulk material.[73] In double-layer transistors based on SWCNT films, tuning the gate voltage resulted in enhanced $S$.[74] Similarly, in the ZnO-based ion-gated transistor, it was observed that the TE performance of a gate-tuned 2D electron gas system formed at the surface of ZnO is significantly higher than that of the bulk sample.[75] A very large $S$ of ~50 × $10^3$ μV/K in Si/Au-doped SL film ($t_f$ = 3000 Å, evaporation method) is speculated to be due to the quantum effect.[76]Theoretical investigation shows that the quantum confinement effect is sensitive to the layer thickness of 2D SnTe. A $zT$ of ~3 is predicted for PbTe/SnTe/PbTe heterostructures at higher temperatures (700 K).[77] Optimal TE performance is observed in monolayer SnTe and bulk like performance for 6 layers.[78] More representative examples are presented in Section 2.9.

## 2.3 Thermoelectric film fabrication methods

### 2.3.1 Pulsed laser deposition (PLD)

PLD is a versatile high growth rate thin-film fabrication technique, which allows us to use multiple elements to produce diverse structures and morphologies of low dimensionality (Figure 4a).[79] The PLD technique provides precise control of the stoichiometry of the films since it can grow a multi-compound film consistent with the intended composition.[11] Despite the ease of experimental set-up, the phenomenon of film growth in PLD is a very complicated process. It entails all physical processes: high power laser and material (target) interaction, instantaneous heating of all elements on the surface of the target to their evaporation temperature, formation of a plasma flume containing ions, atoms, electrons, clusters, molecules, and globules, subsequent transfer of the ablated material to the substrate via plasma plumes, and finally, nucleation and growth of the film on the substrate. The ablation rate is largely dependent on the fluencies of the laser. The fluencies of laser pulses dictate how much laser power density is concentrated on the target region inside each pulse ($J/m^2$). Since the transfer of energy from flumes to the substrate occurs and elevates its temperature, PLD requires a low substrate surface temperature.

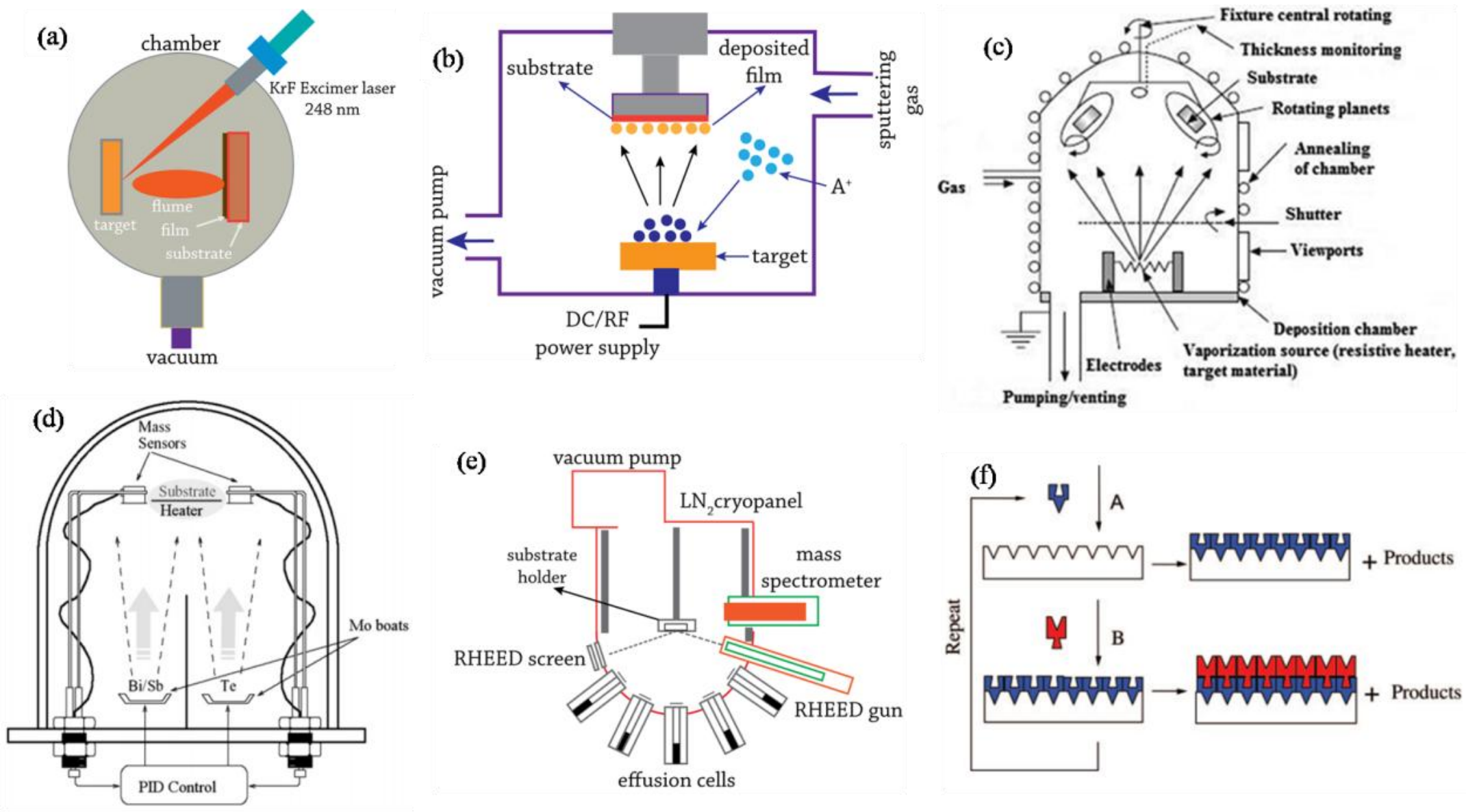

Figure 4: Film fabrication techniques. (a) schematic diagram of PLD; (b) sputtering system; (c) schematics of vacuum evaporation/physical vapor deposition systems with the substrate mounted on a planetary rotation system to give films of uniform thickness. Reproduced with permission from ref.[80] (d) co-evaporation system. Reproduced with permission from ref.[81] (e) schematic diagram of an MBE growth chamber, showing the effusion cells, the substrate stage, and the arrangement of the RHEED system; (f) schematic representation of ALD using self-limiting surface chemistry and an AB binary reaction sequence. Adapted with permission from ref.[82]

### 2.3.2 Sputtering

In the sputtering process, the charged particles bombard the surface of a solid (called target, Figure 4b) and eject the particles/atoms/molecules. These species are finally deposited on a substrate yielding films.[83] The sputtering process is divided into four categories: direct current (DC), alternating current (AC), reactive, and magnetron sputtering processes. Magnetron sputtering is widely used in the preparation of TE films. The films fabricated by magnetron sputtering are dense, uniform, and typically adhere to the substrate strongly owing to good binding force. Furthermore, the TE films can be deposited at relatively low temperatures and the microstructure can be well regulated by adjusting the kinetic energy of bombarding ions. Tellurium, due to its volatility, easily reduces at high sputtering temperatures, negatively impacting the TE property of the materials especially charge carrier concentration in $Bi_2Te_3$ and $Sb_2Te_3$.[84]

### 2.3.3 Vacuum evaporation technique

The general mechanism behind vacuum evaporation technique (commonly called physical vapor deposition technique) is to change the phase of the material from solid phase to vapor phase and convert again to solid phase on a specified substrate under vacuum or controlled atmospheric condition (Figure 4c-d).[80, 85] The vacuum evaporation technique is used in the preparation of TE films, where source evaporation, substrate, and vacuum process play important roles in the optimization of the film performance. Vacuum evaporation technique includes co-evaporation and flash evaporation.

### 2.3.4 Molecular beam epitaxy (MBE)

MBE is a high precision evaporation technique to grow single crystal high quality thin films, SLs, QWs, and complex structures. It can be compared to a more advanced version of vacuum evaporation process. In this growth method, growth of materials takes place under ultra-high-vacuum ($10^{-8}$ - $10^{-12}$ Torr) conditions on a heated crystalline substrate by the interaction of adsorbed species supplied by atomic or molecular beams (from elemental effusion sources, Figure 4e). Because there are no interactions inside or between the element beams, the film growth is influenced only by the beam fluxes and surface reactions, allowing for unprecedented control over thickness, doping concentration, composition, and reproducibility.[12-13] On the negative side, high equipment and maintenance costs, as well as long deposition times, limit the scope of a large-scale application.

### 2.3.5 Atomic layer deposition (ALD)

ALD is a chemical fabrication technique of depositing thin-films on substrates by exposing surfaces of substrates to alternate gaseous species (called precursors, Figure 4f).[15] The TE films obtained via ALD have fewer impurities due to the surface reaction, regulative thickness at angstrom scale, and high quality by the self-regulated reaction on the surface.

### 2.3.6 Electrodeposition

Ever since the first report of the electrodeposition (also called electrochemical deposition) of thin $Bi_2Te_3$ films from acidic baths in 1993,[86] this technique has become very attractive for research activities, precipitating a slew of papers in the years since. Electrodeposition is a process of coating by electrolysis and is analogous to a galvanic cell acting in reverse. The commonly used electrodeposition includes reaction deposition, codeposition, and two-step deposition. Electrodeposition is attractive in the fabrication of TE films for its high deposition rates, low cost, ambient temperature operation, no vacuum requirement, and easy scalability.

### 2.3.7 Cosolvent route to obtain TE films

In principle, one could deposit inorganic semiconductors by directly dissolving them in a solvent, depositing the solution, and drying it to get the required shape. This is one of the most advantageous and efficient ways for the fabrication of large-area thin films with greatly simplified procedures and lower costs. Unfortunately, inorganic semiconductors are generally insoluble due to their strong covalent bonds. The workaround for this hurdle is to create soluble semiconductor precursors that can be transformed into crystalline semiconductors after deposition. The use of hydrazine to create chalcogenidometallate precursors, which can be transformed into crystalline metal chalcogenide semiconductors via mild heat treatment, is one such example.[87] However, hydrazine is highly toxic, explosive, and carcinogenic. Later, it was demonstrated that the binary solvent mixture comprised of 1,2-ethylenediamine (en) and 1,2-ethanedithiol (edtH$_2$) has the excellent ability to rapidly dissolve nine bulk $V_2VI_3$ semiconductors ($V$ = As, Sb, Bi; $VI$ = S, Se, Te) at RT and pressure.[88] This binary solvent approach is particularly appealing because these solvents are much less toxic than hydrazine. This binary solvent approach has since been utilized to make soluble precursors for a wide range of metal chalcogenide semiconductors.[88-91] Since many of the finest TE materials are metal chalcogenides ($Cu_2X$, $Bi_2X_3$, Pb$X$, Sn$X$, etc. where $X$ = S, Se, or Te), a binary solvent approach is a viable solution-phase route to their synthesis.

The cosolvent route can be used to exfoliate and disperse a wide range of 2D layered materials in a mixed solvent of water and alcohols.[92] It's been proven that a mixture of amine and thiol can serve as a general solvent to dissolve a large number of inorganic semiconductors: $Cu_2S$, $Cu_2Se$, $In_2S_3$, $In_2Se_3$, CdS, SnSe, and more. This method allows for the production of high-concentration semiconductor ink (> 200 mg/mL) at RT and pressure, which may then be utilized to uniformly create semiconductor thin films on $SiO_2$/Si wafers, glass, and plastic substrates. Electrical transport studies of the prepared $Cu_2S$ and $Cu_2Se$ thin films show the highest $\sigma$ compared with other solution methods. Besides, the cosolvent approach can also be employed to process many other important materials: $CuIn(S_xSe_{1-x})_2$ ($0 \leq x \leq 1$), SnS, CdSe, ZnSe, and $MoS_2$.[91] More such results are discussed in Section 2.10.

## 2.4 Thin film configurations for TE applications

Based on the heat flow direction, two types of thin film geometries can be identified for TE applications (**Figure 5**a, top image). The planar geometry allows us to obtain high temperature gradient ($\nabla T$) (provided the substrate and film sample have not very high $\kappa$) but shows a small current section (film width × film thickness) and delivers a high voltage in an open circuit ($E_{oc}$) but with low power output. The primary downside this kind of geometry is the high electrical resistance which decreases the electrical power generation. The cross-plane geometry (it is akin to the conventional TE module geometry, **Figure 5**a, lower image) has a small $\nabla T$ (along a step of some 100 nm). The main issue with this configuration is the small $\nabla T$, which results in a low $E_{oc}$ (i.e. low electrical power output).[93]

To achieve high powers, the resistance must be as low as possible (optimal film thickness and nanostructuring) and the $\nabla T$ must be as large as feasible. However, in the current scenario, having both is challenging. As a consequence, TE thin films (modules) are not well suited for high electrical power output, but can be used for microelectronic power production (µW) or temperature sensors. In the case of films fabricated with compounds with free carrier transport, the film thickness has a significant impact on $\sigma$. In the case of thin film structure, for which the film thickness ($t_f$) is in the range order of the carrier mean-free-path (MFP), a $t_f$ decreasing below MFP leads to a lower $\sigma$ because of the carrier scattering at the interface and the surface, which reduces the relaxation time. Therefore, the $t_f$ must be optimized as a function of the carrier MFP in this case. Further, nanostructures and microstructure of films influence the $\sigma$ when the grain sizes are comparable to the carrier MFP. Then it is beneficial to have a value of the carrier MFP different from those of the phonon wavelengths in order to exclusively diffuse only the phonons without impacting the carrier transport.[93]

## 2.5 TEG and TEC design consideration

Apart from the TE material properties ($S$, $\rho$, $\kappa$), the electrical and thermal contacts, and the number and geometry of the TE elements influence the device performance. The working mechanism of a TE device in power generation and cooling mode can be found in ref.[1, 3] The Seebeck voltage produced in a device under a temperature gradient $\Delta T$ is, $V = S\,\Delta T$, which is equal to the sum of the internal voltage drop and the voltage drop on the load. The maximum output power, $P_{max}$ is attained when the internal resistance ($r_i$) of the TEG is equal to the resistance of the load ($r_L$), i.e. $P_{max} = (S\cdot\Delta T)^2/4R$. The output power ($P_o$) per unit area ($A$) is

$$\frac{\boldsymbol{P_o}}{\boldsymbol{A}} = \frac{\boldsymbol{\varepsilon}}{(\boldsymbol{\varepsilon}+\mathbf{1})^2}\frac{\boldsymbol{S^2\Delta T^2}}{(\boldsymbol{A_n}+\boldsymbol{A_p})\left(\frac{\boldsymbol{\rho_n l_n}}{\boldsymbol{A_n}}+\frac{\boldsymbol{\rho_p l_p}}{\boldsymbol{A_p}}\right)} \quad \text{Equation 5}$$

Where $\varepsilon = r_L/r_i$, the subscripts $n$ and $p$ refer to the type of TE leg ($n$ or $p$-type), $l$ is the leg length, $\rho$ is the resistivity, and $A$ is the cross sectional area. Consequently, optimized ratio of length to cross-sectional area is critical to obtain high power output.

In cooling mode, while the maximum temperature difference, $\Delta T_{max}$ is largely determined by the $zT$ of the device (i.e. $\Delta T_{max} = (zT_c)^2/2$, taking into consideration the effects of the electrical and thermal contacts), the maximum heat flow, $Q_{max}$, and the maximum current, $I_{max}$, depend strongly on the size and number of the TE elements ($N$) in the TE cooling module. It has been shown that

$$Q_{max} = \frac{f}{l}\frac{(S_p - S_n)^2 T_c{}^2}{4(\rho_p + \rho_n)} \quad \text{Equation 6}$$

$$\frac{I_{max}}{A} = \frac{fST_c}{Nl\,2\rho} \quad \text{Equation 7}$$

where $f$ is the fraction of area the covered by TE elements, $l$ is the thickness of the TE element and, $p$ and $n$ represent the type of materials, and $\rho$ is the electrical resistivity, and $T_c$ is the temperature of the cold side of the TE module.[94] Therefore, the cooling heat flux ($Q/A$) is $\propto f/l$. Control of $Q/A$ can best be achieved by tuning the $l$, or $f$ rather than the materials properties. For very high $Q/A$, $l$ needs to be reduced since the maximum value of $f$ is l. Further, $l$ also controls the speed of the TE cooler.[95] Hence, small $l$ devices, fabricated via thin-film techniques have high heat flux capability and are very fast. Since the heat is only transferred from the cold to the hot side of the TE cooler, the heat transfer distance is also $l$. Thus, long (i.e., thicker), narrow TE elements, can transfer heat a further distance, but only with low heat fluxes.[96] The best cooling performance is achieved when then the current is approximately $\approx I_{max}$**.** From Equation **6**, it is clear that $Q_{max}$ can be increased by reducing the TE element thickness, which is limited for two reasons. First, difficulties associated with TE elements film fabrication. Second, the electrical constant resistance can have a significant influence on the performance of thin film thermoelectric modules because the scale of the contact resistance itself can be equal to that of the thermoelectric element. With such limitations, the obtained cooling flux is insufficient for the required thermal management systems, especially when the device size gets small.[97-98]

As the thickness of a thin film is typically microns, current moves through the TE leg of an out-plane device (where the current travels perpendicular to the film surface) in a short route with a large cross-sectional area, resulting in a large maximum cooling flux (Equation 6) and low internal resistance. For in-plane devices (where the current travels parallel to the film surface), however, it is the opposite. High cooling flux is critical for a cooler and low internal resistance lowers the heat generated inside the device. Consequently, out-plane structures have been commonly used for TECs.

## 2.6 Thin film TE property characterization techniques

A considerable number of publications on TE films mention the usage of custom-made set-up to evaluate electrical and thermal properties. Therefore various measurement techniques for $\sigma$, $S$, and $\kappa$ were developed. As a result, there is no single consistent measurement method for all kinds of TE thin films.[99-102] The $\sigma$ for thin films can be measured with a four-point probe instrument (Figure 5b). As shown in Figure 5b, four collinear contact points with identical spacing are made on the sample surface. A current, $I$ is feed through the outer contacts (1 and 4) and the resulting voltage drop $V$ is measured across the inner electrodes (3 and 4). If the sample size >> the probe spacing and the thickness of the film ($d$) < ½ of contact spacing, the $\sigma$ of the film is by $\sigma = \frac{ln 2}{\pi d}\frac{I}{V}$. This method can minimize any contact resistance during the measurement when the thickness of the film is very small. For in-plane $\sigma$ and $S$ measurements, the substrate needs to be non-conducting and cross-plane measurement requires specially fabricated device to confine the current and to reduce contact resistance (i.e. minimize the film thickness). This method is not suited for integrated semiconductor layers. Because of very small dimensions, it is quite difficult to place four contacts on top of a conducting layer.

The van der Pauw method (**Figure 5**c) is a very convenient four-probe technique to measure the $\sigma$ of thin film samples of random shape.[103] To use this technique, there are five conditions that must be met:[104] 1**.** The sample must have a flat shape of uniform thickness; 2. No isolated holes in the sample; 3. The sample must be homogeneous and isotropic; 4. All four ohmic contacts must be located at the edges of the sample; 5. Individual ohmic contact must be infinitely small, i.e., should have an area of contact that is at least an order of magnitude less than the total sample area. Further, thickness must be << than the width and length of the sample. One measurement of $\sigma$ by this method consists of two resistance measurements: $r_{ver}$ and $r_{hor}$. For one direction of current, for instance, from contact 1 to 2 (**Figure 5**c) and the voltage is measurement between 3 and 4 to obtain the first resistance: $r_{12,34} = V_{34}/I_{12}$. For accuracy, the contact points are swapped in terms of applied $I$ and measured $V$ to obtain $r_{34,12} = V_{12}/I_{34}$. $r_{\mathrm{ver}}$ is the average of $r_{12,34}$ and $r_{34,12}$. The $r_{\mathrm{hor}}$ is obtained in a similar fashion. A current is passed from contact 1 to the adjacent contact 4, and the voltage between 2 and 3 is measured to get $r_{14,23} = V_{23}/I_{14}$. And again $r_{23,14} = V_{14}/I_{23}$ is also obtained. $r_{\mathrm{hor}}$ is the average of $r_{14,23}$ and $r_{23,14}$. To remove parasitic voltages that may be present during the measurement, it is often necessary to reverse the current polarity and test various voltages created by varied magnitudes of current. Then according to Van der Pauw's theorem for an isotropic conductor[103]:

$$exp^{\left(\frac{\pi d}{\rho} r_{hor}\right)} + exp^{\left(\frac{\pi d}{\rho} r_{ver}\right)} = 1 \quad \text{Equation 8}$$

Where $d$ and $\rho$ are the smaple thickness and resistivity, respectively. If $d$, $r_{\mathrm{hor}}$, and $r_{\mathrm{ver}}$ are known, $\rho$ can be obtained via Equation **8**. For a rotationally symmetric sample, $r_{\mathrm{hor}} \approx r_{\mathrm{ver}}$, in that case $\rho = \pi d(r_{\mathrm{hor}} + r_{\mathrm{ver}})/2{\cdot}\ln 2$. To reduce the measurement error, a circular cloverleaf sample geometry (where the diameter of the inner area is ¼ of the whole sample, Figure **5**c-lower image) is preferred.

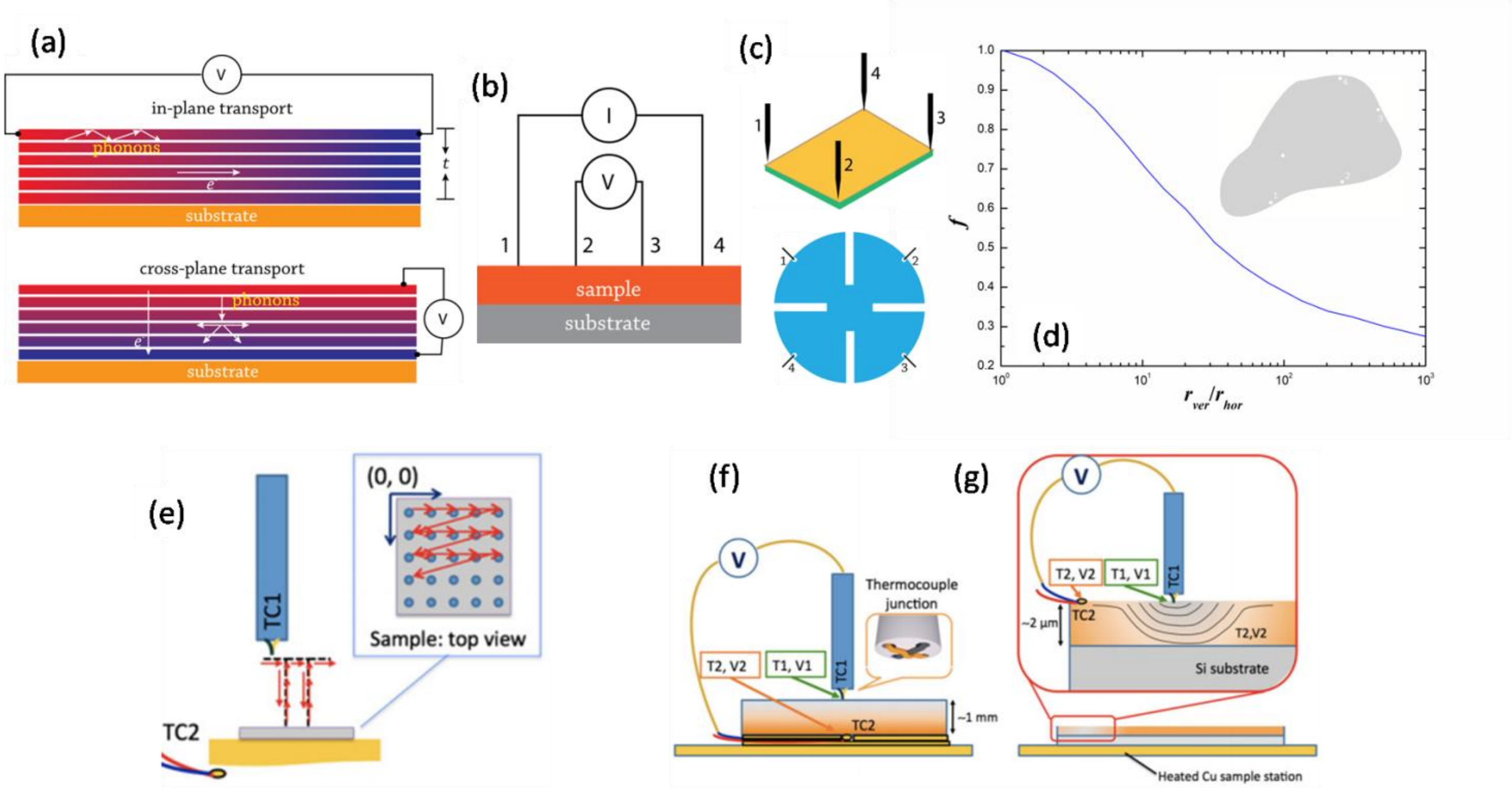

Figure 5: (a) Thin films configurations; (b-c) four probe configurations for $\sigma$ measurement of thin films; (b) conventional collinear four-point probe for a large sample; (c) conventional probing geometry for the van der Pauw method. Lower image: circular cloverleaf geometry; (d) $r$ vs. $f$ in Equation 12. Inset shows sample of arbitrary shape; (e-g) The ''cold tip'' scanning Seebeck coefficient measurement system. (e) imitated sample test process; (f) schematic of the setup for bulk and (g) thin-film samples. Reproduced with permission from ref.[105]

However, in the general case of arbitrary shape, Equation 8 is equivalent to Equation 9

$$\rho = \frac{\pi d}{2ln2} \frac{r_{ver} + r_{hor}}{2} f(r)$$ Equation 9

$$cosh\left(\frac{r-1}{r+1}\frac{ln2}{f}\right) = \frac{1}{2}\exp\left(\frac{ln2}{f}\right)$$ Equation 10

Where $f$ is called geometric factor, which can be determined numerically or graphically.[106-109] $f$ is the only function of $r = \frac{r_{ver}}{r_{hor}}$, which satisfies the Equation 10 and graphically represented in Figure 5d. Thus to determine $\rho$, we need to have $r$ value first, then get the corresponding $f$ value from the graph and calculate $\rho$ from Equation **9**.

A scanning Seebeck measurement system, which can directly detect local variations of $S$ on thin film sample is shown in (**Figure 5**e-g).[105] The distinctive feature of this measurement system is that it does not require substrate removal, which means that the homogeneity of the film sample deposited on an insulating substrate (or multilayer film insulated from a conducting substrate) can be directly characterized.

The cross-plane $\kappa$ and in-plane $\kappa$ have been measured by the 3ω and time-domain thermoreflectance measurements (TDTR) methods and are reported.[110-111] In 3ω, in-plane method needs suspended or supported films.[101] Also, when it comes to measuring the $\kappa$ of anisotropic thin films, the 3ω technique offers a lot of room for improvement, especially in the in-plane direction. 2ω method based on a thermoreflectance signal analysis which enables the $\kappa$ measurement of high thermal resistance film is also reported.[112] Some instruments are designed to simultaneously measure the TE properties and $zT$ of a TE device directly in power generation mode (called Z-meter).[113-115] These instruments are often used to measure thick TE samples under a large $\Delta T$, but they may also be used for thin films provided the electrical and thermal related parasitic effects are minimized.[114]

### 2.7 Effect of substrate types, annealing temperature (and time) on grain size and orientation

It should be noted that in PLD, MBE, sputtering, and thermal evaporation techniques, substrate temperature ($T_{sub}$) affects the amalgamation kinetics and redistribution of dopants, site incorporation (for amphoteric dopants), re-evaporation, surface segregation, and precipitation (at high doping densities), the surface morphology, crystalline growth, the abruptness of doping transitions or the relaxation processes in heterostructure systems.[116-117] The choice of substrate is very important as it allows controlling the deposition temperature. The coefficient of thermal expansion (CTE) of both films and substrate should match to reduce residual stress and improves film-to-substrate adhesion. For example, for amorphous glass, optical quality glass (1.5 mm thick

N-BK7), poly-crystalline alumina, single crystalline sapphire, polyimide (PI), and of $Bi_2T_3$, the CTE are 3.2, 7.1, 7.2, 7.7., 27, and 16.8 (× $10^{-6}$/K) respectively.[118-119] Linear CTE of other TE are reported[120]: $Bi_2Te_{2.8}Se_{0.2}$ (14.85 × $10^{-6}$/K) and $Bi_{0.5}Sb_{1.5}Te_3$ (14.98 × $10^{-6}$/K) both in 200-400 K range; $Bi_2Te_{2.4}Se_{0.6}$ and $Bi_{0.4}Sb_{1.6}Te_3$ TEC weakly changed in 380-600 K and was in the range of (13.93-14.33) × $10^{-6}$/K; PbTe (23.07 × $10^{-6}$/K); GeTe (24.47 × $10^{-6}$/K); nanostructurad $Si_{0.8}Ge_{0.2}$ (4.68 × $10^{-6}$/K) in 300-1200 K range.

CTE of single-layer graphene (SLG) and multi-layer graphene (MLG) thin films floating on a water surface (to avoid substrate effect) was measured and found to be negative, i.e., -5.8 × $10^{-6}$/K and -0.4 × $10^{-6}$/K for SLG and MLG, respectively in the 297-320 K range, indicating in-plane contraction of the films.[121] The transverse vibration of carbon atoms is the reason for thermal contraction.[122-124] In-plane TEC of $MoS_2$ is reported to be 2.48 × $10^{-6}$/K at RT[125]; 0.5 × $10^{-6}$/K at RT.[126]

In comparison to traditional epitaxy, the lack of covalent bonding between the film and the substrate allows for more lattice matching and prevents strain build-up in the films.[127] For example, van der Waals epitaxy (vdWE) of 2D layered materials on a dangling bond-free 2D substrates. Graphene offers several practical benefits for vdW epitaxy, including crystalline ordering, mechanical durability, chemical inactivity, and thermal stability,[128-130] and hence has been used as a substrate to grow BST alloy.[131] One major downside of flexible/PI substrate is the limitations on the film processing temperature, as it cannot withstand very high temperatures and its shrinkage factor is 0.25% at 573 K. Added is the complexity of anisotropic CTEs along the lateral (9 × $10^{-6}$/K) and vertical (110 × $10^{-6}$/K) directions of PI, attributed to the in-plane orientation of the molecular chains in PI.[132] Experimentally, it has been shown that in SnSe, at RT, the TEC along crystallographic *c*-axis is highly negative and it increases with increasing temperature.[133-134] In the same material, numerical study shows that at high temperatures the linear TEC along *a* direction (2.76 × $10^{-5}$/K) < along *b* direction (3.10 × $10^{-5}$/K), and along the *c* direction it is highly negative (-1.97 × $10^{-5}$/K); and volumetric TEC is reported as 3.88 × $10^{-5}$/K.[135] Thus, depending on the film we intend to deposit, substrates with appropriate properties need to be chosen. Another route to suppress the problems (compressive stress) related to TEC mismatch between the film and the substrate is the introduction of an interlayer (having a CTE that is intermediate between the deposited film and the substrate) on the substrate prior to the film deposition. The interlayer acts as a compensatory compliant layer, reducing stress produced by CTE mismatch.[136] Alucone films, for example, were added as interlayers (via molecular layer deposition) to relieve stress caused by TEC mismatch between $Al_2O_3$ films (via ALD) and Teflon fluorinated ethylene propylene (FEP) substrates.[137] $BaF_2$ and $CaF_2$ have well-matched TECs of ~19 × $10^{-6}$/K (close to the value of PbSe) and have been used in the MBE growth of PbSe.[138] In the deposited assembly, (PbSe/$BaF_2$/$CaF_2$/Si(111), $BaF_2$/$CaF_2$ buffer layers played an important role in the high-quality epitaxy of PbSe on Si substrates. The PbSe epilayer was recovered by dissolving $BaF_2$ in deionized water.

The lattice mismatch between the substrate and the depositing film is another aspect to consider while fabricating films. For example, $Bi_2Te_3$ and $Sb_2Te_3$ grown layer by layer via MBE under ideal conditions (following the Frank van der Merwe growth model) can yield potentially perfectly layered films. However, threading dislocations arise when there is a 2.8% lattice mismatch between the substrate and $Sb_2Te_3$ and between $Bi_2Te_3$ and $Sb_2Te_3$. This mismatch can be as large as 22% for GaAs or as small as 2.7% for $BaF_2$, resulting in SLs with different degrees of dislocations.[139] In many cases, the film and substrate assembly (without stand-alone film recovery) is used for electrode bonding and device making. In such cases, the lattice mismatch plays an important role as well; it can cause strain to develop between the layers, causing crystallographic defects to form, resulting in high contact resistance or macroscopic defects like cracking, buckling, and even de-lamination.[140] Further, substrates should have suitable mechanical properties to enable the integration of such substrates with many types of devices. The lattice mismatch $\Delta a/a_{sub}$ is given by:[141] $\Delta a/a_{sub} = (a\text{-}a_{sub}/\ a_{sub}) \times 100\%$. Where $a_{sub}$ is the lattice constant of the substrate. When $|\Delta a/a_{sub}| < 5\%$, the interface between the film and the substrate is completely coherent (providing good adhesion of the film to the substrate). For $Bi_{0.5}Sb_{1.5}Te_3$/Cu combination, $|\Delta a/a_{sub}| \approx 2.4\%$. In a study AlN/PI/Cu film combination has been used as a substrate to deposit (110) oriented $Bi_{0.5}Sb_{1.5}Te_3$.[142] In this case, Cu layer was in direct contact with the film layer. However, there are exceptions. It was shown that MBE growth of $Ge_2Sb_2Te_5$ on Si(111) is possible despite a large lattice-mismatch (~+11%). The possible reason is $Ge_2Sb_2Te_5$(111) epitaxial layers on Si(111) are shown to reversibly switch from the amorphous to the crystalline state during the deposition process, re-establishing the initial epitaxial relationship to the template.[143] Similarly, MBE growth of GeTe thin films on highly lattice-mismatched Si(111) substrate (8% lattice misfit) is reported.[144] Another way to remedy the lattice mismatch problem, a matching layer or a buffer layer can be introduced, as in the case of GeTe/$Sb_2Te_3$/Si(111) system, where the $Sb_2Te_3$ layer function as a matching layer to PLD of GeTe epitaxial layer.[145]

Metallic materials such as Pt, Au, or stainless steel can be utilized as seed layers or substrates for electrochemical deposition. If the deposited film needs to be removed from the substrate for further characterization, stainless steel has the benefit of having a very low adhesion between the film and the substrate.[146]

In $Bi_2Te_3$ the $\sigma$ along the *a,b*-axis is approximately three times larger than that along the *c*-axis.[147] Therefore, it is critical to increase the crystal orientation of $Bi_2Te_3$ to improve its TE properties. Tan et al. have shown that (00l)-textured $Sb_2Te_3$ film performed better, with carrier concentration and $\mu$ ~half and 4.5 times of an ordinary film.[148] Carbon nanotubes (as a bundle of networks) can function as a 3D flexible deposition substrate and enable TE film/CNT network or mesh assembly (hybrid structure) a freestanding structure, which can be transferable to any target substrate for TE device integration. Further, as observed in the instance of $Bi_2Te_3$/SWNCT hybrid structure, a crystallographic connection may exist between the CNT bundle axis and crystal alignment on CNT,[149] which can be utilized to control the crystal orientation.

The size of the crystallite also has a significant impact on the TE performance. When the crystallite size is smaller than  phonon MFP, phonons are scattered at the crystal boundaries, lowering the $\kappa$.[150] In contrast, when the crystallite size is large, $\sigma$ increases because charge carriers can easily pass through crystal boundaries. Post-deposition annealing temperatures ($T_{an}$) also affect the TE properties significantly. Thermal annealing is the most convenient approach to control the crystal orientation and crystallite size by adjusting the annealing temperature and time as many studies show discussed in the subsequent section.

### 2.8 $Bi_2(Te,Se)_3$ and $(Bi,Sb)_2Te_3$ based TE films

In PLD, the important deposition parameters such as substrate temperature ($T_{sub}$), partial pressure, and laser fluence affect the chemical composition (defect formation) and the crystalline phase of the deposited films.[151] It is very important to maintain the stoichiometry of the reaction within the solubility limits of the Bi/Sb−Te phase diagrams to attain high $zT$ values in $Bi_2Te_3$ films.[10] Chemical composition and the degree of crystallinity of the films decide the $\sigma$ and $S$.

Symeou et al. have reported the annealing temperature effect on $Bi_{0.5}Sb_{1.5}Te_3$ films deposited on silica and Kapton substrates via PLD. Thin films of $Bi_{0.5}Sb_{1.5}Te_3$ were first prepared by RT PLD and subjected to annealing in vacuum.[152] The as-prepared films were amorphous but the annealing process enhanced their crystallinity and texture. Thin films of $Bi_{0.5}Sb_{1.5}Te_3$ grown on fused silica and annealed at 350 °C for 16 h showed a $\sigma$ of 869.5/Ω-cm, $S$ ~210 μV/K, and $PF$ ~37.5 μW/cm-K$^2$ at 380 K; $Bi_{0.5}Sb_{1.5}Te_3$ films grown on Kapton and annealed at 250 °C for 5 h displayed 26 μW/cm-K$^2$ at 390 K.[152]

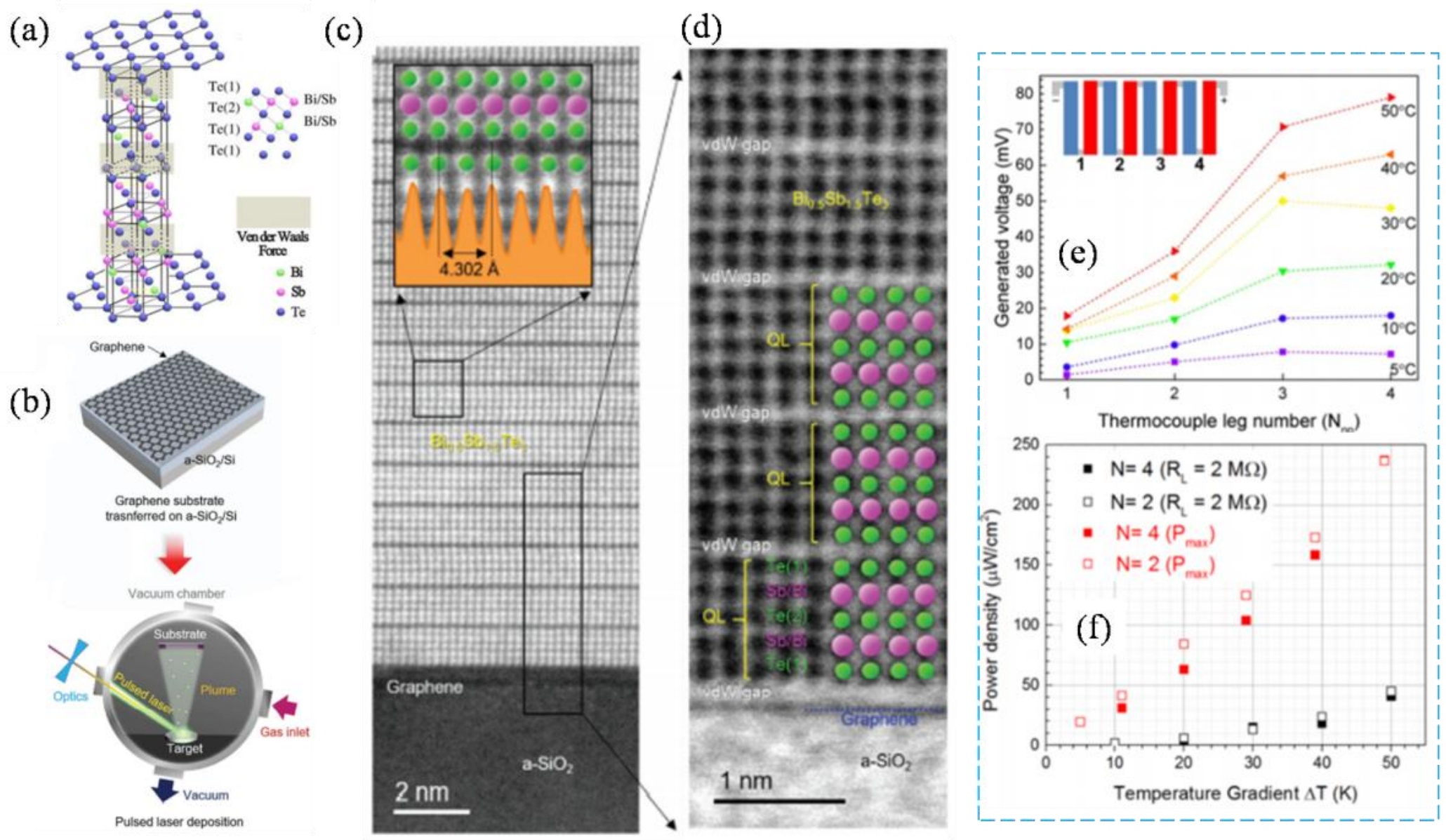

Figure 6: (a) Schematic illustration of the (*00l*) oriented $Bi_{0.5}Sb_{1.5}Te_3$ layered structure. Reproduced with permission from ref.[153] b) graphene substrate after the chemical removal of residual PMMA; lower image: deposition of BST thin film on graphene/a-$SiO_2$/Si via PLD; (c) cross-sectional HAADF-STEM image of vdWE BST film; (d) BF-STEM image for the same region of (c) showing a clear line of demarcation between graphene and $SiO_2$. Box in (c) shows line profiles of intensity traces for atomic columns along [010] and [001]. Adapted with permission from ref.[131] (e) Seebeck voltage vs. connected thermocouple leg number (N) generated at different Δ*T*s; (f) power density vs. different Δ*T*s. Reproduced with permission from ref.[154]

Graphene as a substrate can promote epitaxial growth of 2D layered materials via vdW epitaxy. Further, wafer-scale single-crystalline graphene is already accessible, and it can be readily transferred to any substrate. In comparison to BST QL grown on an a-$SiO_2$/Si substrate, BST QL deposited on graphene exhibited better performance despite a considerable lattice mismatch (-3.15%) with BST.[131] First, graphene was synthesized on a Cu foil via a thermal CVD method and then transferred onto a-$SiO_2$(300 nm)/Si. Using $Bi_{0.5}Sb_{1.5}Te_3$ as a target in the PLD, BST films were deposited on graphene/a-$SiO_2$/Si and a-$SiO_2$/Si (Figure 6b). In addition to well-arranged atomic stacking along [00l] of QLs of BST, a clear line of demarcation between $SiO_2$/Si and graphene was also observed (Figure 6c-d). In terms of TE properties, BST grown on graphene fared better in comparison to the film deposited on a-$SiO_2$/Si. As a result, a *PF* of ≈ 4.68 mW/m-K$^2$ (equivalent to *PF* of single crystal) was observed in the vdW epitaxy BST/graphene film.

$Sb_2Te_3$ and $Bi_2Te_3$ thin films showing *S* of 624 and -78 μV/K respectively were prepared by PLD.[154] These films deposited on glass substrates had smooth surfaces with small crystal grains and blurry boundaries. The avg. grain sizes of 9.85 and 29.33 nm were observed for $Sb_2Te_3$ and $Bi_2Te_3$ films, respectively. In this method, films were prepared at different $T_{sub}$ and post-annealing treatment free. Defects formed during the PLD growth enhanced the charge carrier concentration, (1-7) × $10^{21}$/cm$^3$. TE power generator consisting of four pairs of *n*-type $Bi_2Te_3$ and *p*-type $Sb_2Te_3$ legs connected in series generated a maximum voltage of 50 mV and a power density of ~120 μW/cm$^2$ at $\Delta T$ of 30 K (Figure 6e-f). Nanostructured $Bi_2Te_3$ (*n*-type) thin films were deposited on $SiO_2$/Si(100) substrates by PLD. The $Bi_2Te_3$ films with high (001) orientation and a layered structure exhibited *PFs* of 18.2-24.3 μW/cm-K$^2$ when the substrate temperature ($T_{sub}$) was between 220 and 340 °C.[155] Bassi et al., have applied PLD technique to prepare $Bi_2Te_3$ films on Si or cleaved mica with different morphologies at the micrometer/nanometer scale.[156] Films with a layered $Bi_2Te_3$ structure showed the best properties, with *S* ranging from -175 to -250 μV/K, $\sigma$ ~ 714-666/Ω-cm, $\mu$ = 100 cm$^2$/V-s, $\kappa$ ~ 1.6-1.7 W/m-K, and *PF* in the range of 20-45 μW/cm-K$^2$ with an expected $zT$ > 1.5. Further, it was found that the higher deposition temperatures (> 620 K) favor the hexagonal $Bi_2Te_3$(001) formation. Using $Bi_2Se_2Te$, a single crystal target, Le et al. prepared *n*-type nanocrystalline Bi–Se–Te ($Bi_3Se_2Te$) thin films on ($SiO_2$/Si) via PLD and obtained high *n* of ~$10^{20}$ cm$^{-3}$, $\sigma$ of 1747.5/Ω-cm at RT, *S* of -68.8 μV/K at RT, and optimal *PF* of 8.3 μW/cm-K$^2$ at RT.[157] Pulsed laser deposited (on $SiO_2$/Si) *n*-type $Bi_2Te_{2.7}Se_{0.3}$ and $Bi_2Te_3$ showed RT *PF*s of 3179 μW/m-K$^2$ and 2765 μW/m-K$^2$ respectively. For the same samples, RT $\kappa$ were 1.05 W/m-K and 1.28 W/m-K. The RT $zT$ of the $Bi_2Te_{2.7}Se_{0.3}$ was 0.92.[158] Different types of highly oriented and twinned $Bi_xSb_{2-x}Te_3$ nanocrystal films were fabricated by a large-area PLD technique on insulated silicon substrates at various deposition temperatures from 673 K to 883 K.[159] The thermally induced spontaneous growth of highly oriented and twinned nanoassemblies, not only helped in modifying carrier transport across interfaces, but also aided in effective phonon scattering. The presence of the nonbasal- and basal-plane twins across the hexagonal BiSbTe nanocrystals evidently contributed to the high $\sigma$ of ~2700/Ω-cm and the *PF* of ~25 μW/cm-$K^2$ as well as the relatively low $\kappa$ of ~1.1 W/m-K in the nanostructured films.[159]

$Bi_2Te_3$ films ($t_f$ = 1 μ) have been deposited (via DC magnetron sputtering) on four types of substrates: glass, alumina, sapphire, and polyimide and subjected to three heat treatment processes: non-substrate heating (condition A), substrate heating during film deposition (condition B), and thermal annealing after film deposition (condition C).[160] Among the heat treatment conditions, the most obvious variation in the films due to the substrate types emerged in condition C. The $Bi_2Te_3$ film deposited on the glass substrate had an almost perfect crystal orientation (with a relatively large crystallite size of 50 nm, Figure 7n), demonstrating a high TE performance with a *PF* of 27.3 μW/cm-$K^2$ and estimated $zT$ = 0.70. The $Bi_2Te_3$ film deposited on the polyimide, on the other hand, did not have the preferred crystal orientation and had a small crystallite size of 20 nm, resulting in a dismal low TE performance, with *PF* = 6.6 μW/cm-$K^2$ and estimated $zT$ = 0.34. For the films deposited on polyimide, the lattice constants along both axes were lower than the corresponding standard values owing to the high coefficient of thermal expansion ($27 \times 10^{-6}$/K) and shrinkage of polyimide (Figure 7o). The differences in the crystallize size, morphology, and orientation were explained on the basis of changes in the length of the $Bi_2Te_3$ unit cell due to thermal expansion (lattice match/mismatch of film and substrates), solid-phase growth, and the high oxygen composition of the film (Figure 7a-l).

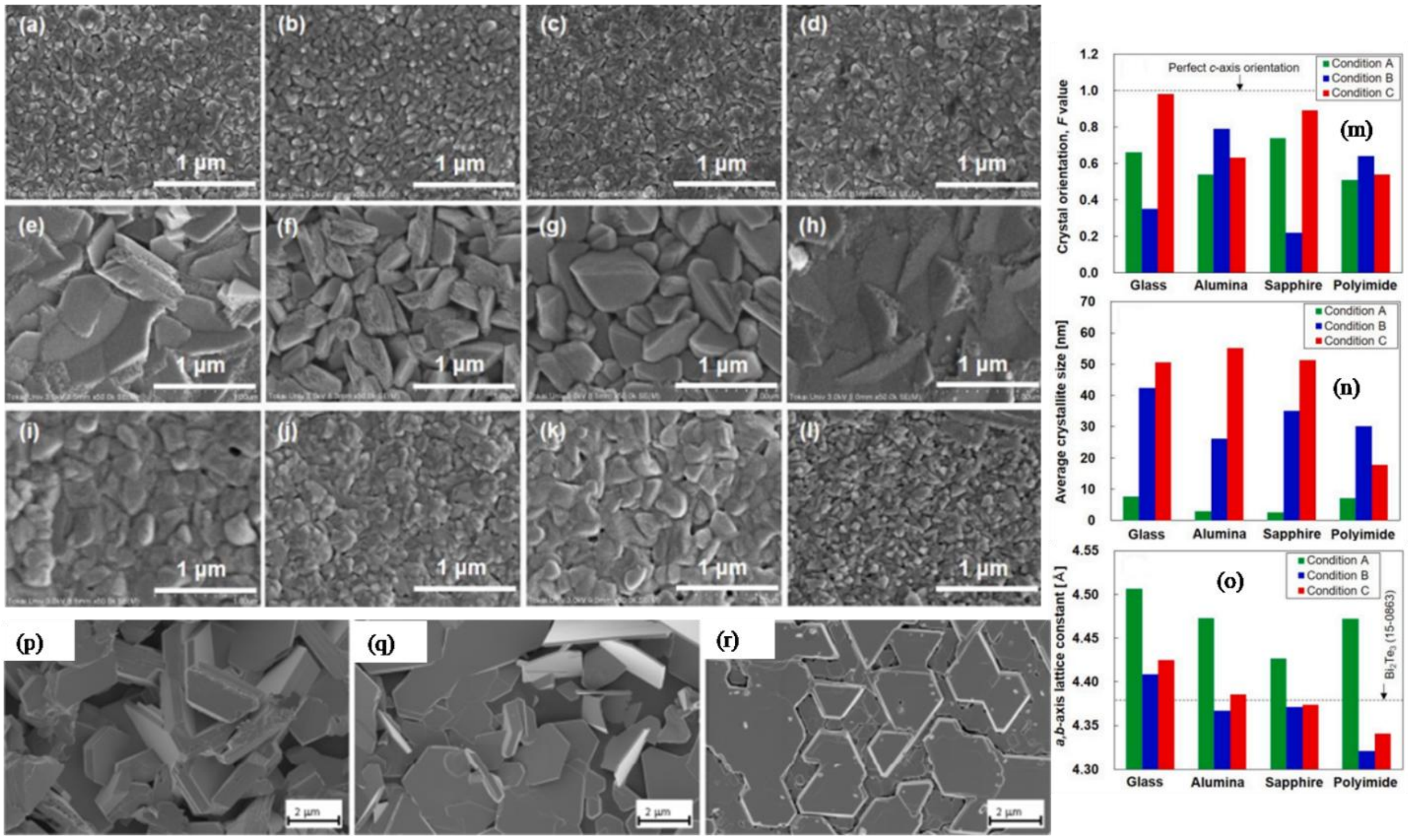

Figure 7: SEM images showing the surface morphologies of the $Bi_2Te_3$ thin films. Images (a-d) show morphologies of films under condition A deposited on glass, alumina, sapphire, and polyimide, respectively. Images (e-h) indicate the same under condition B. Images (i-l) show the same under condition C. (m) crystal orientation *F*; (n) crystallite size; (o) *a*,*b*-axis lattice constant. Adapted with permission from ref.[160] (p-r) scanning electron micrographs of $Bi_2Te_3$ films grown on different substrates at two different magnifications: (p) $Bi_2Te_3$/Si (100), (q) $Bi_2Te_3$/Si (111) and (r) $Bi_2Te_3$/sapphire (100). Adapted with permission from ref.[161]

Diaz et al. have studied the substrate effect on $Bi_2Te_3$ crystal orientation.[161] Figure 7p-r show the SEM micrographs of the $Bi_2Te_3$ layers deposited (via vapor deposition) on Si(100), Si (111), and sapphire (001). The *a*-plane (100) orientation of films produced on Si appears to be random, but those developed on sapphire are aligned along defined directions. This is to be anticipated, given Si(100) has a square symmetry, but sapphire (001), Si(111), and $Bi_2Te_3$ have hexagonal symmetry. The grain size of the crystals were found to be 339.7, 87.3 and 141.7 nm for layers deposited on Si(100), Si(111) and sapphire, respectively.

Working pressure during DC magnetron sputtering can affect the microstructural, elemental composition, and crystal structure of the $Bi_2Te_3$ films. Somdock et al. obtained $Bi_2Te_3$ thin films (1 μm) having a stoichiometric composition and highly (00l) orientation structures by using the proper working pressure of $1.8 \times 10^{-2}$ mbar with pre-heating the substrate at 350 $^{o}$C.[162] The growth mechanism was thought to involve the formation of (00l) seed layer and the diffusion of sputtered particles. The substrate was pre-heated sufficiently to promote (00l) orientation seed layer, which later acted as a template to induce preferential direction growth of (00l) orientation. Optimum pressure and pre-heating conditions resulted in enhanced carrier concentration ($7.87 \times 10^{20}/cm^3$) and $\mu$ (118.0 $cm^2$/V-s) leading to enhanced $\sigma$ ($14.90 \times 10^3$/Ω-cm) in *a*-*b* plane.

Park and coworkers studied the out-of-plane *S*, $\kappa$, and $\sigma$ of *p*-type $Bi_2Te_3/Bi_{0.5}Sb_{1.5}Te_3$ (BT/BST) SL films and *p*-BST film in the temperature range of 77-500 K.[163] The films ($t_f$ ~200 nm) were deposited on $SiO_2$/Si(001) using a radio-frequency (RF) magnetron sputtering system and then characterized in the out-of-plane direction. The $\sigma$ of the BT/BST SL films decreased from ~1084 to ~854/Ω-cm as the temperature increased from 77 to 500 K, which is ~4-5.5 times larger than the value for the BST thin film. The *S* values of the BST film and BT/BST SL film were determined to be 33-133 and 33-123 μV/K in the temperature range of 77-500 K, respectively. As a result, the maximum *PF*s of the BST and the BT/BST SL films were 3.3 and 14.4 μW/cm-$K^2$, respectively. The $\kappa$ values for the same films were measured to be ~0.26-0.58 and 0.34-0.62 W/m-K, respectively in the 77-500 K range. Due to a synergistic combination of the energy filtering effect (due to band alignment between the *n*-BT and *p*-BST in SL structure, Figure 8d) and low interfacial resistance of the BT/BST SL structure, a high *zT* of 1.44 was obtained at 400 K for the 200 nm thick *p*-type film (43% higher compared to the pristine *p*-BST films of the same thickness).

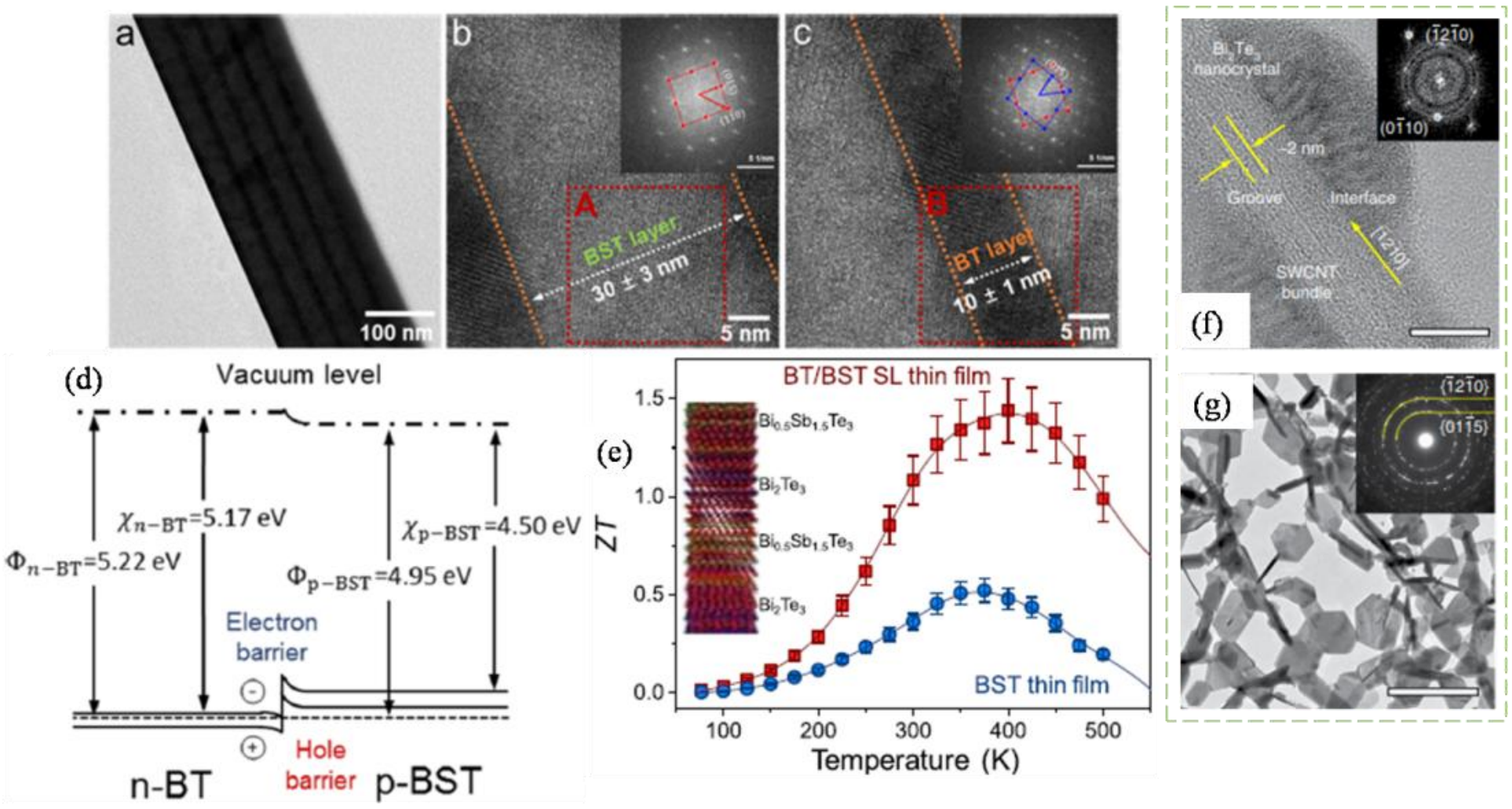

Figure 8: Structural properties and SL configuration of the *p*-BT/BST SL thin film. (a) cross-sectional TEM image of the *p*-BT/BST thin film; (b, c) enlarged HR-TEM images of the BST (bright contrast) and BT (dark contrast) layers. Insets (in b, c) show the FFT patterns of the BST

and BT layers, respectively; (d) energy band diagram of the *p*-BT/BST SL film; (e) out-of-plane *zT* for the BST film and the BT/BST SL film in the temperature range of 77−500 K. Adopted with permission from ref.[163] (f) clear and sharp interface between $Bi_2Te_3$ nanocrystals and the SWCNT bundle. Inset: FFT pattern of the $Bi_2Te_3$. Scale bar: 10 nm; (g) a hybrid material obtained after ~120 s deposition. Inset: corresponding SAD pattern. Scale bar: 400 nm. Adapted with permission from ref.[149]

The *n*-type $Bi_2Te_3$ film ($t_f$ = ~1 μm) prepared by RF and DC co-sputtering (with Te and Bi as targets) on BK7 glass and then annealed at 400 °C showed $\sigma$ of 3.14 × $10^4$ to 5.51 × $10^4$ /Ω-cm and *S* of 49-177 μV/K, culminating in a maximum *PF* of 3.28 × $10^{-3}$ W/m-$K^2$ at 275 °C.[164] Following a simple magnetron co-sputtering method, a highly (00*l*)-oriented $Bi_2Te_3$ thin films ($t_f$ = tens of nm) with in-plane layered columnar nanostructures were fabricated on quartz.[100] The *PF* had reached 33.7 μW/cm-$K^2$ (with max. $\sigma$ = 840/Ω-cm, and *S* = -200 μV/K) and the corss-plane $\kappa$ was reported to be 0.86 W/m-K at 300 K. Bottner et al. fabricated *n*-type $Bi_2Te_3$ and *p*-type $(Bi,Sb)_2Te_3$ thin films on metalized Si substrates by sputtering. *PF* of 30 μW/cm-$K^2$ (for *n*-type) and 40 μW/cm-$K^2$ (for *p*-type) were achieved using optimum annealing conditions.[165] Similarly, the *PF* of the ion beam sputtered *n*-type $Bi_xSb_{2-x}Te_3$ film ($t_f$ = 710 nm) was improved upon post-deposition-annealing at $T_{an}$ = 400 °C.[166] The *PF* was nearly quadrupled at $T_{an}$ = 200 °C, reaching 3.1 μW/cm-$K^2$. Magnetron co-sputtering ($Bi_2Te_3$ and Te as targets) was used to assemble layer-structured $Bi_2Te_3$ on a CNT scaffold to create a flexible (*000l*)-textured $Bi_2Te_3$–SWCNT hybrid film ($t_f$ = 600 nm).[149] This film exhibited a *PF* of 16 μW/cm-$K^2$ at 300 K and 11 μW/cm-$K^2$ at 473 K. A *zT* value of 0.89 was obtained with the in-plane $\kappa_l$ of ~0.26 W/m-K at 300 K. Clean and clear $Bi_2Te_3$/SWCNT interface was observed between the $Bi_2Te_3$ nanocrystals and the SWCNTs (Figure 8f). In the hybrid structure, crystal orientation, interfaces, and nanopore structures contributed to improving the TE performance.

$Bi_xTe_y$ thin films deposited on PI substrate (via RF magnetron sputtering) and annealed under $N_2$ at temperatures from 250 to 400 °C have shown improved crystallinity, hardness, and TE properties.[167] The annealing temperature enhanced the crystallinity and film density for the temperature range of 250-300 °C (Figure 9a-d). However, due to re-evaporation of Te, the crystal structure of $Bi_2Te_3$ almost converted to BiTe after annealing the films above 350 °C. The maximum hardness of 2.30 GPa was observed by annealing the films at 300 °C. The highest *PF* of 11.45 μW/cm-$K^2$ at 300 °C with the carrier concentration and $\mu$ of 6.15 × $10^{20}$/$cm^3$ and 34 $cm^2$/V-s, respectively, were achieved for the films annealed at 400 °C.

The effects of hydrogen addition during RF magnetron sputtering on the surface morphology, crystal structure, elemental composition, and TE properties of the $Bi_2Te_3$ thin films ($t_f$ = 0.7-1 μm) have been reported.[168] It was found that the ratio, Te/(Bi + Te) of the film decreased with increasing mixing ratio (MR) $H_2/(H_2 + Ar)$, indicating that Te atoms evaporated from the film surface by a chemical reaction between $H_2$ and Te. The oxygen concentration inside the films decreased as the $H_2/(H_2 + Ar)$ ratio increased, resulting in an increased *PF* of 9.0 μW/cm-K$^2$ at a MR of 10%. For pure Ar, a rough surface with densely arranged irregular crystal grains with sharp edges (Figure 9e), for 10% MR (Figure 9f), densely arranged isotropic granular grains (size 100 nm) with voids, and for MR of 15%, slightly smaller rounded crystal grains were observed. These changes in the surface morphology indicate that the surface atoms, especially those located at the edges of the grains, evaporated when the proportion of $H_2$ gas was increased.

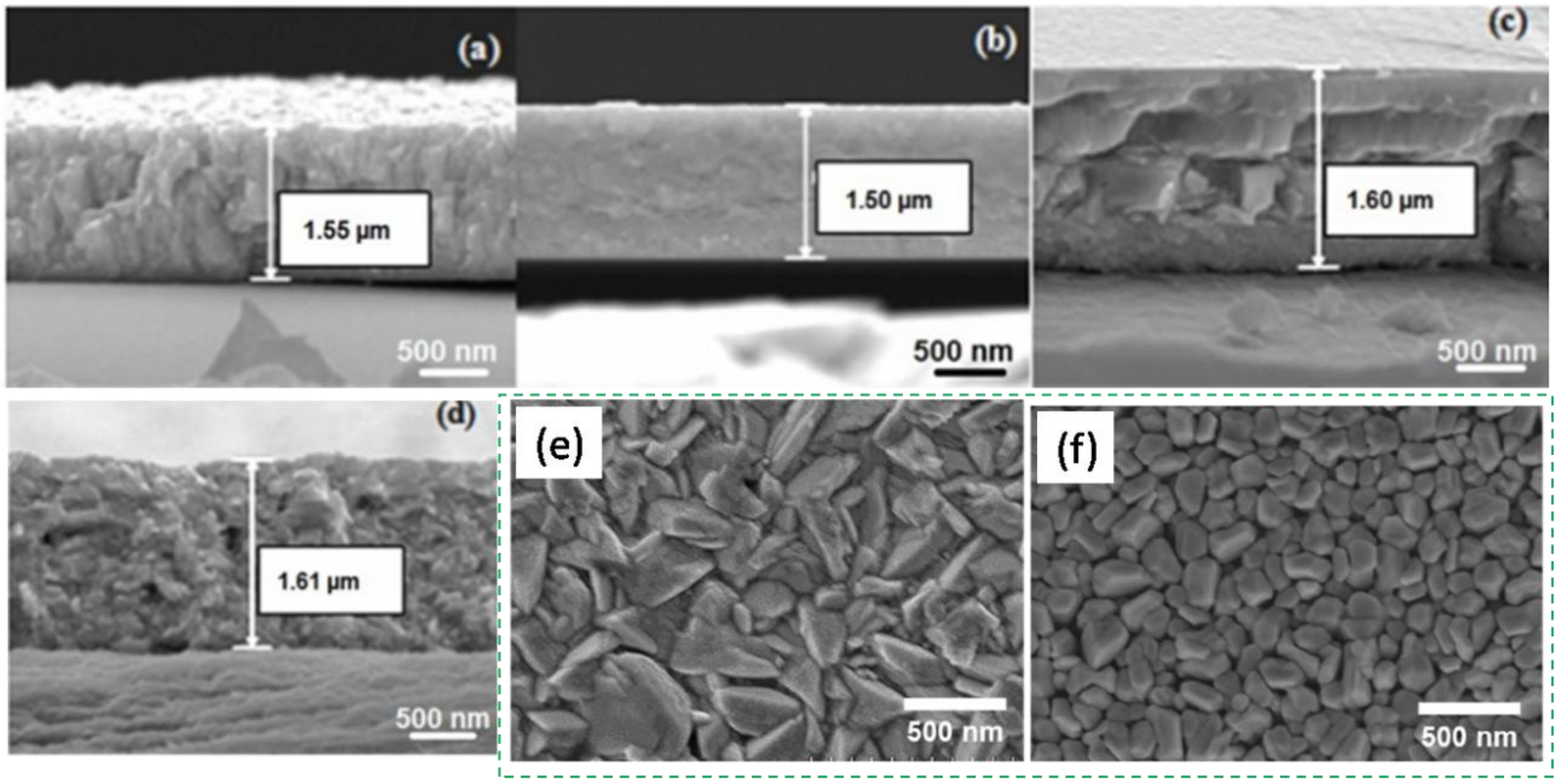

Figure 9: $Bi_xTe_y$ thin films deposited on PI. cross-section of (a) as-deposited and annealed at different temperatures: (b) 250 °C, (c) 300 °C, (d) 350 °C. Adapted with permission from ref.[167] SEM images showing the surface morphologies of the $Bi_2Te_3$ thin films for different $H_2/(H_2 + Ar)$ ratio. (a) 0%; (b) 10%. Reproduced with permission from ref.[168]

A 10 μm-thick $Bi_2Te_3$ film deposited on a PI substrate by DC magnetron sputtering and annealed at 250 °C has shown a *PF* of 13.5 × 10$^{-4}$ W/m-K$^2$ due to high $\sigma$ and high *S*; annealing improved the crystalline structure, boosted carrier mobility, and lowered the carrier concentration.[84] The proposed mechanism of defect generation during annealing is shown in Figure 10b. The evaporation of Te generates holes, and the carrier concentration of $Bi_2Te_3$ films would be reduced by the compensation of holes. The output power of single-leg, thick $Bi_2Te_3$ film annealed at 250 °C was 0.98 μW at a $\Delta T$ = 50 °C (Figure 10c-d).

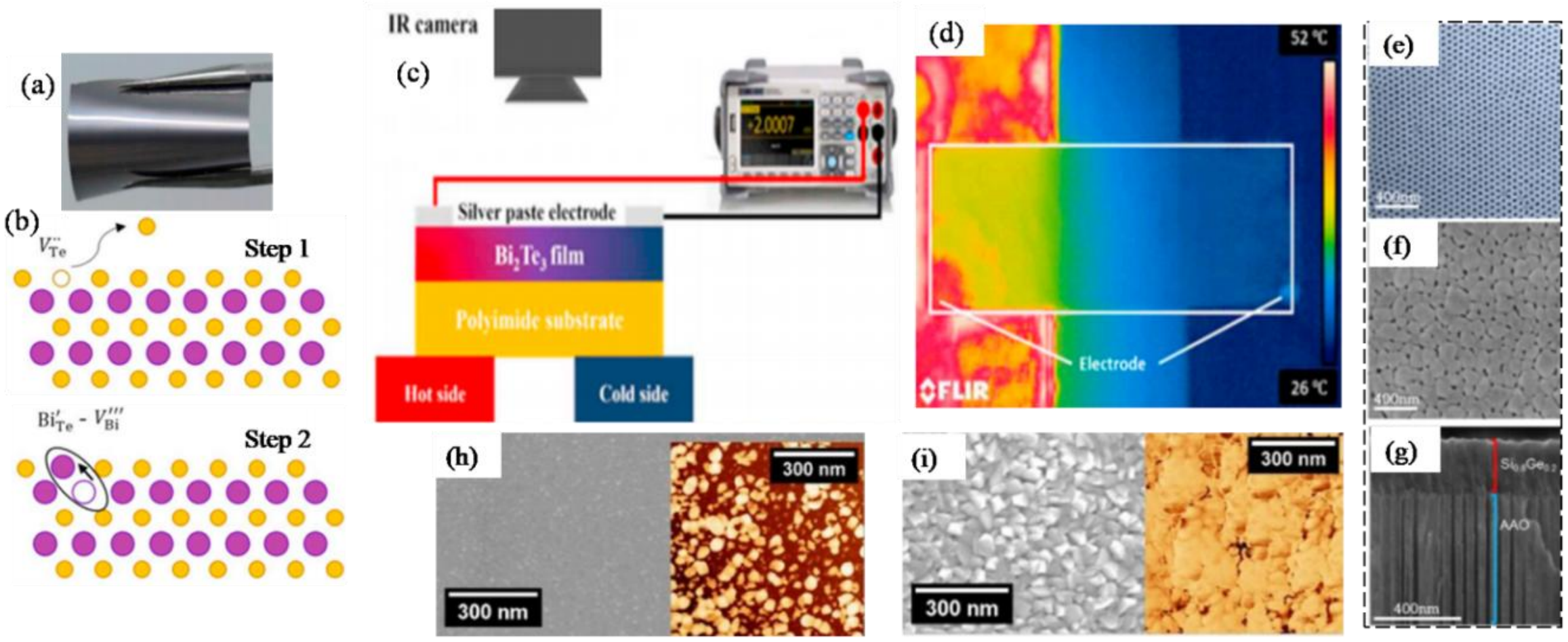

Figure 10: (a) Thick $Bi_2Te_3$ film deposited on a flexible substrate; (b) schematics of defect generation in $Bi_2Te_3$ atomic structure from the annealing process. In step 1, $V_{Te}^{\cdot\cdot}$ is formed by the volatile of Te atoms on surface. In step 2, a $Bi_{Te}'$-$V_{Bi}'''$ pair is formed by the migration of Bi atom from an adjacent Bi site; (c) characterization of the power output of a single-leg thick $Bi_2Te_3$ film; (d) IR top-view image of $Bi_2Te_3$ film taken during characterization. Reproduced with permission from ref.[84] Right most frame (e-g): SEM images of (e) porous alumina substrate template with 31 ± 4 nm diameter; (f) sputtered $Si_{0.8}Ge_{0.2}$ nano-meshed film; (g) lateral view of these sample, where the $Si_{0.8}Ge_{0.2}$ films and the alumina matrix can be observed. Reproduced with permission from ref.[169] SEM images and AFM phase images of (h) 1 min and (i) 5 min Bi-Te films. Reproduced with permission from ref.[170]

Large-area $Si_{0.8}Ge_{0.2}$ nanomeshed (nanoporous or holey silicon or SiGe membrane) films were fabricated on highly oriented porous alumina matrices (with pore diameters: 436 ± 16 nm to 31 ± 4 nm) via DC sputtering of $Si_{0.8}Ge_{0.2}$.[169] The as-formed $Si_{0.8}Ge_{0.2}$ films replicated the porous alumina structure, yielding the nanomeshed films (Figure 10e-g). The $\kappa$ of the films was reduced as the diameter of the pores became smaller, resulting in $\kappa$ = 1.54 W/m-K (294 ± 5 nm nanopore film), down to a low $\kappa$ = 0.55 W/m-K (31 ± 4 nm nanopore film). The plausible reason for the drop in $\kappa$ is ascribed to an alloying effect and phonon boundary scattering on the upper/lower boundaries within the nanomeshes. The *PF* of the $Si_{0.8}Ge_{0.2}$ films varied from 445 μW/m-$K^2$ (for 294 ± 5 nm pore nano-mesh) to ~65 μW/m-$K^2$ (for 31 ± 4 nm pore nano-mesh) and the highest *zT* for these nano-mesh was < 0.1 at RT. A series of Bi -Te films of different thickness (few nm to 370 nm) were deposited on polymer substrates at RT via DC magnetron sputtering.[170] An island growth mode was observed and there was a phase of film growth where the layer just partially coated the substrate, with only random connections between the islands (Figure 10h-i). In this partially covered region, the coating exhibited an extremely high *S*. The *PF* (5.72 μW/cm-K$^2$) increased significantly in very thin films attributed to a quasi-decoupling of *S* and $\sigma$. Such RT sputtering on a polymer substrate is very suitable for large area devices.

The $Bi_{0.5}Sb_{1.5}Te_3$ film containing interfaces among $Bi_{0.5}Sb_{1.5}Te_3$, Te, and $Sb_2Te_3$ exhibited optimized *PF* and $\kappa$ values of 23.2 μW/cm-K$^2$ and 0.8 W/m-K, respectively, at 300 K.[171] The film was prepared through the magnetron co-sputtering method using $Bi_{0.5}Sb_{1.5}Te_3$ and Te as targets. A prototype TEG fabricated using this *p*-type $Bi_{0.5}Sb_{1.5}Te_3$-based heterostructure film and a normal *n*-type $Bi_2Te_3$ film yielded a power density of ~897.8 μW/cm$^2$ at a $\Delta T$ = 40 K (Figure **11**a-b). Similarly, the $Bi_2Te_3$ film ($t_f$ = tens of μm) deposited on cellulose fiber (CF) using a magnetron co-sputtering method exhibited *PF*s of 2.50 to 3.75 μW/cm-K$^2$ from RT to 473 K and a *zT* of 0.38 at 473 K.[172] The enhanced TE performance is attributed to significantly reduced $\kappa_l$ (0.27 at 330 K; ~0.5 at 480 K) owing to the strong phonon scattering effect and enhanced *PF*. 12 pairs of *n*- and *p*-type TE legs on the double sides of the CFs-paper sheet constituted the prototype of the flexible TE generator, which showed $V_{out}$ = 0.144 V at $\Delta T$ = 50 K.

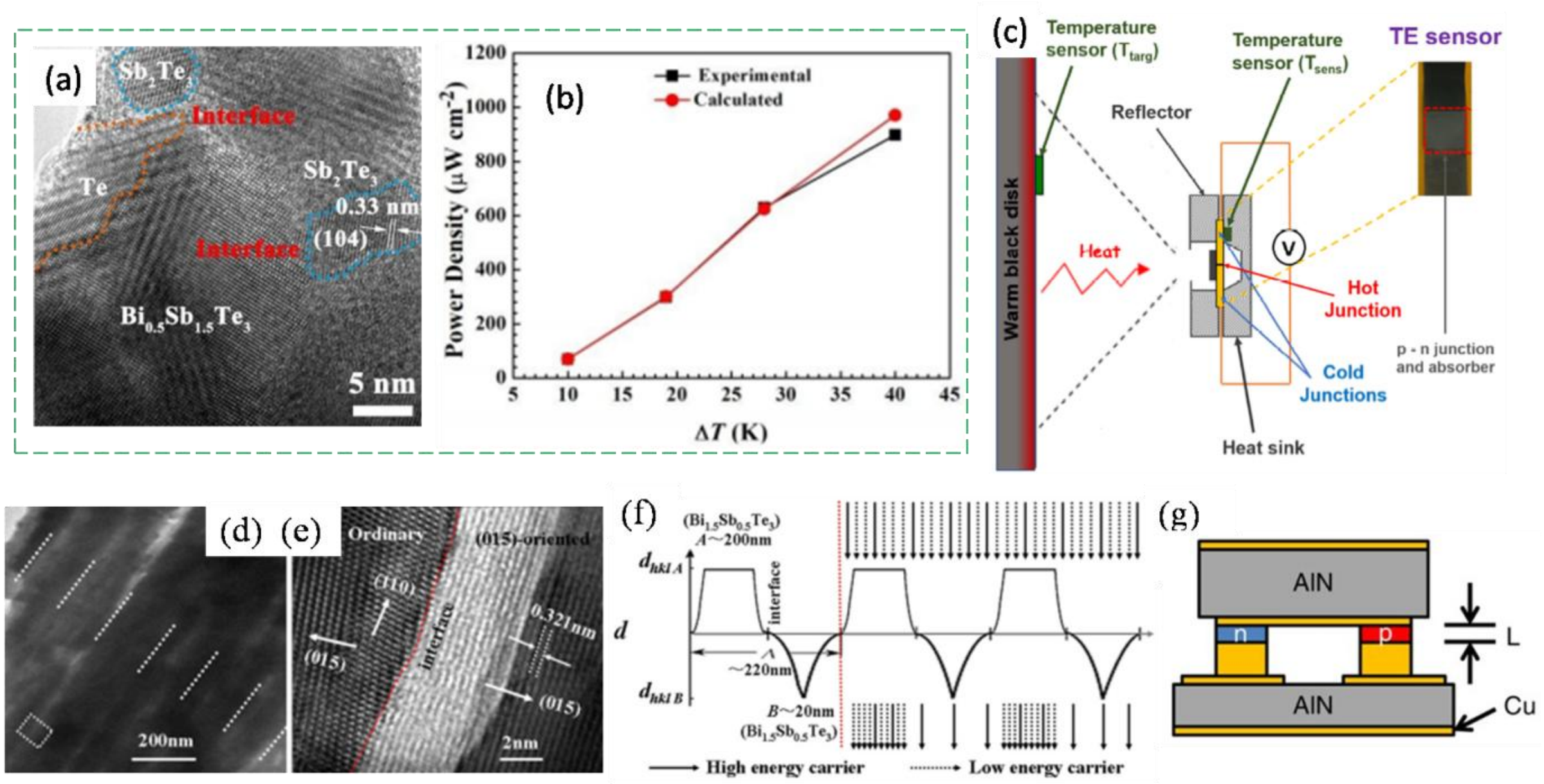

Figure 11: (a) HRTEM image showing interfaces in $Bi_{0.5}Sb_{1.5}Te_3$-film; (b) calculated and experimental electrical power density tested at various $\Delta T$. Adapted with permission from ref.[171] (c) scheme of operation of the *p-n* TE device used as radiation sensor. Adapted with permission from ref.[173] (d) TEM image of $Bi_{1.5}Sb_{0.5}Te_3$ multilayered film; (e) HRTEM image of selected area marked by the square in (d) (The ordinary nanolayer structure marked by the dotted lines, as shown in (d), and the (015)-oriented thin-layers located between the dotted lines); (f) the in-plane transport mechanism of the (015)- preferential $Bi_{1.5}Sb_{0.5}Te_3$ multilayered film with alternating

stress field. Reproduced with permission from ref.[174] (g) illustration of thin-film-based TE module, showing top and bottom AlN headers, Cu traces and *n*-type and *p*-type active elements. *L* represents the length of the active elements (8 μm in this case). Reproduced with permission from ref.[175]

Thermal co-evaporation method allows independent control of temperature and therefore the partial vapor pressures of components such as Bi and Te. Goncalves and coworkers have studied the influence of deposition parameters (evaporation rates of Bi and Te and $T_{sub}$s) to establish a correlation between the growth conditions and the TE properties of the $Bi_2Te_3$ films.[10] It was shown that films with high *PF*s are obtained when their composition is 35-40% Bi, 65-60% Te. Highly uniform $Bi_2Te_3$ thin films were prepared through the two-step thermal vapor process with a single evaporation source. A *zT* value of ~0.51 was achieved in the sample at 400 K.[176] *p*-$Sb_2Te_3$ and *n*-$Bi_2Te_3$ films ($t_f$ = 400 nm) have been deposited on borosilicate glass and Kapton PI substrate via thermal co-evaporation.[173] The $\sigma/S$ for *p*- and *n*-type films were (320/Ω-cm)/194 μV/K at 298 K and (500/Ω-cm)/-233 μV/K at 298 K respectively. The *p*- and *n*-type films exhibited *PF*s of 23 μW/cm-$K^2$ and 59 μW/cm-$K^2$ respectively at 373 K. A flexible thermopile based on *p-n* films was demonstrated as a possible thermal sensing element. This device showed a responsivity of 0.05 V/W and a specific detectivity of $1.6 \times 10^7$ cm √Hz/W (Figure **11**c).

Doping and interface engineering strategies have been employed in the fabrication of Ag-modified *p*-type $Bi_{0.5}Sb_{1.5}Te_3$ film ($t_f$ = 750 nm), which was deposited on PI substrate via magnetron sputtering method.[177] The *S* (128.8 μV/K at RT) and *σ* (740/Ω-cm at RT) were found to be decoupled, resulting in their simultaneous increment, leading to an improved *PF* of 12.4 μW/cm-$K^2$ at 300 K. The structural defects generated by Ag atoms in the cation sublattice facilitated charge transport in the film matrix and the heterostructure interfaces between the $Bi_{0.5}Sb_{1.5}Te_3$ and $Sb_2Te_3$ presumably aided in improving the *S*. A flexible TEG consisting of four legs built using the $Ag_{0.005}Bi_{0.5}Sb_{1.5}Te_3$ films has shown an output voltage of ~31.2 mV and a power density of ~1.4 mW/$cm^2$ at a $\Delta T$ of 60 K.

Multilayered $Bi_{1.5}Sb_{0.5}Te_3$ film composed of (015)-oriented and ordinary nanolayers fabricated via thermal co-evaporation technique has exhibited *S* of 242 μV/K, *PF* of 3.98 mW/m-$K^2$ at 120 8 C, and a low *κ* of 0.91 W/m-K at RT, resulting in a RT *zT* of 1.28.[174] The lattice misfit value for (015)-oriented/ordinary $Bi_{1.5}Sb_{0.5}Te_3$ interface was about 6%. As a result, a coherent interface arising from crystal lattice stresses in both layers may accommodate the mismatch. The proposed in-plane transport mechanism of the multilayered $Bi_{1.5}Sb_{0.5}Te_3$ film with alternating stress field is shown in Figure 11f. The movements of a large number of charge carriers are restricted in (015)-preferential 200 nm-thick nanolayers and a few carriers with high energy can pass through and along the interfaces and ordinary 20 nm-thick nanolayers (Figure 11d-e). Thin nanolayers with a (015)-oriented plane can provide channels for easier carrier flow, resulting in increased carrier $\mu$ and a higher *PF*.

Other methods such as chemical vapor deposition (CVD) and metal–organic chemical vapor deposition (MOCVD) have also been employed in the fabrication of TE films. In *p*-type $Bi_2Te_3/Sb_2Te_3$ SLs prepared via CVD technique, a $zT$ of ~2.4 was achieved.[64] The $\kappa_l$ values of the SL structures showed a minimum of ~0.22 W/m-K, which was 2.2 smaller than the $\kappa_l$ of $Bi_{0.5}Sb_{1.5}Te_3$ alloy (~0.49 W/m-K). Whereas, the TE performance of *n*-type $Bi_2Te_3/Bi_2Te_{2.83}Se_{0.17}$ SLs was dismal ($zT$ ~1.4 at 300 K) as compared to the *p*-type due to high $\kappa_l$. Due to the absence of mirror-like SL interfaces, the phonon interference was not appreciable to reduce $\kappa_l$. The enhancement was due to optimizing the transport of phonons and electrons in the SLs. The prototype device based on these films has shown a significant cooling effect (32 K at ~300 K).

The $Bi_2Te_3$ films ($t_f$ = 2 μm) were deposited on (001) GaAs substrates by modified MOCVD. The *S* of the film was about -225 μV/K at RT and the $\sigma$ was 312- 384/Ω-cm, resulting in a maximum *PF* of ~18.6 μW/cm-K$^2$.[178] *p*-type $Bi_{0.4}Sb_{1.6}Te_3$ and *n*-type $Bi_2Te_3$ films ($t_f$ = 4 μm) were fabricated by MOCVD at a deposition temperature of 400 $^{o}$C.[179] The *S*'s of the films were 230 μV/K and -180 μV/K for *p* and *n*-type respectively; the *PF* were reported to be 15 and 25 μW/cm-K$^2$ for the same films. The prepared planar-type TE device containing 20 *p*/*n* pairs generated an estimated power of 1.3 μW at a $\Delta T \approx 45$ K.[179] TE cooling modules made of *p*-type 10/50 Å $Bi_2Te_3/Sb_2Te_3$ SL and *n*-type $\delta$-doped $Bi_2Te_{3-x}Se_x$ (both grown heteroepitaxially using MOCVD) have exhibited cooling fluxes ($q_{max}$) of 258 W/cm$^2$ (**Figure 11**g).[175] The individual $zT$ of *n* and *p* type materials used in the fabrication of cooling devices were 1.4 and 1.5, respectively. The *n*-type $Bi_2Te_3$ and *p*-type $Bi_{0.4}Sb_{1.6}Te_{3.0}$ films were prepared via MOCVD.

$Bi_2Te_3$ thin films and $Bi_2Te_3/Bi_2(Te_{0.88}Se_{0.12})_3$ SLs were grown via MBE on $BaF_2$ substrates with periods of 12 and 6 nm, respectively.[180] The thin films and SL showed carrier densities between 2.8 × $10^{19}$ and 4.4 × $10^{19}$ $cm^{-3}$, $\mu$ of 110 $cm^2$/V-s, $\sigma$s of 400–1000/Ω-cm, *PF*s between 28 and 35 μW/cm-$K^2$, and the dislocation density of ~2 × $10^{10}$/$cm^2$. The $\kappa_l$ varied between 1.60 W/m-K for $Bi_2Te_3$ thin films and 1.01 W/m-K for a SL with a period of 10 nm. The phonons are more strongly scattered on surfaces due to the small thicknesses of the films; the $\kappa$ of SL is reduced by a factor of 1.3 compared to single layer thin films and bulk materials with the same Se content. Also, the $\kappa$ decreased with decreasing SL period. The *n*-type polycrystalline $Bi_2Te_3$ films ($t_f$ ~600 nm) were prepared by the MBE method. The *S* of the films was -180 μV/K.[181] A crystalline $Bi_{0.4}Sb_{1.6}Te_3$ thin film was fabricated by MBE at a $T_{sub}$ of 280 $^oC$.[182] XRD and HRTEM analysis showed that a small amount of Bi had been incorporated into Sb lattice sites, resulting in substituent doping. Larger grain size, lower carrier concentration, and lower mobility were observed for the highly crystalline $Bi_{0.4}Sb_{1.6}Te_3$ film than for the $Sb_2Te_3$ film. The lower carrier concentration of $Bi_{0.4}Sb_{1.6}Te_3$ was due to increased resistance resulting from antisite defects. In the range 323 K to 423 K, the temperature-dependent $\sigma$ of the $Bi_{0.4}Sb_{1.6}Te_3$ film increases nearly exponentially with increasing temperature, in contrast with the $Sb_2Te_3$ film, the $\sigma$ of which is inversely proportional to temperature. This is due to the introduction of the Bi atoms, which effectively lower the energy level of impurities, resulting in higher activation energy (43.2 meV) for conductivity in $Bi_{0.4}Sb_{1.6}Te_3$ film.

The $Bi_2(Se_xTe_{1-x})_3$ films with various concentrations of Se and Te and thicknesses of 350-500 nm were prepared by MBE on GaAs substrates and the dependence of $\kappa$ on film composition was investigated. It was observed that the $Bi_2(Se_xTe_{1-x})_3$ ternary alloy films have much lower $\kappa$ values compared to those of $Bi_2Se_3$ and $Bi_2Te_3$ films.[183] It is reported that, in $Bi_2Te_3$ films (on GaAs (001) substrates, by MBE), the electron concentration decreased by over an order of magnitude when the thickness of film increased from 10 nm-1000 nm.[184] A change in the carrier concentration with the film thickness was also observed in $Bi_2Se_3$.[185] Similarly, in $Sb_2Te_3$ thin films (grown on Si(111) substrate via MBE), the carrier concentration, $\mu$ and $\sigma$ are reported to be sensitive to thickness.[186] With the increase in film thickness, the grain size is also increasing. The 28-nm-thick film has the highest carrier concentration, 2.493 × $10^{19}$/$cm^3$, and the value decreases to 1.701 × $10^{19}$/$cm^3$ when the film thickness increases to 67 nm. The $\mu$ increases from 107 $cm^2$/V-s (in 28 nm film) to 229 $cm^2$/V-s (in 67 nm film) is related to the growth of microcrystals and decrease of grain boundary density in thicker films. For the same films, the $\sigma$ increases from 425.7/Ω-cm to 1036/Ω-cm.[186]

In $(Bi,Sb)_2$ $(Te,Se)_3$ based alloys, inherent point defects generated during the crystal growth from the stoichiometric melt. Negatively charged antisite defects: $Sb_{Te}^{'}$, and $Bi_{Te}^{'}$, are formed when excess cation atoms occupy the vacant anion sites giving rise to *p*-type conduction.[187] Contrarily, the anion vacancies ( $V_{Te}^{\circ\circ}$and $V_{Se}^{\circ\circ}$) carrying positive charges are generated if the empty anion sites are retained, which is compensated by electrons.[188] Many of the properties of V–VI compounds depend on the types of impurity/defects which provide charge carriers for the materials.[189] Therefore, discerning their origin and the effect on the carrier transport properties of films will contribute to better understanding of the effect of doping. In bulk $Bi_2Te_3$ crystal growth, near the solid–liquid equilibrium temperature, the concentration of antisite defects or vacancies strongly depends on the composition. However, thin film growth is far from the equilibrium conditions, therefore the composition of Bi/Te can be tuned by growth temperature (i.e. $T_{sub}$) to generate defects.[190] It is shown that in $Bi_2Te_3$ thin films (with different $E_F$ levels) fabricated by MBE, the variation of the electronic structure can be attributed to the antisite defects that are controllable with the growth temperature. Angle-resolved photoemission spectra (ARPES) and extended X-ray-absorption fine-structure (EXAFS) results provided direct evidence for the substitution of Bi for the Te(1) site, which was the factor for the shift of the $E_F$ or the position of the Dirac point in $Bi_2Te_3$.[190]

The influence of $T_{sub}$ and film thickness ($t_f$) on the transformation mechanism of intrinsic point defects in $Bi_2Te_3$ films is reported.[191] The *n*-type $Bi_2Te_3$ single crystalline films with different $t_f$ were grown epitaxially on sapphire (000*1*) substrates at different $T_{sub}$s. The formation of point defects was analyzed by the combined use of ARPES and electronic transport measurements. The negatively charged vacancies, $V_{Te}^{\circ\circ}$, initially the dominant intrinsic defects, transformed gradually during the growth process into positively charged anti-site defects, $Bi_{Te}^{'}$, induced by thermal annealing from a continuously heated substrate. From the surface of the film into the inner strata of the film, the density of $V_{Te}^{\circ\circ}$ decreased while the density of $Bi_{Te}^{'}$ increased, leading to a gradient of vacancies and anti-site defects along the film growth direction (Figure **12**). As a result, the electron density in $Bi_2Te_3$ films decreased monotonically with increasing $t_f$. Elevating $T_{sub}$ led to a more significant in situ annealing effect and triggered intrinsic excitations that deteriorated electronic transport properties. The thinnest $Bi_2Te_3$ film (16 nm) grown at $T_{sub}$ = 518 K had the highest electron concentration of 2.03 × $10^{20}$ $cm^{-3}$ and also the maximum room temperature *PF* of 1.6 mW/m-$K^2$ of all the films.

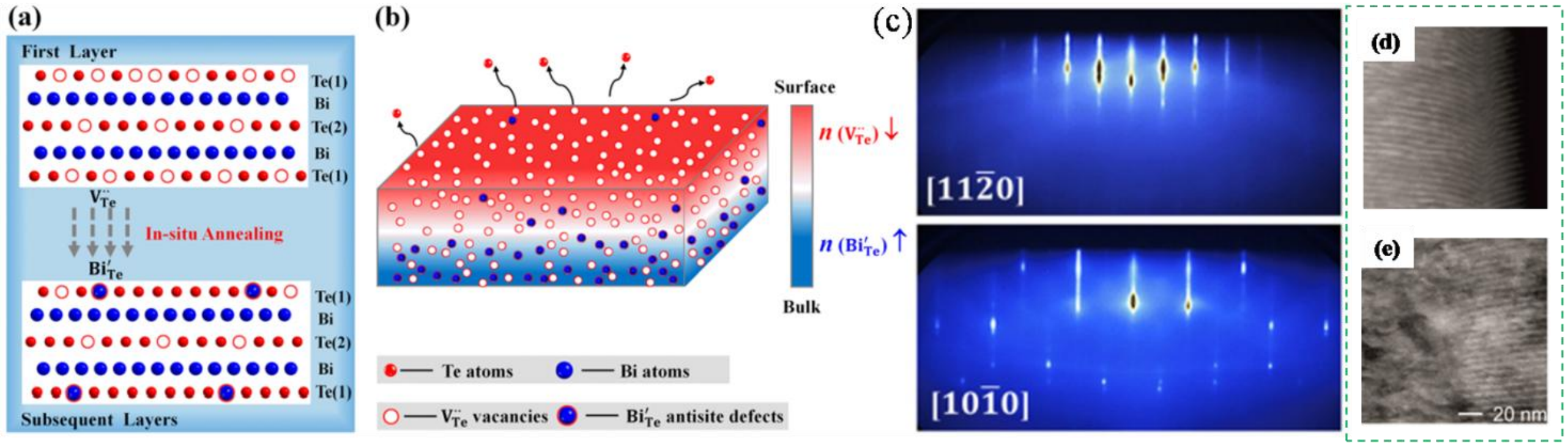

Figure 12: In situ transformation of point defects along the thickness direction of the $Bi_2Te_3$ single crystalline films. (a) transformation from $V_{Te}^{\circ\circ}$ vacancies to$Bi_{Te}^{'}$ antisite defects during the film growth; (b) the gradient in the distribution of $V_{Te}^{\circ\circ}$ and $Bi_{Te}^{'}$ along the thickness direction; (c) RHEED patterns along the [10$\bar{1}$0] and [11$\bar{2}$0] azimuthal directions. Reproduced with permission from ref.[191] Right side frame (d-e) HAADF-STEM micrographs for ex situ heating experiments. (d) as-deposited SL; (e) SL heated for 1 h at 250 °C followed by heat treatment for 1 h at 300 °C shows a strong fading of the SL structure. Reproduced with the permission from ref.[139]

$Bi_2Te_3$ thin films were grown on muscovite mica substrates by MBE. The topographic and structural analysis revealed that the $Bi_2Te_3$ thin films have atomically smooth large-area terraces and a high crystalline quality. A phase coherence length of 277 nm for a 6 nm thin film and a high surface mobility of 0.58 $m^2$/V-s for a 4 nm thin film were observed.[192] Investigation into the temperature stability and quality of 1 nm $Bi_2Te_3$/5 nm $Sb_2Te_3$ SLs (MBE grown) using in situ and ex situ X-ray diffraction and TEM shows that the SL structures are not stable against interdiffusion of the components upon heating.[139] The micro and nanostructural changes occurred at temperatures as low as 200 °C; the interdiffusion started next to the SL defects. Most of the samples exhibited no features of SL after being heated further (Figure 12d-e). At 300 °C due to the interdiffusion of $Bi_2Te_3$ and $Sb_2Te_3$ layers, a thermodynamically stable $Sb_{1.66}Bi_{0.33}Te_3$ alloy formed, implying that the structural integrity of $Bi_2Te_3$/$Sb_2Te_3$ SLs is not stable for a long time above 200 °C.

It is reported that MBE grown Te crystal-embedded $Bi_2Te_3$ (i.e. Te– $Bi_2Te_3$) thin film can be formed by simple annealing of Te-rich Bi/Te multilayered film structure.[193] Modulations in structure and composition were observed at the boundaries between the two phases of Te and $Bi_2Te_3$. Furthermore, the samples contained regularly shaped nanometer-scale $Bi_2Te_3$ single grains. The $\sigma$ at the interface Te/$Bi_2Te_3$ was high because of the low electrical barrier at RT and the additional contribution of the topological surface states (TSS). As a result, while the $S$ and the $\sigma$ remained high, the $\kappa_l$ was suppressed to a low value, < 0.25 W/m-K in the 300-400 K range, due to phonon scattering at the Te/$Bi_2Te_3$ interface. A $zT$ value of 2.27 at 375 K was obtained for the Te– $Bi_2Te_3$ thin film.

$Bi_2Te_3$ thin films were deposited by ALD using $BiCl_3$ and $(Et_3Si)_2Te$ at 160-300 °C.[194] The $\sigma$ of the films grown on glass were between $10^4$ and $10^3$/Ω-cm and a $S$ of ~-180 µV/K at RT for the film deposited at 160 °C. The ALD growth mechanism of $Sb_2Te_3$ and $Bi_2Te_3$ multilayer composite films is reported.[195] Trimethylsilyl telluride ($(Me_3Si)_2Te$), bismuth trichloride ($BiCl_3$) and antimony trichloride ($SbCl_3$) were used as the precursors for tellurides, Bi, and Sb respectively. No TE properties are reported.

Akin to the variation of $n$ and $\mu$ with film thickness,[184-186] parameters such as $S$ and $\sigma$ also vary with film thickness. As previously stated, the film thickness is a critical factor that influences the maximum output power ($P_{max} = (S\cdot\Delta T)^2 tw/4\rho l$) of the TEG. Where $\rho$ is the resistivity, $l$ is the length, $t$ is the thickness, and $w$ is the width. Therefore, with respect to device geometry, in addition to a large $PF$, a large $t$, which contributes to reducing the internal resistance is required. A detailed understanding of the TE film performance in relation to its thickness (and temperature) is important as this information guarantees the selection of material, film thickness, and fabrication methods when it comes to RTG/TEC design.

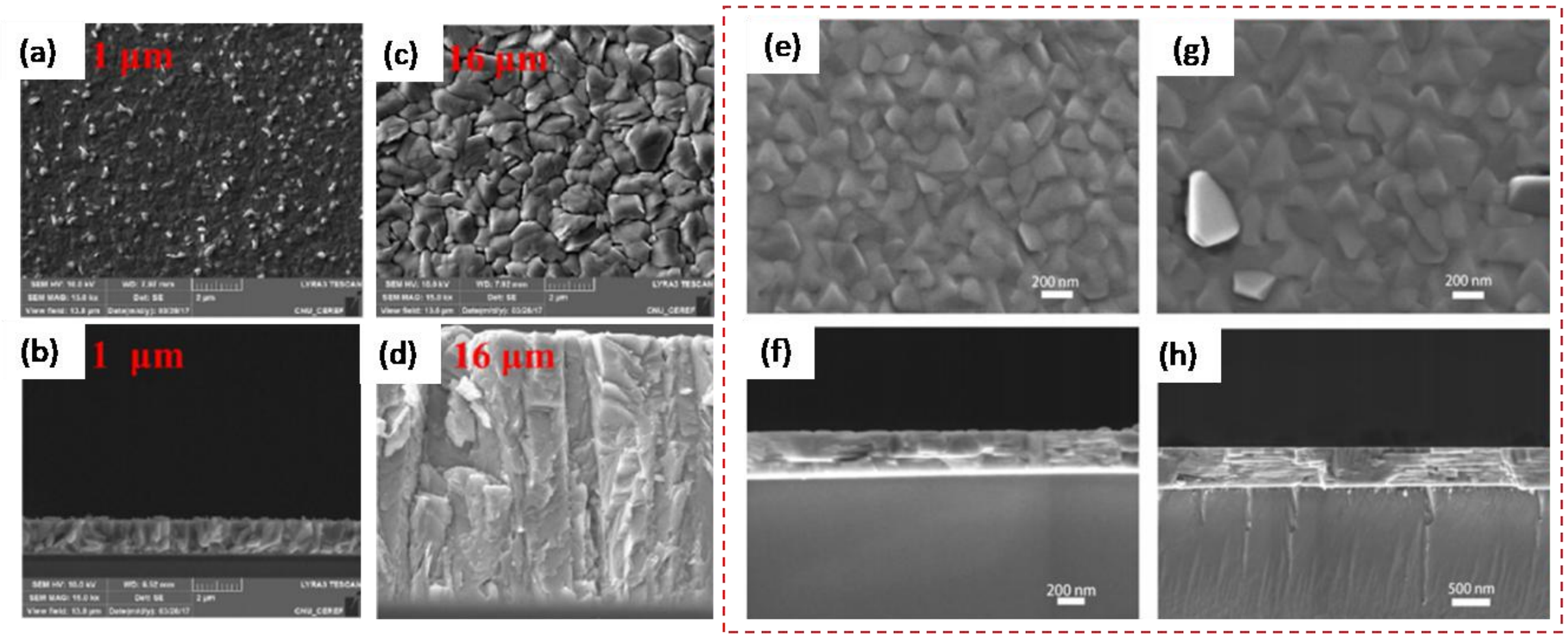

Figure 13: (a & c) top surface and (b & d) cross-sectional SEM images of 1 and 16-µm-thick $Sb_2Te_3$ film samples. Adapted with permission from ref.[196] Right side frame (e-h) FESEM

images of the surface (e & g) and cross section (f & h) of co-sputtered $Bi_{0.5}Sb_{1.5}Te_3$ films of 240 nm and 500 nm thick samples. Adapted with permission from ref.[197]

Investigation into the thickness dependent TE performance of MBE grown $Bi_{0.64}Sb_{1.36}Te_3$ films (6-200 nm) shows that the $\sigma$ of the 6 nm and 15 nm-thick films were higher than in the 200 nm-thick film due to the highly mobile topological surface state (TSS) conduction channel, implying the contribution of the TSS starts to diminish as the thickness increases.[198] As a consequence of the enhanced $\sigma$ and the suppressed bipolar effect in transport properties for the 6 nm thick film, a *PF* of ~2 mW/m-$K^2$ was achieved at RT. Films of $Sb_2Te_3$ (thicknesses of 1, 6, 10, and 16 μm, Figure **13**a-d) were deposited on $SiO_2$ at $T_{sub}$ = 250 °C by the co-evaporation method.[196] The microstructure revealed grains of a thin columnar type, whose preferred orientation shifted from (1010) to (015) as the film grew (110). This texture shift had a substantial influence on the films' mobility (due to anisotropy) and, as a result, their overall TE characteristics. The *n*, $\mu$, $\rho$, and *S* for 1 μm and 16 μm thick samples were $0.83 \times 10^{19}/cm^3$, 480 $cm^2$/V-s, 16 μΩ-m, 230 μV/K; and $1.57 \times 10^{19}/cm^3$, 160 $cm^2$/V-s, 24 μΩ-m, 200 μV/K respectively (Figure 14a-c). The highest *PF* of 3.3 mW/m-$K^2$ was observed for the 1 μm-thick $Sb_2Te_3$ film. In a similar study, (*00l*)-oriented $Bi_{0.5}Sb_{1.5}Te_3$ and Te targets were used to deposit BST films (thicknesses: 50 to 500 nm, Figure **13**e-h) on the *c*-axis sapphire.[197] It was revealed that there exists an optimal thickness (~240 nm), which exhibits the peak *PF* of 29 μW/cm-$K^2$. The grain size and stoichiometric ratio varied as the film thickness increased, affecting the carrier concentration, $\sigma$, and *S* (Figure 14d-f).

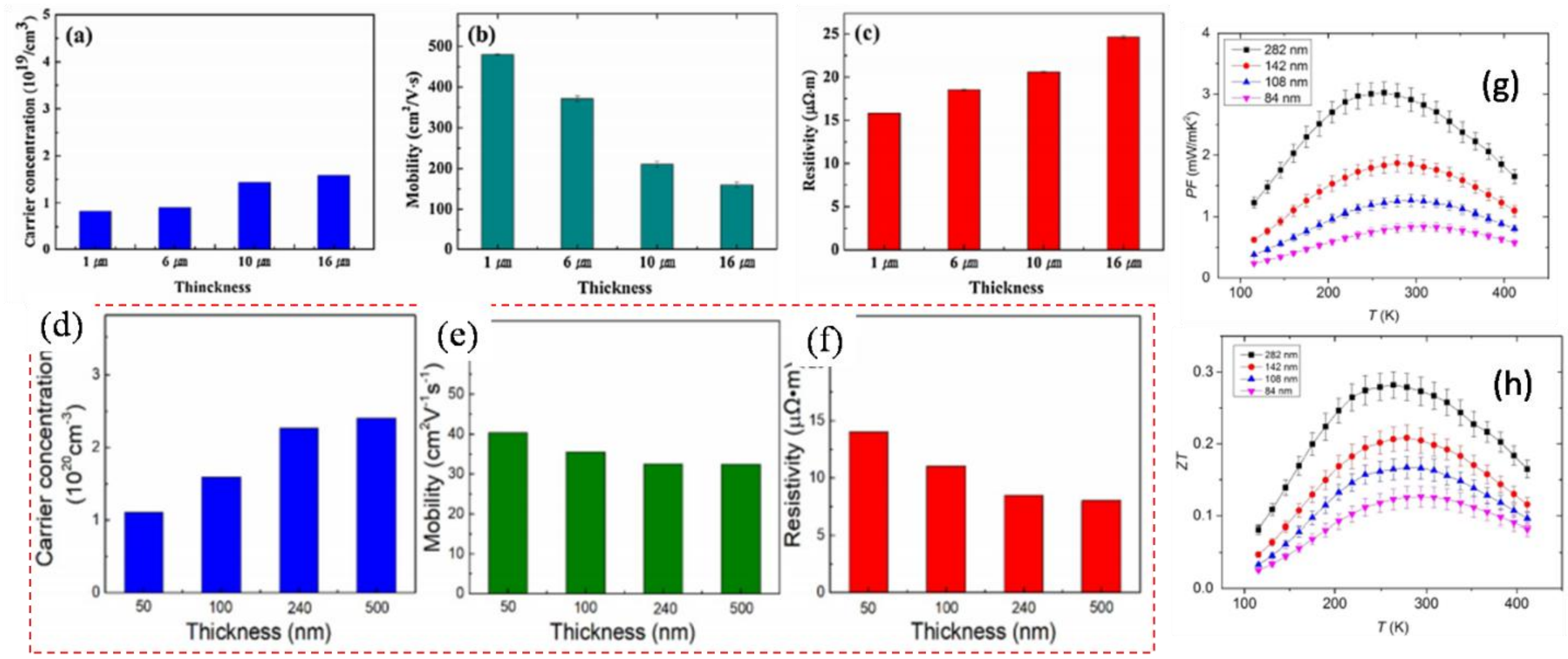

Figure 14: Film thickness dependence of $Sb_2Te_3$ electrical properties: (a) carrier concentration, (b) mobility, (c) resistivity. Adapted with permission from ref.[196] Lower image. (d-f) Carrier concentration, mobility, and resistivity as a function of (*00l*)-oriented $Bi_{0.5}Sb_{1.5}Te_3$ film thickness. Adapted with permission from ref.[197] Right most images: *PF* (g) and *zT* (h) of the $Bi_{87}Sb_{13}$ nanofilms in the temperature range from 110 K to ~420 K. Adapted with permission from ref.[199]

In-situ investigation into the annealing effect on the in-plane TE properties of polycrystalline $Bi_{87}Sb_{13}$ layers (via evaporation method) of different thicknesses shows that the $\sigma$, $\kappa$ and absolute *S* decrease with decreasing film thickness, resulting in the largest TE efficiency for the 282 nm-thick sample.[199] The optimum applicable temperature window for the 282 nm film was between 230 K and 270 K, where the largest TE efficiency was obtained with a *PF* of 3.02 mW/m-$K^2$ and a *zT* of 0.282 at 263 K (Figure 14, g and h). Study on the thickness and temperature dependence of $\rho$ in *p*-type $Bi_{0.5}Sb_{1.5}Te_3$ thin films (ranging from 80 to 320 nm)[200] has revealed that that the $\rho$ has a near linear relationship with the 1/thickness, which agrees with Tellier's model[201].

(*00l*)-oriented $Bi_{0.5}Sb_{1.5}Te_3$ thin films (20, 40, and 140 nm) with a multi-layered nanostructure were fabricated by a simple magnetron sputtering method.[153] It was found that the $\sigma$ increases with the thickness and the *S* decreases with the film thickness. For 20, 40, and 140 nm films, the *n* and $\mu$ at RT were 6.4, 7.5, 8.1 ($\times 10^{19}$ $cm^3$) and 66.4, 127.7, 128 $cm^2$/V-s; $\sigma$ and *S* were 7.9, 14.6, 16.7 ($9 \times 10^4$/Ω-m) and 209, 191.3, 163 μV/K. The maximum *PF* (53 μW/cm-$K^2$ at 310 K) was obtained for a 40 nm film. Similarly, in Bi-Te films prepared via the co-evaporation method, with an increase in film thickness, *S* fluctuated slightly (-182 to -202 μV/K), $\rho$ increased (due to sharp drop in $\mu$), and *PF* decreased, i.e. 2.8 mW/m-K$^2$ for the 1 μm thick film and 1.5 mW/m-K$^2$ for the 18 μm thick Bi-Te film.[202] For $Sb_2Te_3$ films fabricated by different methods, the observed parameters: thickness, *n*, $\mu$, and $\sigma$ are MBE[186]: 121 nm, 2.1 × $10^{19}$/cm$^3$, 305 cm$^2$/V-s, and 1035/Ω-cm; ALD[203]: 128 nm, 0.24 × $10^{19}$/cm$^3$, 270 cm$^2$/V-s, 104/Ω-cm; RF magnetron sputtering[204]: 240 nm, 7.71 × $10^{19}$/cm$^3$, 95 cm$^2$/V-s, 1170/Ω-cm; thermal co-evaporation:[205] 1 μ, 2.4 × $10^{19}$/cm$^3$, 186 cm$^2$/V-s, 714/Ω-cm.

An electrochemical (EChem) deposition method was used to fabricate freestanding TE film ($t_f$ = ~10 μm) of $Bi_{1.5}Sb_{0.5}Te_3$ ternary compound into an epoxy template with predefined periodic 3D nanostructures.[206] Optimized electroplating conditions (i.e., precursors, current-voltage, and pulse duration) for the 3D nanoconfined geometry facilitated the dense and uniform filling of $Bi_{1.5}Sb_{0.5}Te_3$ into the template over a large area (625 mm$^2$). The 3D nanostructure resulted in a decrease in $\kappa$, from 1.14 to 0.89 W/m-K, at 350 K while maintaining a constant $\sigma$ of 64.450/Ω-m and *S* of -144 μV/K. Consequently, the 3D nanostructured $Bi_{1.5}Sb_{0.5}Te_3$ film showed a *zT* value of 0.56 at 400 K, which was ~50% higher than the value for an ordinary $Bi_{1.5}Sb_{0.5}Te_3$ film due to the suppression of $\kappa_l$ without a significant deterioration in *S* or $\sigma$.

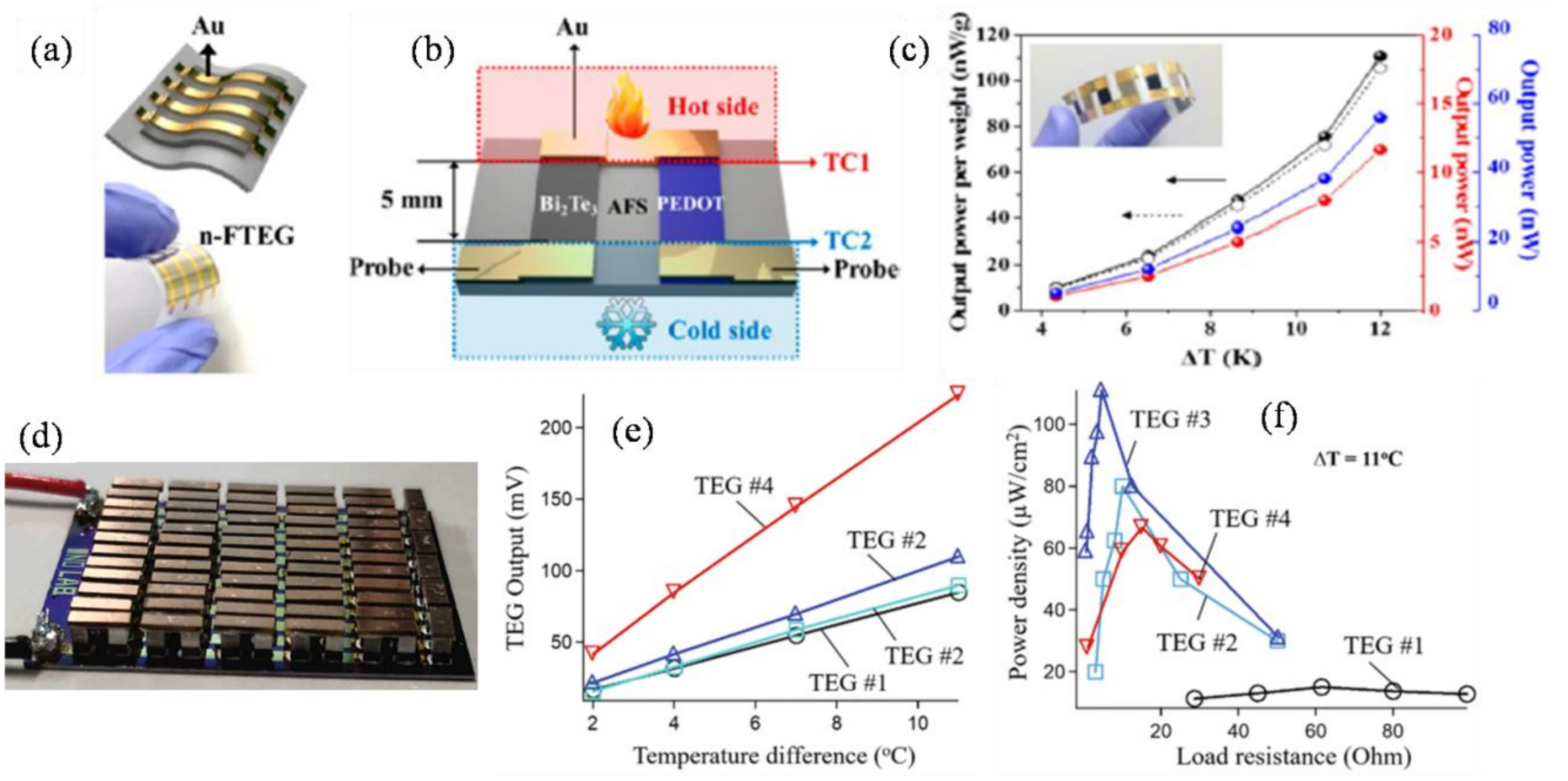

Figure 15: (a) FTEG with thermally evaporated gold electrode; (b) *pn*-FTEG with measurement points of temperature and *S*; (c) output power per weight of 1-couple (filled black circle) and 5-couples (open black circle) and output power of 1-couple (red circle) and 5-couples (blue circle) of *pn*-FTEGs over a temperature gradient ($\Delta T$). Adapted with permission from ref.[207] Bottom images. (d) 66 pairs completely fabricated device; (e) TEG output voltage as a function of temperature difference; (f) power density of the fabricated devices at $\Delta T$ = 11 $^{o}$C. Adapted with permission from ref.[21]

Li et al. reported the electrodeposition of thick $Bi_2Te_3$ films ($t_f$ = 100-500 μm) on a sputtered gold substrate.[208] The films had polycrystalline $Bi_2Te_3$ hexagonal unit cells with an avg. crystallite size of ~10-30 nm. The $S$ of as-deposited $Bi_2Te_3$ thick film was -68 μV/K before annealing, and increased to -125 μV/K after annealing at 300 °C for 2 h in an inert atmosphere. Compact, uniform, and stoichiometric *n*-type $Bi_2Te_3$ films ($t_f$ = 800 μm) were prepared by pulse electrodeposition on 1 $cm^2$ Ni substrates at an average film growth rate of 50 μm/h. At 300 K, the $S$ and $\sigma$ of the films were -80 μV/K and 330/Ω-cm respectively.[209] *n*-type $Bi_2Te_3$ films were prepared by electrodeposition and then transferred to flexible substrates. This film ($t_f$ = 2−3 μm, prepared at a deposition voltage of 0.02 V) had a $\sigma$ of 691/Ω-cm and a *PF* of 1473 μW/m-$K^2$.[207] Based on this *n*-type $Bi_2Te_3$ film, a prototype *p*-*n*-type flexible TEG (*pn*-FTEG) was fabricated using *p*-type poly(3,4-ethylenedioxythiophene). The *pn*-FTEG (5-couples) generated a $V_{out}$ of 5 mV at $\Delta T$ = 12 K with a output power of 56 nW (Figure 15a-c). The Nickel (0.7 at%) doped BiTe-based very thick film ($t_f$ = 2 mm) electrodeposited on a $SiO_2$/Si wafer and annealed at 250 °C has shown $S$, $\sigma$, *PF*, $\kappa$, and $zT$ of -143 μV/K, 975/Ω-cm, 20.5 μW/cm-$K^2$, 0.76 W/m-K, and 0.78 respectively.[21] A *n* type electrodeposited thick film (from this study) and *p* type $Bi_{0.3}Sb_{1.7}Te_3$ synthesized by the hot pressing method were used to make a 28 pairs device, which measured a maximum output power of 444.1 μW and power density of 111 μW/$cm^2$ at $\Delta T$ = 11 °C (Figure 15d-f). A slope type TEG fabricated using *n* & *p*-type $Bi_2Te_3$ films (1 μm-thick annealed at 250 °C) having $S$ = -80 μV/K (*n*-type) and 100 μV/K (*p*-type) and *PF* of 5.2 μW/cm-$K^2$ (*n*-type) and 8.8 μW/cm-$K^2$ (*p*-type) respectively has generated an open circuit voltage, $V_{oc} \approx$ 7.2 mV, and a maximum power, $P_{max} \approx$ 18.3 nW at a $\Delta T \approx$ 15 K. The films were obtained via electrochemical deposition on a stainless steel plate.[210]

It is shown that just by changing the volume ratio of ethylene glycol (EG) in the Bi-Te co-electrodeposition solution, either *n*-type or *p*-type Bi-Te alloy can be obtained.[211] With an increase in the vol% of EG, the (Bi/Te) ratio in the deposited Bi-Te alloy increased linearly with a slope of 0.463. When EG is not added, the deposited material was of *n*-type (Bi/Te ≈ 1.66/3.34) and changed to *p*-type (Bi/Te ≈ 2.06/2.94) when the EG content reached 20% v/v. TEGs fabricated (Figure 16a-e)[211] using electrodeposited *n*-type (without EG) and *p*-type (30 vol% of EG) Bi-Te alloys using polymer SU-8 mold substrate (Figure 16f) is shown in Figure 16h. The mold, which was obtained via photolithographic technique, served as a template for the final device. Similarly, it is shown that by controlling the electrolyte composition of the deposition bath and the current density used in electroplating, either *p*- or *n*-type Bi—Te alloy microposts can be produced.[212] Using this technique, Bi-Te alloy microposts of high aspect ratio (heights of up to 500 μm; diameter of 150 μm) were produced (Figure 16j).

A Bi-Te based Π-type TEG was fabricated via electrochemical deposition which delivered a open circuit voltage of 17.6 mV and 0.96 μW of power.[213] EChem methods allow us to grow *p*- and *n*-legs in 100 μm scale with high aspect ratio (height and cross section, Figure 16g).[214] More such devices and their output characteristics are listed in Table 2.

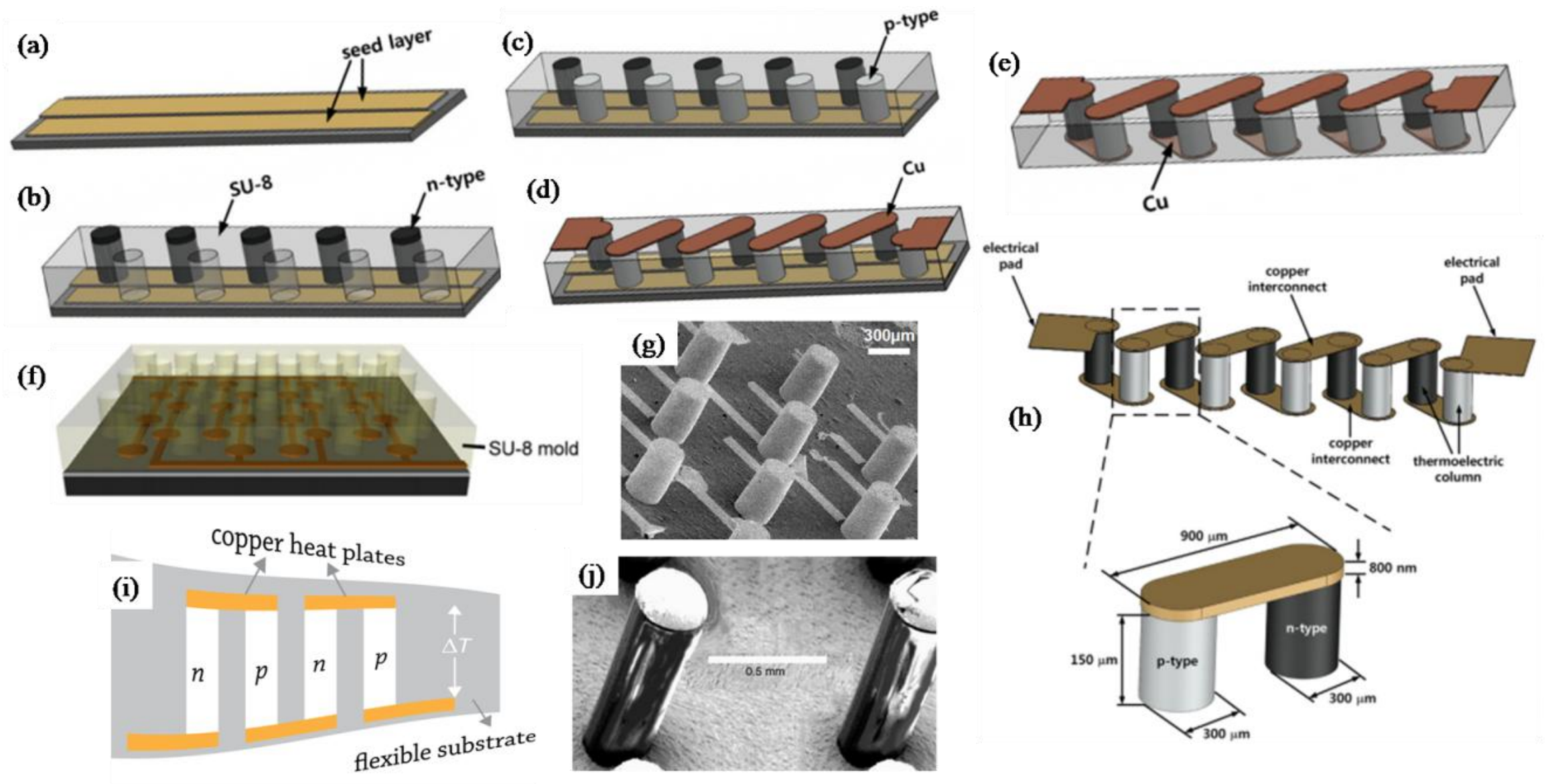

Figure 16: TEG microfabrication process sequence. (a) Ti/Au conducting seed layer deposition and patterning; (b) photolithography using SU-8 and electrodeposition of *n*-type Bi-Te; (c) electrodeposition of *p*-type Bi-Te material and chemical mechanical polishing (CMP) of the top surface; (d) topside Cu deposition, and (e) removal of Si substrate and Ti/Au seed layer by CMP process and bottom side Cu deposition; (f) polymer SU-8 mold substrate.[215] (g) electrochemically deposited $Bi_2Te_3$ thermolegs (high aspect ratio) using SU-8 mold.[214] (h) configuration of a TEG having 5-TCs and an individual TC with dimensions. Adapted with permission from ref.[211] (i) cartoon image of self supported π type flexible RTG.[216] (j) SEM image of bismuth-telluride microprobes with a height of 750 μm.[212]

The cationic surfactant, cetyltrimethylammonium bromide (CTAB) was employed in the electrodeposition of $Sb_2Te_3$ film (on Au/Ni/Si substrate) comprising Te nanodots 10-20 nm in size. This film exhibited a *PF* of 716.0 μW/m-K$^2$ after annealing at 200 $^o$C. Further, the film was smooth, dense, and strongly adhered to the substrate in addition to having a hardness of 1.953 GPa in comparison to 1.429 GPa of the film prepared without CTAB.[217]

## 2.9 PbTe films and superlattice structures and GeTe films

Theoretical prediction of higher *zT*s in semiconductor SLs under decreasing QW width has stirred tremendous interest in IV–VI semiconductor compounds, which are recognized as promising TE materials, and has encouraged further research in that direction.[71, 218]

The TE properties of $Pb_{1-x}Eu_xTe$/PbTe multiple-quantum-well (MQW) structures grown by molecular beam epitaxy (MBE) on $BaF_2$ substrates are reported.[219] In this study, numerous $Pb_{1-x}Eu_xTe$/PbTe MQW samples were grown with PbTe well widths in the range of 10-50 Å and $Pb_{1-x}Eu_xTe$ barrier widths in the 300-600 Å range.

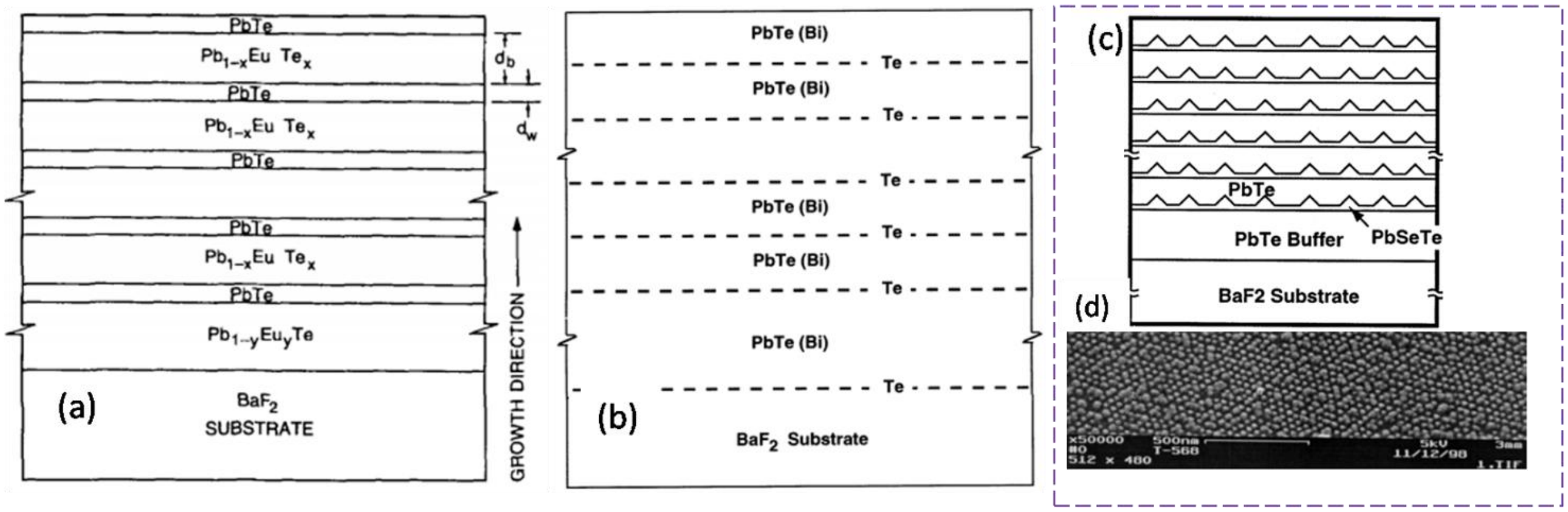

Figure 17: (a) Schematic cross section of the PbEu-chalcogenide multiple quantum-well structure. Adapted with permission from ref.[219] (b) schematic cross section of the PbTe/Te SL structure. Reproduced with permission from ref.[220] (c) a cross section of a $PbSe_{0.98}Te_{0.02}$/PbTe QDSL structure along with ~200-nm-thick Bi-doped PbTe buffer layer and the (111) $BaF_2$ substrate; (d) Field-emission SEM image of QDSL structure (altered). Reproduced with permission from ref.[221]

An illustrative cross section of the $Pb_{0.927}Eu_{0.073}Te$/PbTe MQW structure along with the $Pb_{0.958}Eu_{0.042}Te$ buffer layer and the $BaF_2$ substrate is shown in Figure **17**a. The thicknesses of the PbTe quantum well ($d_w$) and the $Pb_{0.927}Eu_{0.073}Te$ barrier layer ($d_b$) were tuned by computer programmed shutter times. The measured $S$ at 300 K of $Pb_{0.927}Eu_{0.073}Te$/PbTe MQW samples with near-optimum carrier concentrations were about two times larger than those of high-quality bulk PbTe for quantum well widths in the 17-23 Å range. At 300 K, electron mobility of a sample with a well width of 19 Å was 15-20% higher than lower-doped single-layer MBE grown PbTe layers, an indicative of modulation doping. The combined enhancements of the $S$ and $\mu$ resulted in the RT $zT$ of PbTe from ~0.45 for bulk material to > 1.2 in the best MQW. Similarly, in PbTe/Te SL structures[220] grown by MBE (Bi-doped PbTe as buffer layer and $BaF_2$ as substrate, Figure **17**b), the *PF*s for the PbTe/Te SL films were significantly higher (for the same carrier concentration) although their mobilities were lower than the bulk or homogeneous PbTe film. Experimental in-plane *zT* values for the Pb-chalcogenide film were increased from ~0.37 for homogeneous PbTe to 0.52 for the PbTe/Te SL structures. It was inferred that the mechanism underlying the improved *zT* and *S* data for the PbTe/Te SL structures involves a resonance carrier scattering (RSC) mechanism, which explains why both shorter and longer period thicknesses are ineffective. Since the distance between the Te adsorbed layers was much shorter than the MFP of the conducting electrons, the quantum effects are supposedly involved. The inclusion of the Te layers might have improved the scattering parameter (*r*) with a more favorable scattering mechanism (i.e. enhancing the *S* and *zT*). The same research group observed a higher *zT* of ~0.9 in MBE deposited $PbSe_xTe_{1-x}$/PbTe quantum-dot superlattice (QDSL structure, Figure **17**c-d).[221] Quantum dots (QD) are 0D structures. Their confinement effect is more intense than QW and quantum wires. The energy of an electron confined in a small volume by a potential barrier of QD is strongly quantized (the energy spectrum is discrete). For the QDs, the conduction band offsets and/or strain between the surrounding material and the QD acts as a confining potential. In such structures, the electronic DOS, (*dN*/*dE*) takes the shape of a Dirac delta function,[221]

$$\frac{dN}{dE} \propto \sum_{\varepsilon_i} \delta(E - \varepsilon_i) \qquad \text{Equation 11}$$

Where $\varepsilon_i$ is the discrete energy levels and $\delta$ is the Dirac function. Figure **17**d shows individual QD distributed vastly in planar arrays. The large effective mass induced by the band structure modification owing to quantum confinement results in a larger *S*. Further, the large lattice mismatch plays a vital role in the formation of QD. In the presence of QD, the $PbSe_{0.98}Te_{0.02}$ layers become more stable by releasing the strain. The factors aided in enhancing the *zT* were determined to be 1) Adsorbed or precipitated extra Te provided a more favorable scattering mechanism; 2) The $PbSe_{0.98}Te_{0.02}$ islands or dots embedded in PbTe matrix resulted in electron confinement in QDs; 3) Depressed $\kappa_l$.

The *zT*s, ranging from 1.6 at 300 K to *zT* = 3 at 550 K, are reported for Bi-doped *n*-PbSeTe/PbTe QDSL samples grown by MBE. Initial experiments have also shown that *zT* ~1.1 at 300 K is attainable in Na-doped *p*-PbSeTe/PbTe QDSL samples.[222] To ascertain the performance of QDSLs in TE devices, TE cooling test devices were fabricated from MBE grown PbSeTe/PbTe QDSL material and their cooling performance was tested.[223] 43.7 K of cooling below RT was measured, even though one leg was a zero $zT$ gold wire. On the other hand, the conventional $(Bi,Sb)_2(Se,Te)_3$ solid solution alloy material in the same test could only achieve 30.8 K of cooling under the identical conditions (i.e. same $T_h$, and ~aspect ratio).

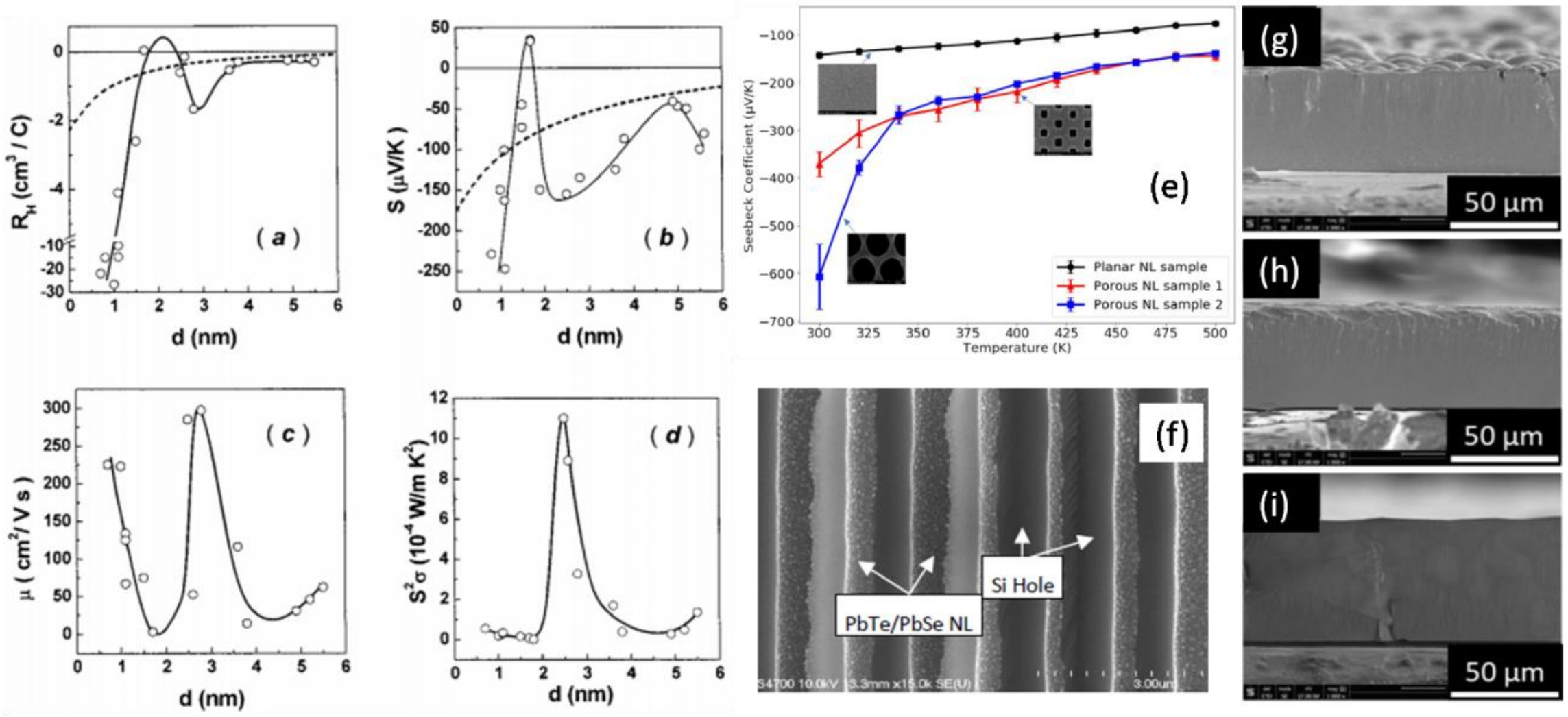

Figure 18: RT dependencies of the (a) Hall coefficient $R_H$; (b) $S$; (c) charge carrier $\mu$; and (d) $PF$ on the SnTe film thickness $d$ in $n$-PbTe/$p$-SnTe/$n$-PbTe heterostructures. The dotted lines are the results of the theoretical calculations based on a three-layer model. Adapted with permission from ref.[224] (e) plot of $S$ as a function of temperature comparing (i) ALD PbTe/PbSe (10/10 nm) nanolaminates deposited on planar Si substrates, (ii) ALD deposition on microporous Si templates with small square shaped pores, and (iii) porous Si templates with thinner pore wall and larger pore diameter; (f) cross-sectional images of PbTe/PbSe nanolaminates with period of 10 nm. Adapted with permission from ref.[225] cross sectional SEM images of PbTe electrodeposited at different applied potential: (g) -0.9; (h) -0.95; (i) -1.0 V. Adapted with permission from ref.[226]

In SLs, the dependence of $\kappa$ on period length is investigated by molecular dynamics simulation.[48] For perfectly lattice matched SLs, a minimum is observed when the period length is on the order of the effective phonon MFP. Lattice mismatch leads to diffusive phonon scattering and cancels the phonon Bragg reflection conditions, which eliminates the minimum $\kappa$. If the lattice constants of the two alternating layers differ by 4%, the $\kappa$ decreases monotonically with decreasing period length.[48] Beyer et al. reported the in-plane $\kappa$ of $n$-PbTe/$PbSe_{0.20}Te_{0.80}$ SL at RT using the bridge method.[227] The $\kappa$ was found to be reduced by more than 25% (from 2.31 to 1.73 W/m-K) when the PbTe layer thickness was changed from 13.4 nm to 2.3 nm while the thickness of $PbSe_{0.20}Te_{0.80}$ was confined to 1.8 nm. This study shows that smaller SL thickness (period) means introducing more interfaces and therefore buttressing the phonon scattering. According to studies, the cross-plane $\kappa$ of a SL is more sensitive to the SL period. With a decreasing SL period, $\kappa$ decreases first and then increases, which is observed in several types of SLs: $Bi_2Te_3/Sb_2Te_3$,[65] $PbTe/PbTe_{0.75}Se_{0.75}$,[228] and Si/Ge.[229]

Period independent $\kappa$ has been observed in $(PbTe)_{1-x}/(PbSe)_x$ nanodot superlattices (NDSLs)[230] with a wide range of periods: 5 nm $\leq h \leq$ 50 nm; compositions: $0.15 \leq x \leq 0.25$; growth temperatures: 550 K $\leq T_g \leq$ 620 K, and growth rates: 1 mm/h $\leq R \leq$ 4 mm/h. It was concluded that short MFPs of phonons (in PbTe) and small acoustic impedance mismatch at the PbTe/PbSe interface is the reason.[230] It is reported that the thermal treatment of 1 μm thick *n*-type PbTe doped with In (PbTe:In) films in the oxygen atmosphere led to two processes.[231] The PbTe:In films were prepared by PVD on a 100 μm thick PI substrate. The inversion of the $\sigma$ from *n*- to *p*-type in the films occurred after thermal annealing in oxygen ($T_{an}$ = 400 °C). This result is caused by In segregation at the film's surface and the generation of Pb vacancies (acceptor states) within the grains. Second, photoconductivity in PbTe:In films, attributed to oxygen diffusion along the grain boundaries and creation of potential relief on GBs. Rogacheva et al. studied the thickness ($d$) dependences of the $S$, $\sigma$, and Hall coefficients of PbTe and PbS epitaxial thin films ($d$ = 5 - 200 nm) deposited on (001) KCl substrates by thermal evaporation.[232] It was proposed that the oxidation of the films in air at 300 K induced a change in carrier type from *n* to *p* in films with $d$ < 125 nm (for PbTe films) and 110 nm (for PbS films). Similarly, studies on the thickness dependencies of the TE properties of *n*-PbTe/*p*-SnTe/*n*-PbTe heterostructures (with varying SnTe QW width of ~0.5-6.0 nm and a fixed PbTe barrier layer thickness) show that the thickness dependencies of the $S$, $\sigma$, the Hall coefficient, $\mu$, and *PF* are distinctly nonmonotonic (Figure 18a-d).[224] The size quantization of the energy spectrum of the hole gas in a SnTe QW is thought to be the reason for the observed effect. The oscillatory dependence of the transport properties of (100)/KCl/PbSe/EuS epitaxial structures[233] on the PbSe QW thickness (3 < $d$ < 200 nm) was revealed to be due to the quantization of electron energy into discrete sub bands as a result of 2D confinement in the growth direction.

2D *n*-type $PbTe/PbTe_{0.75}Se_{0.25}$ structures were prepared via an evaporation process using three sources—PbTe, PbSe, and $Bi_2Te_3$, where $Bi_2Te_3$ furnished Bi for *n*-type doping.[228] The films, which were grown on (111)-oriented $BaF_2$ substrates, had continuous PbTe and $PbTe_{0.75}Se_{0.25}$ alternating layers with a true 2D SL structure and ranged in thickness from 1 to 4 μm. The extrinsic *figure-of-merit*, *ZTe* (for the cross-plane heat and electric transport only) for a $PbTe/PbTe_{0.75}Se_{0.25}$ SL with a period of ~50 Å was measured to be 0.63 at RT. The intrinsic *ZT* (*ZTi*, a mix of in-plane and cross-plane properties) for the same SL but with a 55 Å period was ~1.75 at 425 K.

PbTe/PbSe nanolaminates were grown on planar Si wafers and porous Si templates via ALD.[225] A maximum *S* of -574.24 at 300 K was observed for the sample grown on porous Si membranes with round pores. The existence of a regular array of etched out pores, which offer extra periodic structure of the nanolaminate sample, is attributed to the increased *S* (Figure 18e-f). It has been shown[234] that the phonon thermal transport pathways in ALD produced PbTe–PbSe nanostructured, polycrystalline TE thin films are determined by: a) thickness dependent crystalline quality, where structural defect densities increase in thinner films due to the ALD growth processes; b) point defects inherent in the ALD growth process (e.g., oxygen defects); and c) compositional-driven defects (point defects and phase boundaries between the PbTe and PbSe), which give rise to additional phonon scattering events that reduce the cross plane $\kappa$ below that of the corresponding ALD-grown PbTe and PbSe films. Further, the phonon scattering and resulting $\kappa$'s in these ALD-grown PbTe–PbSe TE materials are triggered by extrinsic defect scattering processes as opposed to inherent phonon–phonon scattering processes observed in PbTe or PbSe phonon spectra.[234] PbTe thin films on Si substrates were prepared by an ALD using lead (II) bis (2,2,6,6-tetramethyl-3, 5-heptanedionato) and (trimethylsilyl) telluride as precursors at deposition temperature as low as 170 °C.[235] It was found that the formation of a PbTe thin film on the Si substrates was strongly dependent on the growth temperature.

The epitaxial electrodeposition of PbTe films on (111) indium phosphide single crystals in an acidic nitrate bath was described by Beaunier et al.[236] The electrolyte consisted of 0.05 M $Pb(NO_3)_2$ and 0.001 M $TeO_2$ in 0.1 M $HNO_3$ with and without 0.5 M $Cd(NO_3)_2$. The working electrode potential, measured against a saturated sulfate electrode, was monitored by a potentiostat. The as-deposited PbTe films on (111) InP had poor epitaxy due to a large lattice mismatch between PbTe and InP (9%) compared to the PbSe/InP system (4.4%). It was also discovered that adding 0.5 M $Cd(NO_3)_2$ to the electrolytes changed the film morphology from dendritic growth to homogeneous dense growth. The TE properties of the films are not reported. Ivanova et al. studied the potentiostatic electrodeposition of PbTe films onto *n*-Si(1 0 0) wafers utilizing a solution of 0.05 M $Pb(NO_3)_2$, 0.001 M $TeO_2$, and 0.1 M $HNO_3$.[237] The deposition potential was fixed at −0.38 V vs. Ag/AgCl. The obtained PbTe thin film had [200] preferred crystal orientation.

The *p*-type Te-embedded PbTe nanocrystallline thick films ($t_f$ = 50 μm, Figure 18g-i) have been prepared via electrodeposion (on Au/Cu and epoxy), where the grain size of the PbTe and Te phases are controlled by adjusting the applied potential and subsequent thermal treatment.[226] Although *S* was reduced (125 μV/K), enhanced $\sigma$ (116/Ω-cm) contributed to a *PF* of 183 μW/cm-$K^2$ in the film containing 100 nm size grains.

GeTe, a PbTe analogue, has lately recognized as a potential replacement for traditional PbTe.[35, 238-240] At RT, GeTe exhibit a rhombohedral structure ($\alpha$-GeTe) and undergoes a ferroelectric phase transition from the low-temperature $\alpha$-GeTe to cubic structure $\beta$-GeTe at ~700 K[241] because with an increase in temperature, the effective size of Ge rises faster than Te[242] and the ratio of the effective radii is greater than 0.41, which is the criterion for the formation of NaCl lattices. The amorphous structure of GeTe is thought to have a randomised rhombohedral structure that is near to the crystalline state locally but lacks long-range regularity.[241] Due to the presence of a high concentration of Ge vacancies,[243] undoped rhombohedral GeTe is a typical degenerate *p*-type semiconductor with intrinsically high hole concentration ($10^{21}$/$cm^3$ at RT), very low *S* (~30 μV/ K) that results in a low *zT*. To compensate for this deficiency, In, Bi, or Sb doping, as well as Pb alloying on the Ge site and Se alloying on the Te site, have been shown to reduce hole concentration and enhanced *zT*. Codoping of Bi and In in GeTe resulted in elevated TE performance (*zT* ~0.85 in 550 K–773 K for $Ge_{0.93}Bi_{0.05}In_{0.02}Te$) of GeTe by the combined effects of nanostructuring, reduction of carrier density and creation of resonant levels.[244] Similarly, *zT* ~1.3 at 650 K for 2 mol% In doping in GeTe[245]; *zT* of ~2 for Sb-doping and Se alloying in GeTe (i.e $Ge_{0.9}Sb_{0.1}Te_{0.88}Se_{0.12}$),[239] and Y-doped GeTe[246] are reported. The majority of GeTe studies focus on lowering its excessive carrier concentration and bringing it into the optimum range.[246]

It is demonstrated that epitaxial growth of GeTe thin films on Si(111) substrate is possible despite a large lattice-mismatch (8.3%).[144] A Ge/Te flux ratio of 0.4 in the temperature interval 220–270 $^oC$ was used in the fabrication. At lower $T_{sub}$s films were polycrystalline, but at higher $T_{sub}$s, impinging Ge and Te adatoms did not adhere to the surface. The epitaxial layers crystallized in the rhombohedrally distorted rock-salt structure, $\alpha$-GeTe, and grew exclusively [0001]- oriented, aligning their high symmetry in-plane directions with those of the Si(111) substrate, $\alpha$-GeTe $<10\bar{1}0>$ || Si$<11\bar{2}>$ and $\alpha$-GeTe $<10\bar{2}0>$ || Si$<1\bar{1}0>$. However, the deposited $\alpha$-GeTe layers were slightly tensely strained in-plane, due to the thermal mismatch during cooling down and/or a deviation from stoichiometry (Ge substoichiometry). RHEED patterns of Si, GeTe during growth stage, and as-deposited GeTe is shown in Figure 19a-c.

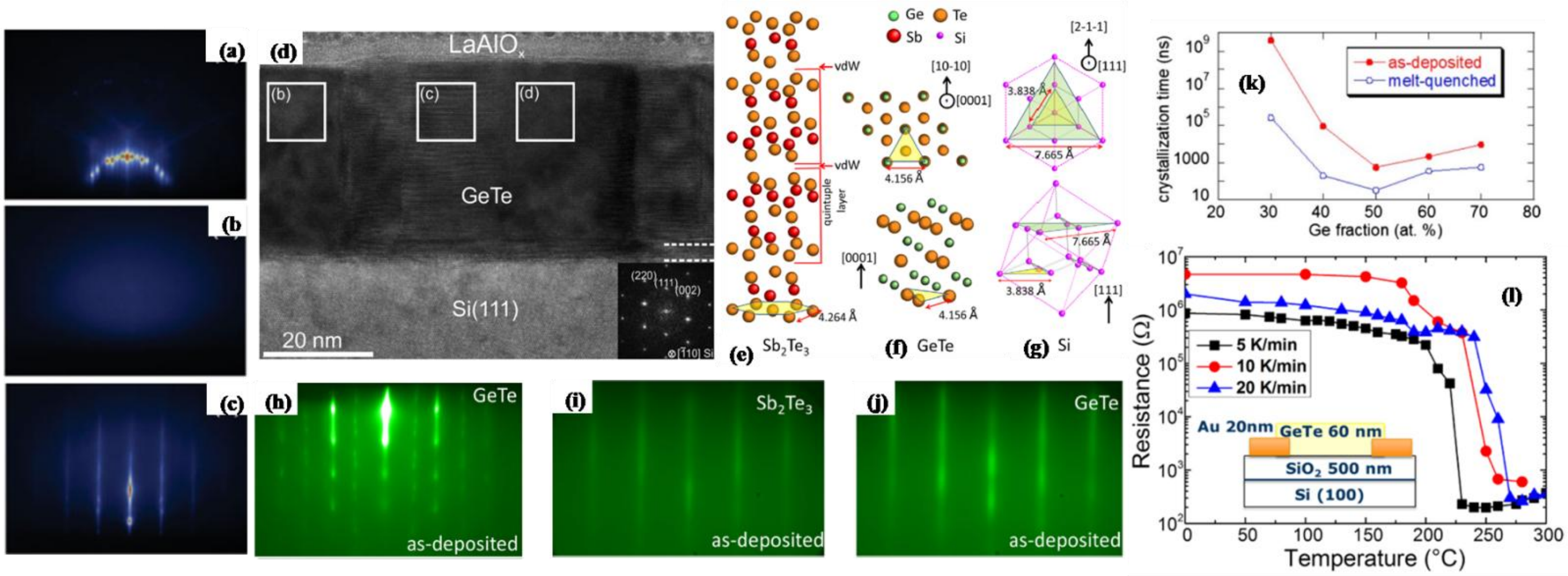

Figure 19: RHEED patterns (a) Si(111) surface; (b) the amorphous transition in the initial GeTe growth stages; (c) the final surface of the GeTe epitaxial layer. Adapted with the permission from ref.[144] (d) HRTEM image of a GeTe thin film deposited on $Sb_2Te_3$-buffered Si(111) substrate. The two parallel dashed lines mark the buffer layer (4 nm thick). The layer above GeTe is the capping $LaAlO_x$ amorphous layer. For brevity, FFT images of domains marked with rectangles are not shown; (e) crystal structure of $Sb_2Te_3$; (f) GeTe trigonal cell; (g) cubic Si. The transparent colored (yellow and green) hexagon and triangles mark the lattice constants of each corresponding crystal; (h-j) RHEED patterns (h) GeTe layer on Si(111); (i) as-grown $Sb_2Te_3$ film; (j) as-grown GeTe film on $Sb_2Te_3$-buffered Si(111). Adapted with the permission from ref.[145] (k) crystallization time as a function of Ge fraction for as-deposited, amorphous and melt-quenched, amorphous Ge–Te materials. Adapted with the permission from ref.[247] Resistance as a function of temperature for 60 nm thick as-deposited GeTe films. The inset shows a schematic view of samples used for resistance measurements.[248]

Epitaxial GeTe thin films have been deposited on $Sb_2Te_3$-buffered Si(111) substrates via PLD. The GeTe crystal and the Si(111) substrate have an in-plane lattice mismatch of 8.3% (Figure 19f-g). Therefore, an epitaxial $Sb_2Te_3$ thin layer (4 nm) was first deposited on Si(111) as a seeding layer and then a GeTe thin film was epitaxially grown on the $Sb_2Te_3$ thin layer.[145] The *a*-lattice parameter of $Sb_2Te_3$ differs by 2.5% from that of GeTe in the trigonal lattice (Figure 19e-f). Because PLD growth of $Sb_2Te_3$ on Si(111) implies a 2D growth mode,[249] $Sb_2Te_3$ material is an ideal seeding layer for GeTe thin film deposition. As a results, a distinct growth mode of GeTe is predicted on $Sb_2Te_3$-buffered Si(111) substrate. TE properties are not reported.

An investigation into the crystallization times of Ge–Te materials with different Ge concentrations (29.5– 72.4 at%) has shown a very strong dependence of the crystallization time on the composition for as-deposited, amorphous films (obtained via sputtering, $t_f$ = 30 nm) with a minimum (recrystallization time ~30 ns, Figure 19k) for the stoichiometric composition of GeTe. Further, crystallization times were depended on the substrate, capping layer, and film thickness.[247]

The study on the crystallization kinetics of GeTe film during PLD on Si substrate indicates that the critical crystallization (amorphous to the rhombohedral structure) temperature ($T_c$) lies between 220 and 240 °C, which is higher than the $T_c$ of GeTe films produced by magnetron sputtering.[248] The prepared films were annealed at different $T_{an}$ and heating rates. Independent of heating rate, the resistance of (~$10^6$ –$10^7$ Ω) in the as-deposited films decreased progressively with increasing temperature up to 200–220 °C due to a thermally activated conduction mechanism (Figure 19l). The precipitous drop in resistivity is closely linked to the amorphous to crystalline (rhombohedral) phase transition, which promoted an increase in $\mu$ rather than an increase in $n$. However, the resistances in the crystalline state did not show any dependence on annealing temperature, $T_{an}$. Sn atom (larger radius than Ge atom) was used as a dopant to reduce the temperature necessary to form a cubic GeTe structure at RT.[250] The Sn doped GeTe films were DC magnetron co-sputtered (Sn and GeTe targets) at RT and then annealed. The variation of electrical resistivity as a function of temperature for Sn doped GeTe thin films showed semiconducting behavior. From beyond 400 K, the resistivity drops sharply. $S$ values were seen to decrease in accordance with resistivity fluctuation. The Sn addition modifies the energy band structure, influencing carrier concentration and effective mass. A high $PF$ of $2.789 \times 10^3$ μW/m-K$^2$ at 300 K was observed for the amorphous films and $1.423 \times 10^3$ μW/m-K$^2$ at 718 K for the crystalline film.

## 2.10 SnSe, SnTe, $Cu_{2-x}Se$ films

SnSe has garnered a lot of attention among the 2D layered chalcogenides because of its chemical stability and low toxicity.[251-252] SnSe and $SnSe_2$ are the two phases of tin selenide. Another phase, $Sn_2Se_3$, has been reported by some researchers.[253-254] SnSe has a good TE performance at temperatures above 700 K, regardless of single crystals or nanocrystals,[255] and may therefore be utilized for energy harvesting in a high temperature regime.[28] Further, its applications are versatile: photodetector,[256] gas sensing,[257-258] phase change memory,[253] photocatalytic,[259] solar cells,[260-261] anode material for battery,[262] supercapacitor,[263] and topological insulator (TI).[264] These applications rely heavily on the optical, electrical, and microstructural characteristics of SnSe.[265] Material synthesis/deposition methods also play an essential role in obtaining high-quality materials with desired properties.[266-268]

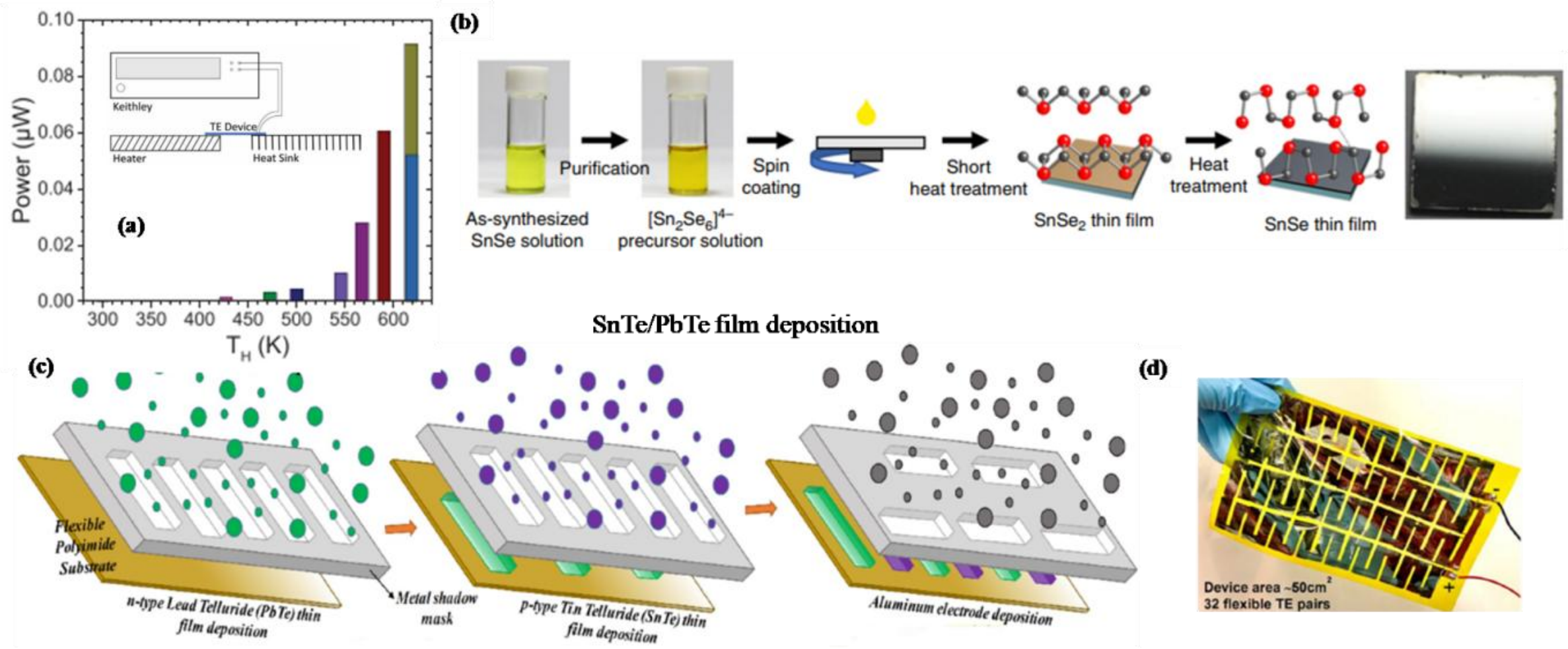

**Figure 20:** 1 μm-thick SnSe films (a) Peak power outputs of the device. Inset: schematic of device testing setup. Adapted with permission from ref.[269] (b) solution-processed fabrication of highly textured SnSe thin films. The photograph shows the fabricated film had a mirror-like reflection, indicating high uniformity. Adapted with permission from ref.[270] Bottom image. (c) schematic diagram shows the steps involved in TEG fabrication; (d) flexible SnTe–PbTe TEG with 32 TE pairs. Adapted with permission from ref.[271]

SnSe films of ~1 μm thickness were deposited on a glass substrate using a thermal evaporation process.[269] A low $\kappa$ of 0.08 W/m-K between 375 and 450 K, $S$ of > 600 μV/K at RT (but drops to ≈140 μV/K at 600 K), and a $zT$ of 0.055 was obtained at 501 K. Factors such as phonon scattering sites and interface effect at the junction between the nanosheets caused the reduction in $\kappa$. A TEG fabricated based on SnSe films, has generated an output power of ~0.09 μW at 618 K on the hot side, **Figure 20**a.[269] Highly textured $p$-type SnSe thin films (1 μm-thick) prepared from steps involving cosolvent route: dissolving SnSe powder in mixed solvents of ethylenediamine and ethanedithiol, spin coating, and optimized heat treatment (400 °C for 9 min, **Figure 20**b) have shown a $PF$ of ~4.27 μW/cm-K$^2$ at 550 K, which is higher than the SnSe single crystals. A $zT$ value of 0.58 at 750 K is anticipated in 400 °C/13-min-heat treated sample assuming the $\kappa$ of the sample is 0.82-0.89 W/m-K.[270] SnSe film ($t_f$ = 50 nm) with (400) texture was deposited on a mica sheet at ~450 $^o$C by low-pressure CVD using commercial SnSe powder.[272] The $\sigma$ of the SnSe film was increased from 8.3 to 21.8/Ω-cm between 300 and 500 K;

the $S$ was 392 μV/K at RT and the $\kappa$ varied between 1.7 and 1.12 W/m-K in the 300 K-525 K range. The $PF$ of the film was increased from 1.3 to 3.1 μW/cm-K$^2$ in the temperature range of 300-550 K, culminating in a $zT$ of ~0.15 at 550 K. Using the obtained SnSe films on a mica sheet, a wearable TE generator (seven units in series) was devised, which generated a voltage of 6.82 mV by using wrist heat and a voltage of 1.03 mV by using solar heat (**Figure 21**a-b). Polycrystalline SnSe films (100 nm) deposited on Si(111) substrates via a single mode microwave plasma chemical vapor deposition (MPCVD) method, have shown improved TE properties[273] such as maximal $S$ of ~627.7 μV/K at 600 K, $\sigma$ of ~9.3 -19.2/Ω-cm (in 550 K-900 K), and 0.92-0.58 W/m-K (in 300 K-900 K), culminating in a $PF$ of 3.98 μW/cm-K$^2$ at 600 K and a $zT$ of 0.33 at 600 K.

It is reported that the physical properties of SnSe thin films can be substantially altered by air annealing at a low temperature of ~480 K.[274] The $S$ reversed its sign from positive to negative (+757 to -427 μV/K as shown in **Figure 21**e) when the annealing time was extended, suggesting that air annealing caused surface oxidation, resulting in the formation of $n$-type semiconductors, which ultimately dominated the thin films' physical properties.

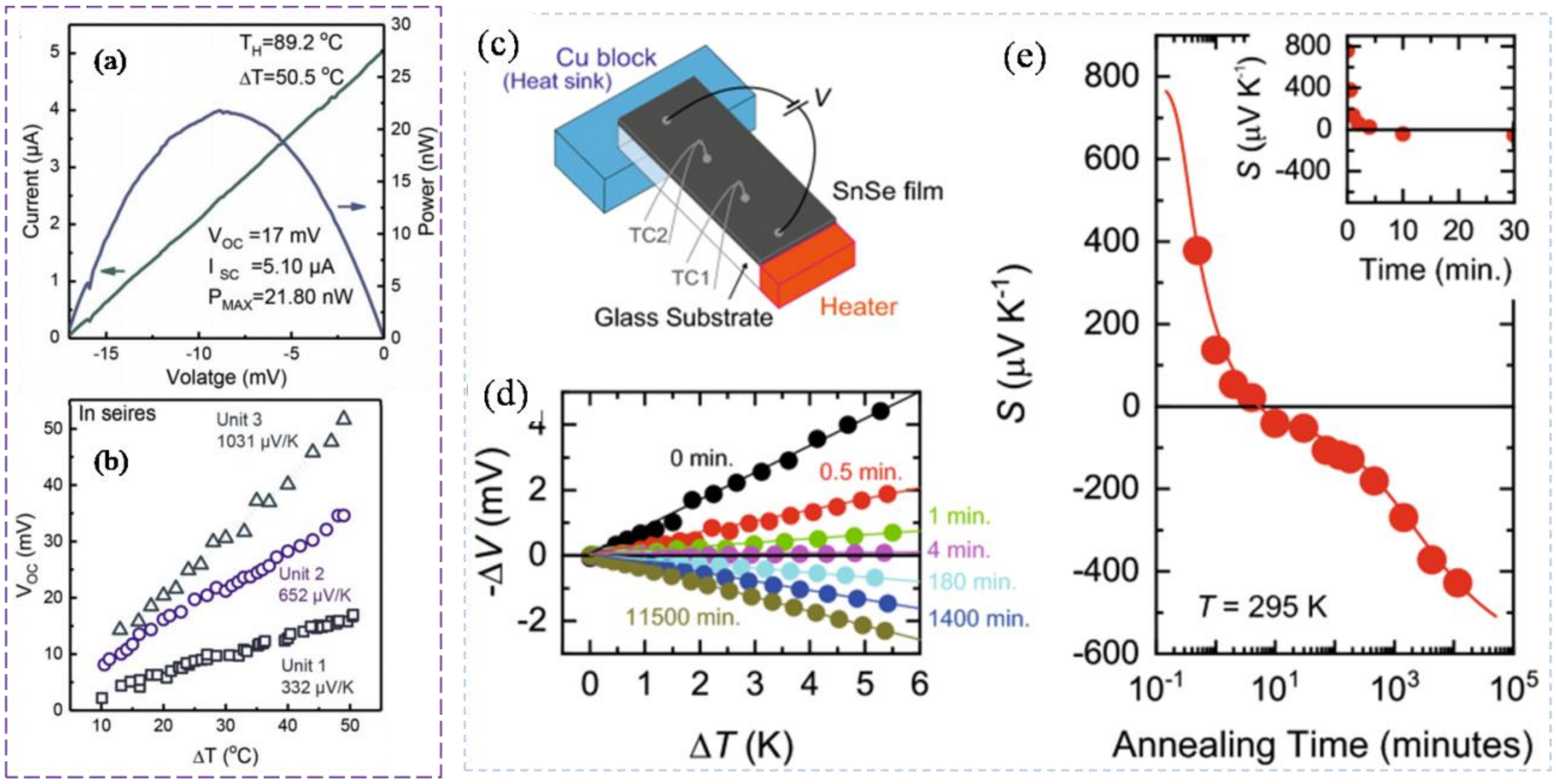

**Figure 21:** (a) TE generator based on SnSe films; (b) output characteristic of a single unit generator; (c) output voltages of single unit, double units and triple units. Reproduced with permission from ref.[272] Right side frame (c) schematic sample configuration for TE measurements. TC1 and TC2 are thermocouples; (d) Δ$V$ vs. Δ$T$ plot of SnSe thin films for different annealing time; (e) RT $S$ as a function of annealing time. Inset: the data for 0 to 30 min of annealing. Reproduced with permission from ref.[274]

In Rock-salt (RS) SnSe (orthorhombic GeS-type structure), because of the anisotropic layered structure, the effective masses of holes along the *a*-axis (i.e., through the layers) are significantly greater than those along the *b*- and *c*-axes (in the layer plane), revealing that much higher hole mobility can be realized in the *b*-*c* plane rather than along the *a*-plane.[275] As a result, *a*-axis oriented films are generally advantageous for a wide range of device applications that rely on lateral carrier movement. However, studies suggest that RS SnSe exists only as a metastable phase under high pressure or strain[276] and expected to be a topological insulator.[45, 264] To get materials with specified characteristics, it is critical to select appropriate substrates and regulate the film growth kinetics. *a*-axis oriented SnSe thin films grown by PLD on sapphire substrates possess a high *S* and an ultralow $\kappa$.[277] The SnSe thin films grown on $SiO_2$/Si substrate (via PLD) followed by thermal annealing have shown best crystal quality, uniform, and mirror-like surface (*b*-*c* plane with preferred (100) crystalline orientation); *S* and *PF* of 383 μV/K and 15.4 μW/m-$K^2$, respectively.[278] Metastable RS SnSe(111) films were obtained via MBE ($Bi_2Se_3$ as a substrate) with RS SnSe phase stabilized in epitaxial films of ~20 nm thick.[45] The RS-type structure is thermodynamically stable for PbSe at RT and, therefore, a thin RS-type PbSe epitaxial template layer has been utilized as a sacrificial layer while preparing 58% Pb-doped RS-type (Sn,Pb)Se films via a reactive solid-phase epitaxy (R-SPE) technique. A $\mu$ = 290 $cm^2$ /V-s was observed at RT. *p*- to *n*-type transition occurred at Pb ≥ 61% and the obtained (Sn,Pb)Se films had $\mu$ = 340 $cm^2$ /V-s.[279] More findings on SnSe films are listed in the Table **1**.

Films of *p*-type-SnTe ($t_f$ = 100 nm) and *n*-type PbTe ($t_f$ = 100 nm) exhibiting *zT* of ~0.3 and ~0.2 at 550 K respectively have been prepared via RT physical vapor deposition on PI.[271] The *S* of the thin films showed a maximum of 36 μV/K and 140 μV/K at 550 K and *PF* of 1.4 μW/cm-$K^2$ and 1.1 μW/cm-$K^2$ for SnTe and PbTe, respectively. The $\kappa$ of SnTe and PbTe were 0.31 W/m-K and 0.36 W/m-K at 300 K. Using these films, a TEG (SnTe/PbTe) with 4 *p*-*n* pairs interconnected by aluminum film (50 nm-thick) was fabricated (**Figure 20**c). This device with an efficiency of 3% yielded a maximum $V_o$ and power density of 250 mV and 8.4 mW/$cm^2$ at a $\Delta T$ of 120 $^oC$. Using the films, a flexible thermal touch sensor was also devised (with responsivity of 0.29 μV/W over a wide temperature range) based on the technique of heat transfer from the finger to the sensor surface that creates a $\Delta T$ on the material interface resulting in the impulse current generation.

Thin films of metal chalcogenide semiconductors: $Cu_{2-x}Se_yS_{1-y}$ and Ag-doped- $Cu_{2-x}Se_yS_{1-y}$ have been prepared via spin coating the soluble precursors prepared with thiol–amine solvent mixtures and a subsequent optimum annealing process. The resulting films have shown TE characteristics comparable to those of $Cu_{2-x}Se$ alloys obtained via conventional synthesis methods.[90] Following the cosolvent approach, $Cu_2Se$ ink solution was prepared by completely dissolving $Cu_2Se$ powder in the mixture of ethylenediamine and ethane dithiol for several minutes at RT.[280] The formulated ink was used to deposit smooth and crystalline thin films on $Al_2O_3$ or plastic substrates via spin coating. The $Cu_2Se$ thin films (both annealed at 703 K) exhibited *PF* of 0.62 mWm-K$^2$ at 684 K on an $Al_2O_3$ substrate and 0.46 mWm-K$^2$ at 664 K on a flexible PI substrate, and the RT $\kappa$ of all films were 1.3-1.5 W/m-K. Similarly, $Cu_2Se$ thin films were prepared by dissolving $Cu_2Se$ powder in ethylenediamine and ethane dithiol mixture. The $Cu_2Se$ solution was spin coated onto glass substrates, annealed at 350 °C, and then soaked in a Cu ion solution. A *PF* of 653 μW/m-K$^2$ was observed at RT.

## 2.11 Skutterudite TE films

Obtaining Co-Sb thin films with good crystalline purity is still a challenge. As a result, $CoSb_3$ thin films' TE characteristics have not yet surpassed those of bulk materials due to the presence of impurity phases, larger lattice imperfections, and higher defect concentrations that are commonly present in the Co-Sb thin films.[281-283]

Co-excess and Sb-excess $CoSb_3$ based thin films were fabricated by RF and DC magnetron co-sputtering technique.[284] It was observed that when the Sb content is 82.7%, the *S* is ~26 μV/K. The *PF*s. for the Co-rich and Sb-rich samples were $2.4 \times 10^{-4}$ W/m-K$^2$ and $6.9 \times 10^{-4}$ W/m-K$^2$ respectively, while for the pristine sample it was ~$0.6 \times 10^{-4}$ W/m-K$^2$.

Ag doped $CoSb_3$ thin films were deposited directly on the heated substrate by a magnetron sputtering technique.[285] Due to the improved crystallinity and Ag filling, the *S* and $\sigma$ were enhanced. The *PF* increased to $2.97 \times 10^{-4}$ W/m-K$^2$ which was a ~100% improvement after Ag filling. Indium and Yb-doped $CoSb_3$ thin films were prepared by PLD.[286] Process optimization studies indicated that for the growth of single-phase skutterudite films, a very narrow process window exists. The $\sigma$ and *S* measurement (within the range 300–700 K) revealed an irreversible change (surface roughness of the films or trace secondary phases) on the first heating cycle in Ar ambient. The Yb filled $CoSb_3$ skutterudite TE thin films were prepared by DC magnetron sputtering.[287] The $\kappa$ of Yb filled skutterudite thin film was much less compared with the bulk Yb filled $CoSb_3$ skutterudite. A maximum $zT$ of 0.48 was observed at 700 K for the 130 nm-thick film.

A series of Indium (In) filled $CoSb_3$ thin films were prepared by a multistep co-sputtering method with various In content.[288] Both the $\sigma$ and $S$ of the $CoSb_3$ based thin film increased with an increase in the In filling. The In filled thin film with a single $CoSb_3$ phase and a well-organized nanostructure caused the large reduction of the $\kappa_l$. As a result, the $zT$ value of the thin film was enhanced from 0.05 (unfilled) to 0.56 with the In 14% filling. Co-Sb thin films were deposited on flexible substrates at room temperature by RF magnetron sputtering with different sputtering power.[289] The atomic ratio of Co/Sb in the Co-Sb thin film was ~1:3 when the sputtering power was 55 W. The prepared film was annealed at various temperatures (from 443 K to 593 K). It was revealed that all the thin films showed $n$-type conductivity and the $CoSb_3$ thin films annealed at 493-593 K were of polycrystalline with (310) preferential orientation. The $S$ of $CoSb_3$ thin films annealed at 543 K increased with increasing the measuring temperature (323-473 K), and the maximum $S$ was −88 μV/K. Similarly, $CoSb_3$ thin films were prepared by DC magnetron co-sputtering and the influence of composition and annealing temperature on the TE properties and micro-structure were studied. [290] The annealed $CoSb_3$ thin film (with slight Sb-excess) showed a single $CoSb_3$ phase when the annealing temperature was < 523 K. All the annealed $CoSb_3$ thin films had nano-sized crystallites (~25 nm). The highest $\sigma$ of $3.9 \times 10^4$/Ω-m and largest $S$ were obtained at RT when the annealing temperature was 523 K. This sample exhibited a maximum $PF$ of $1.47 \times 10^{-4}$ W/m-$K^2$ at 453 K. The $CoSb_3$ skutterudite thin film deposited by using RF co-sputtering at $T_{sub} \approx 200$ °C exhibited a $PF$ of 4.1 μW/cm-$K^2$.[291] The $n$-type $CoSb_3$ thin films prepared using DC sputtering and annealing at 200 °C have shown a $S$ of -250 μV/K.[292] 30-nm thick Co–Sb films were prepared by molecular beam deposition via two different methods (co-deposition on heated substrates or RT deposition followed by a post-annealing step).[293] Films from both methods displayed bipolar conduction leading to a poor $PF$; at low temperatures, variable range hopping (VRH) was identified as the dominant conduction mechanism. $CoSb_3$ thin films were deposited on conducting glass substrates, fluorine-doped tin oxide (FTO) by electrodeposition at different bath temperatures (60°C, 70°C, and 80°C), and the influence of bath temperature on the structure, morphology and electrical properties of films was investigated.[294] Different morphologies: branched nano-flakes, nano-needles evolved as the bath temperature increased; the best crystallinity was observed at a bath temperature of 80 °C. The *PF* increased from 19 to 31 μW/cm-K$^2$, as the compact needle-like microstructures provided better conduction channels. Yadav et al. have studied the deposition potential controlled structural and TE behavior of electrodeposited $CoSb_3$ thin films and observed a highest room temperature *PF* of 7.06 ± 0.09 μW/cm-K$^2$ at a deposition potential of -0.97 V.[295]

Ion implantation[296-297] is used for modifying the electronic and TE properties of materials. In ion implantation, doping can be controlled precisely to enhance the $\sigma$ (increasing the carrier concentration). Furthermore, it can have an effect on the $\kappa_l$ by causing crystal defects and disorders. By ion implantation, the void can be filled or occupied by the Co or Sb sites that can act as electron donors or acceptors, leading to *p* and *n*-type semiconductors.[298] Ion implantations by Fe,[283, 299-300] Ni[301] or light ions (such as B or C ions) are expected to occupy the Co or Sb sites, whereas metals such as Th, Ur, La,[302-303] Eu[304] or Ce[305-306] ions are expected to occupy voids and rattle inside, which is expected to induce a rattling effect (and reduce $\kappa_l$). Ion implantation technique was used to substitute Fe for the Co site in a single phase $CoSb_3$ film.[307] Thin films of single-phase $CoSb_3$ were deposited onto Si(100) substrates via PLD (with a polycrystalline $CoSb_3$ as the target).[307] Then, these films were implanted by 120 keV Fe-ions with three different fluencies. The *S* varied with the fluences in the temperature range of 300 K to 420 K, and was found to be highest (i.e., 254 μV/K) at 420 K for the film implanted with 1 × $10^{15}$ ions/cm$^2$. The Fe substitution on the Co site led to the creation of vacancies, giving rise to a high concentration of holes. The high *S* and high $\sigma$ resulted in the highest *PF* of 700 μW/m-K$^2$ at 420 K for the film implanted with 1 × $10^{15}$ ions/cm$^2$.

Table 1: Periodic structure, growth method (& steps), electrical conductivity ($\sigma$) etc, Seebeck coefficient ($S$), power factor ($PF = S^2\sigma$), thermal conductivity ($\kappa$ or $\kappa_l$), and $zT$ of different material films reported in the literature. Units of charge carrier concentration, $n$ is cm$^{-3}$; Units of mobility, $\mu$ is cm$^2$/V-s. Abbreviations: EChem = Electrochemical (electrodeposition); MS = Magnetron Sputtering; RFMS = Radio Frequency Magnetron Sputtering; DCMS = Direct Current Magnetron Sputtering; $T_{sub}$ = substrate temperature; $T_{an}$ = annealing temperature; $t_f$ = film thickness; RT = room temperature.

| Film, periodic structure | Growth method/substrate | $t_f$ | $\sigma$ ( 1/Ω-cm) &/or $\mu$, $n$ | $S$ μV/K | $PF$ μW/cm-K$^2$ | $\kappa$ W/m-K | $zT$ | Ref. |
|---|---|---|---|---|---|---|---|---|
| | **Si-Ge and PbTe periodic structures and films** | | | | | | | |
| Si(20 Å)/Ge(20 Å) | MBE/Si-on-insulator | | | | 16 | in-plane | $z_{3D}T = 0.1$ | 308 |
| SL | (SOI) (001) | | | | | ~5 | at RT | |
| $Si/Si_{0.9}Ge_{0.1}$ (B-doped) | MBE/SOI (001) | 3.5 μm | $n = 5 \times 10^{19}$ | | | 9.6 at RT | | 60 |
| $Ge/Si_{1-x}Ge_x$ *p*-$Ge/Si_{0.25}Ge_{0.75}$ QW width: (65 Å) | LEPECVD/SOI (001) | | 279 | 255 | 18.2 | 4.0 | 0.135 at RT | 309 |
| *n*-type strained Si(5 Å)/Ge(7 Å) SL | MBE/Si | 120 nm; 100 periods | | -500 | | 1-2 cross-plane | 0.11 at 300 K | 310 |
| *n*-type $Si_{0.7}Ge_{0.3}$ (11 nm)/Si (17 nm) SL | MBE/Si | | | | 25 at RT in-plane | ~2.5 at RT cross-plane | expected~ 0.08−0.1 at RT | 311 |
| *p*-type Si-Ge-Au-B | MBE/sapphire | Si: 1 nm; Ge: 0.1 nm; Au: 1 nm; B: 1 nm | 47.80 | 170 | | 1.09 | 1.38 for $Si_{65}Ge_{27}Au_4B_3$ at 1100 K | 312 |
| $Si_{1-x}Ge_xAu_y$ SL | MBE/sapphire | ~200 nm | | 502 | 40 | 0.72 ( $x$ = 0.4 to 0.6), cross-plane | expected: 1 | 313 |
| *p*-type B-doped Si/Ge quantum dot films | MS/Si | 132 nm | 333; $n = 10^{20}$ $\mu = 1200$ | 850 at 473 K | | | | 314 |
| Bi-doped *n*-type PbTe/ $PbSe_{0.20}Te_{0.80}$ SLs ($t_f$ = 1.8 nm) & *p*-type PbTe SL ($t_f$ < 1 nm) | MBE/$BaF_2$(111); different number of periods (*P*s) | | *n*-type: $\mu$ = 1087-1218 (*P*s of 155-660); *p*-type: $\mu$ = 604-672 (*P*s of 54-460) | | *n*-type SLs: ~26.5 for *P*s of 660-155; *p*-type: 25 at 300 K (*P*s of 54-460) | In-plane *n*-type: 1.7,1.93, 2.3; *p*-type: 2.32 for *P*s = 460 | *n*-type: 0.45 (*P*s = 155); *p*-type: 0.33 (*P*s = 460) | 227 |
| *n*-type PbTe | RF sputtering/glass; $T_{sub}$ = 300 °C | 2-3 μm | 1388; $n = 6 \times 10^{19}$ | 30-60 at RT | ~15 | | | 315 |
| *n*-PbTe (a) *n*-EuTe/PbTe (0.3 nm/11 nm) (b) *n*-EuTe/PbTe (0.3 nm /6.2 nm) (c) | hot wall epitaxy/$BaF_2$ (111) & KCl(001) | | $\sigma/\mu$: (a) 352/ 1410; (b) 238/ 880;(c) 357/ 446; $n$ = (1.6, 1.7, 5) $\times 10^{18}$ | | | | | 316 |
| PbTe/PbSe nanolaminate | ALD/planar Si & porous Si membranes | PbTe: 10 nm; PbSe: 10 nm | | −574.24 ± 20.79 at RT for porous Si | | | | 317 |
| Ag ion implanted PbTe | thermal evaporation-ion implantation/quartz | 340 nm to 360 nm | | 273, 278, 297, 344 & 315 for 0, 1,5,10, & 14 at% of Ag | | | | 318 |
| | evaporation/glass | 300 nm | | 788, 1101, | 0.18-0.7 | | | 319 |
| $PbTe_{100-x}Se_x$ | | | | 921 & 780 for $x$ = 0, 6, 10 & 15 | | | | |
| **$Bi_2(Te,Se)_3$ and $(Bi,Sb)_2Te_3$ based films** | | | | | | | | |
| *n*-type PbTe films | evaporation/mica $T_{an}$ = 623 K | | -450 to -165; $n = 8 \times 10^{16}$ to $3 \times 10^{18}$ | 25-1100 | | 1.9-2.3 | | 320 |
| $Sb_2Te_3$ and $Bi_2Te_3$ thin films | PLD/glass; different $T_{sub}$s | | $Sb_2Te_3$: 512 $Bi_2Te_3$ :33.3 $\times 10^3$ | 413 & -52 resp. | $Sb_2Te_3$: 32 $Bi_2Te_3$: 83 | | | 154 |
| $Bi_2Te_3$ superassembly | PLD/($SiO_2$/Si) | | 73 at RT; $\mu$ = 25.9 | -119 at RT | 1.03 | | | 321 |
| *n*-$Bi_2Te_3$ and *p* -$Sb_2Te_3$ | MOCVD/pyrex & Si substrates | 0.3–7 μm | 1111 & 2857 resp. for *n*- and *p*-types | -210 & 110 resp. for *n*- and *p*-types | | | | 322 |
| (000*l*) *p*-$Bi_{0.5}Sb_{1.5}Te_3$ | magnetron co-sputtering/glass; different $T_{sub}$s (523-723 K) | 605–680 nm | 800 ($T_{sub}$ = 723 K film); $n = 7.9 \times 10^{19}$; $\mu$ = 62 | 219 ($T_{sub}$ = 723 K film) | 38 | | | 323 |
| *n*-type $Bi_2Se_3$ | MOCVD/pyrex | | $\mu$ = 247; $n = 2 \times 10^{19}$ | 120 at RT | | | | 324 |
| Nanocomposited $Bi_2Te_3$:Si | co-sputtering/ glass | | $1.41 \times 10^4$ | 51.3 at RT | 37 | | | 325 |
| *p*-$Bi_{0.5}Sb_{1.5}Te_3$ *n*-$Bi_2Se_{0.3}Se_{2.7}$ | magnetron co-sputtering/ $SiO_2$ | 2 μm for both types | for *p*-: 590; $n = 5.1 \times 10^{19}$; $\mu$ = 73; for *p*-:650; $n = 6.8 \times 10^{19}$; $\mu$ = 62 | 207<br>-196 | | 0.96 (*p*-)<br>0.91 (*n*-) | 0.79 (*p*-)<br>0.82 (*n*-)<br>at RT for both | 326 |
| *n*-type $Bi_2Te_3$ | Ion beam sputtering/BK7 glass; $T_{an}$ = 300 °C | | drops from 720 to 392 | surges from -90 to -168 | 11 | | | 327 |
| *n*-type $Bi_2Te_3$ | RFMS/quartz glass wafer | 1.5 μm | $1.8 \times 10^3$ at RT | max. -85 at 450 K | 8.8 to 11.4 | | | 328 |
| *n*-type $Bi_2Te_3$ | magnetron co-sputtering/quartz glass; ($Bi_2Te_3$ and Te targets ) | | max. 912 | max. -92.4 | 7.08 at 373 K | | | 329 |
| *n*-type Pb doped $Bi_2Te_3$ | RFMS/($SiO_2$/Si) (0.38 at% Pb) | | as deposited /annealed: 1730/720 | as-deposited/annealed: -120.3/-173 | as-deposited/annealed: 2.5/2.15 | | | 330 |
| *n*-type $Bi_2Te_3$ | RF magnetron co-sputtering/ PI; different $T_{sub}$ | 1-2 μm | | max: -193 | 9.7 at 32 °C for $T_{sub}$ = 300 °C sample | | | 331 |
| Bi-Te | RFMS/PI | 1.3 μm | $6.25 \times 10^2$ - $8 \times 10^2$ | -120 in 50 - 300 °C range | 12 at 195 °C (at 57 | | | 332 |
| | | | | | at% Te) | | | |
| Te-rich $Bi_2Te_3$ | RFMS/PI<br><br>$T_{an}$s: 250 to 400 °C | 1.50-1.65 µm | 50 × $10^2$ at 200 °C for sample at $T_{an}$= 400 °C; $\mu$ = 34.03 at $T_{an}$= 400 °C | 71.41 for sample at $T_{an}$ = 300 °C | 11.45 at 300 °C for sample at $T_{an}$= 400 °C | | | 167 |
| *p*-type $Bi_{0.5}Sb_{1.5}Te_3$ | MS/PI<br><br>($Bi_{0.5}Sb_{1.5}Te_3$ and Te as targets) | 420 nm | max. 9.2 × $10^2$; $n$ = 2.73 × $10^{19}$; $\mu$ = 157.8 | | ~23.2 at RT; peak *PF*: ~25 at 360 K | ~0.8 at RT | | 171 |
| *p*-type $Sb_2Te_3$ and *n*-type $Bi_2Te_3$ | RFMS/glass;<br><br>different $T_{an}$s | 1 µm for both types | at RT 1077 (ST, $T_{an}$ = 400 °C); 519 (BT, $T_{an}$ = 300 °C) | at RT ST and BT: 140 and -140 resp. both at $T_{an}$ = 300 °C | ST: 12.7; BT: 10.2, $T_{an}$ = 300 °C samples | | ST: 0.48 BT: 0.60 both at RT | 333 |
| $Bi_2Te_3$/CNT scaffold | MS/CNT scaffold; Te & $Bi_2Te_3$ targets | 20 nm | 0.73-0.86 × $10^2$ at RT | -142 to -147 at RT | 1.6 | 0.19 at RT | 0.25 at RT | 334 |
| *p*-$Sb_2Te_3$ *n*-$Bi_2Te_3$ | co-evaporation/ glass | 700 nm for both types | 961(*p*) & 769 (*n*) $\mu$ = 173; 75 | 171; -228 | 28 & 39 resp. | | | 335 |
| *n*-type Bi–Te | co-evaporation/ MgO & glass | 1 µm | 353.4 $\mu$ = 125 | -228 | MgO: 21.1 glass: 18.4 | | | 336 |
| $Bi_{1.1}Te_{3.0}$/$Sb_2Te_3$ multilayer system | two electron gun evaporators/Si & AlN cross-plane meas. | 10 nm for both types | 127.2 for 5 BT/6 ST layers; 125.5 for 19 BT/ 20 ST layers | | | 0.62 & 0.51 for 11 &29 layers resp. | | 337 |
| $Bi_2Te_3$ | co-evaporation/ PI | ~1 µm | 1000; $n$= $3\times10^{19}$ to $20\times10^{19}$ & $\mu$ = 80-170 | 250 | 48.7 | 1.3 at RT | expected *zT* of ~1 at 300 K | 10 |
| $Bi_2Te_3$ (BT) and $Sb_2Te_3$ (ST) | thermal evaporat ion/glass<br><br>different $T_{sub}$s | 400 nm | | max. -148 (BT); 158 (ST) at $T_{sub}$= 230 °C for both | max. 14.7 (BT, $T_{sub}$ = 260 °C); 13.8 (ST, $T_{sub}$ = 290 °C) | | | 338 |
| *n*-type $Bi_2Te_{3-x}Se_x$ ($x$ = 0 to 1.3) | CVD/($SiO_2$/Si); $Bi_2Te_3$ & $Bi_2Se_3$ single source precursor | | $\mu$ = < 10 | -140 at 500 K for $x$ = 0.8 | 7.6 at 550 K | | 0.34 for $Bi_2Te_{1.7}Se_{1.3}$ at 500 K | 339 |
| $Bi_2Te_3$ (BT) & $Sb_2Te_3$ (ST) | thermal co-evaporation/ ($SiO_2$/Si); $T_{sub}$= 533 K; $T_{sub}$ = 503 K | | 531 (BT) 775 (ST) $\mu$ = 113 (BT) and 303(ST) | BT: −208 ST: 160; | 23 (BT); 19.8 (ST) | | | 340 |
| $Bi_2Te_3$ different morphologies | PVD/glass | | flakes: 57 multilayer: 63 nanowire:22.6 | flakes: -65.7 multilayer: -64.7; NW: - | | | | 341 |
| | | | ordered NW | 118; ordered NW:-150 | | | | |
| $Bi_2Te_3$ (BT) & $Sb_2Te_3$ (ST) | PECVD/ graphene/$SiO_2$/Si and $SiO_2$/Si; different deposition temperatures | | graphene: $\mu$ = 115 & 453 for BT & BST resp. at dep. temp 300 °C | | graphene: 35.2 & 36 for BT & BST at dep. temp 300 °C | | | 342 |
| $Sb_2Te_3$-$In_2Te_3$; $In_2Te_3$; $Sb_2Te_3$ | thermal evaporation/soda lime glass | 350 ± 10 nm | 71.46 at 320 K for 75 % $Sb_2Te_3$-$In_2Te_3$ | max. 237 (for $In_2Te_3$); 148 for $Sb_2Te_3$-$In_2Te_3$ | 0.58 for 25% $Sb_2Te_3$-$In_2Te_3$ at 320 K; 1.18 at 450 K | | | 343 |
| *n* –$Bi_2Te_{2\ 2.7}Se_{0.3}$ | thermal evaporation /glass | 2 μm | $\mu$ = 140 | | 45 | $\kappa_l$ = 0.5 | 1.6, at 300 K | 344 |
| (0 0 1)- *p*-type $Bi_{0.48}Sb_{1.52}Te_3$ & *n*-type $Bi_2Se_{0.3}Te_{2.7}$ | evaporation/PI | 1 & 0.4 μm | 780 (*p*-) & 170 (*n*-type) at 400 K | 35.3 (*p*-) -151 (*n*-) at 400 K | 0.97 (*p*-type) 3.8 (*n*-type) at 400 K | | | 345 |
| *p*-$Sb_2Te_3$ & *n*-$Bi_2Te_3$ | RT MBE/$BaF_2$ or ($SiO_2$/Si); $T_{an}$ = 250 °C | ~1 μm for both types | 1696 & 338; $\mu$ = 402 & 80 resp. ; *n* = 2.6 × $10^{19}$ for both | for *p*- & *n*-: 130 & -153 resp. | 29 and 8 resp. (in-plane) | | | 346 |
| $Bi_{0.5}Sb_{1.5}Te_3$ | DC MS sputtering/Cu | 500 nm | 400 | 150 at RT | 12 at 200 °C | 1 at RT | 0.73 | 142 |
| *n*-$Bi_2$(Te,Se)$_3$ nanowire array | co-evaporation/ $SiO_2$ | 1 μm | 490; *n* = 3.2 × $10^{19}$; $\mu$ = 97 | -189 | | 0.52 | 1.01 at RT | 347 |
| *n*-type $Bi_2Te_3$ layers | EChem/Ni | 600 μm | 400 | -200 | 16 | | | 348 |
| $Bi_2Te_3$ | MBE & EChem/($Bi_2Te_3$/ITO) ; 4 nm seed layer | 8 μm | 617.9 at 300 K | at 300 K −49.9 | | | | 349 |
| *p*-type $Sb_2Te_3$ | EChem/stainless steel; different $T_{an}$s | 300 nm | max. 1034 at $T_{an}$ = 300 °C | max. 170 at $T_{an}$ = 200 °C | 13.6 at $T_{an}$ = 300 °C | | | 350 |
| $Bi_{1.5}Sb_{0.5}Te_3$ 3D freestanding film | EChem/epoxy template | 10 μm | 615 at RT | -145 at RT K | 13 at RT | 1.14 to 0.89 | ~ 0.56 at 400 K | 206 |
| *n*-$Bi_2Te_3$ and *p*-$Sb_2Te_3$ | Pulsed EChem/ $SiO_2$/Si; $T_{an}$ = 250 °C | 100 μm each layer | 600; 400 | -150; 170 | 15; 11.6 | | | 351 |
| $Bi_2Te_3$ | RFMS- annealing & EChem/($Bi_2Te_3$/glass ); 1.3 μm seed layer (AES) | 1.3-2.0 μm | AES layer: 674 EChem film: 821; std. ref. 200 | AES layer: -91 to -96; EChem film: -67 to -75 | AES layer: 5.4; EChem film: 4.2; std.ref. 0.5 | | | 352 |
| *n*-$Bi_2Te_3$ and *p*-$Sb_2Te_3$ | MS/PI | 800 nm | for *n*-, 690; *n* = 9.3 × $0^{19}$; $\mu$ = 57.4; for | -177.2 (*n*-) 165.5 | 21.7 (*n*-) 17.5 (*p*-) | | | 353 |
| | | | *p*-, 640; $n$ = 4.7 × $10^{19}$, $\mu$ = 132 | | | | | |
| Bi–Te films | EChem/(Pt/Si) constant voltage (different dep. pot.) | | 2003.6; $\mu$ = 52.5 at RT (both at -35 mV) | -101 at RT (at -35 mV) | 20.15 at RT (at -35 mV) | | | 354 |
| $Bi_2Te_3$ | EChem/Iindium tin oxide (ITO) | | 40 to 100 at RT; $n$ = 8.7 × $10^{20}$ | -5 to -20 | max. 0.050 at 350 K | | | 355 |
| *n*-$Bi_2Te_3$ | solvothermal/drop-casting on ITO/EChem | 10 μm | 122 | -103 | 1.28 | | | 356 |
| *n*-$Bi_2Te_{2.6}Se_{0.4}$ | DCMS/ K9 glass; $T_{an}$ = 300 °C, 40 min in-situ | 600-800 nm | 840 at RT; $n$ = 4.52 × $10^{20}$; $\mu$ = 11.6 | −77.84 at RT | 4.9 at RT | | | 357 |
| **$Cu_2Se$ films** | | | | | | | | |
| *p*-$Cu_{1.9}Se$ & *p*-$Cu_{1.85}Se$ | evaporation/glass; deposition temp: 40 °C ($Cu_{1.9}Se$) & 200 °C ($Cu_{1.85}Se$) | 50-60 nm | at RT, 5× $10^3$ for $Cu_{1.9}Se$; and 2× $10^3$ for $Cu_{1.85}Se$ | | | | | 358 |
| *p*-$Cu_{1.7}Se$ | RT PLD/glass | 40 nm | at RT, 4.5 × $10^3$; $\mu$ = 2; $n$ = 1.4 × $10^{22}$ | | | | | 359 |
| $Cu_xSe$ | RT MS/glass | 225-375 nm | $Cu_{1.87}Se$:714 $Cu_{1.84}Se$:925 $Cu_{1.83}Se$:2105 $Cu_{1.76}Se$:2538; $n$ = 4 × $10^{20}$ to 5 × $10^{21}$ | | | | | 360 |
| $Cu_{1.95}Se$ | electrophoresis deposition (EPD)/ITO; EPD voltage = 30 V; | 30 μm | 130 at RT $n$ = 1.07 × $10^{20}$; $\mu$ = 7.62 | | | | | 361 |
| $Cu_7Se_4$ | reactive evaporation/glass | 350 nm | 1000; $n$ = 8 × $10^{21}$; $\mu_h$ = 2.4 | 6 | | | | 362 |
| $Cu_{1.83}Ag_{0.009}Se_{0.77}S_{0.23}$ | precursor spin coating & heating /quartz or Si; $T_{an}$ = 350 °C | 60-90 nm | 1010 at RT | 51 at RT | | 0.43 at RT | | 90 |
| *β*-$Cu_2Se$ (Cu/Se = 1 to 9) | pulsed hybrid reactive magn. sputtering (PHRMS)/Kapton | 600-850 nm | 1000 | ~120 | 11 (in-plane) (Cu/Se = 2) | $\kappa$ ~0.8 (out of plane) | 0.4 at RT (Cu/Se = 2) | 363 |
| $Cu_{0.68}Se_{0.32}$ | thermal evaporation/ quartz | 322-726 nm | 100 at 250 °C | 319 at RT; 300 at 250 °C | 3.88 at RT; 7.74 at 250 °C | | | 364 |
| $Cu_2Se$ | MS/glass | 100 nm | 4.55 × $10^3$ at | 33.51 at 374 | 2.4 at 368 | 2.14 at | 0.073 at | 365 |
| | | | 310 K | K | K | RT | 374 K | |
| $Cu_2Se$ | cosolvent method/spin coating/glass | several μm | ~250 | 200-250 | 6.53 | < 1 | 0.34 | [366] |
| $Cu_2Se$ | spin-coating/glass; $T_{an}$ = 400 °C | 50-100 nm | 1000 at RT | 80 at RT | | 0.9 at RT | 0.14 at RT | [367] |
| **SnSe films** | | | | | | | | |
| SnSe | PLD/MgO(100), NaCl(100), $SrF_2$ (100), $SiO_2$ | 100 nm for $SiO_2$ substrate; 50 nm for the rest | for all types of substrates $\mu_h$ = 38–60; 17; 22; & 38–50 resp. | | | | | [368] |
| $Sn_{1-x}Ca_xSe$ 0 < $x$ < 1 | PLD/amorphous $SiO_2$ | 100-200 nm | | ~350 for $x$ = 0.16 | 1.7 at RT for $x$ = 0.16 | | | [369] |
| SnSe | reactive evaporation/glass | 180 nm | 0.11 to 0.15 in 4-300 K; $n = 8.7 \times 10^{16}$; $\mu_h$ = 10.8 | 7863 | 7.2 at 42 K | | 1.2 at 42 K | [370] |
| SnSe . | PLD/($SiO_2$/Si)-normal (N) & 80° glancing angle (G) | 730 nm, 300 nm for normal & glancing angle resp | 21 at 300 K and 274 at 580 K (both for G) | max. 193.7 at 477 K (N); 498.5 at 426 K (G) | 18.5 at 478 K for G | max. out-of-plane: ~0.189 at 340 K | | [371] |
| SnSe surface roughness = 23.17 nm | RFMS/glass; $T_{sub}$ = 558 K; Se/Sn ratio:1.09 | | ~11 at 575 K $n = 4.3\times10^{15}$; $\mu$ = 5.95 | 525-300 in 325 -600 K range | 1.4 at 575 K | | | [372] |
| $a$-axis oriented SnSe films | PLD/sapphire($r$-plane) | 400 nm | 28.3 at 800 K | 264 at 800 K | 1.96 at 800 K | 0.35 at 300 K | 0.45 at 800 K | [277] |
| $p$-$Sn_{0.99}Ag_{0.01}Se$ | CVD/mica/ $SiO_2$/Si | 300 nm | 5.14; $n = 4 \times10^{17}$ | 370 | 0.7 at 300 K | | | [373] |
| $p$-$Sn_{0.85}Ca_{0.15}Se$ | PLD/$Bi_2Se_3$ | 100–200 nm | $n = 10^{19}$ | 367 | 1.7 at RT | | | [369] |
| $a$-axis-oriented SnSe film | PLD/$SrTiO_3$ (100) | 1.2-2.6 μm | $n = 3 \times 10^{16}$-$7 \times 10^{17}$ | | | | | [374] |
| $n$-$(SnSe)_{0.66}(SnSe_2)_{0.34}$ | thermal evaporation/ sapphire | ; 330 nm | 23.8 | -255 | 1.6 at 523 K | | | [375] |
| RS-type (Sn,Pb)Se (58% Pb-doped) | Reactive solid-phase epitaxy (R-SPE)/SnSe/PbSe/MgO; RS-PbSe as sacrificial layer | SnSe + PbSe layers = 50−60 nm | $\mu$ = 290; $p$-to $n$-type conversion: at ≥ 61% Pb with highest $\mu$ = 340 | | | | | [279] |
| SnSe | thermal evaporation/glass | 1 μm | 4.7 at 600 K | > 600 at RT; 140 at 600 K | 0.11 at 505 K | 0.08 (375-450 K) | 0.055 at 501 K | [269] |
| | thermal evaporation/ | 100-500 | 5 at 450 K; $n$ | 209 | 0.21 at 450 | | | [376] |
| *n*-type SnSe | BK7 glass; different $T_{an}$ | nm | = 1.69 × $10^{18}$ & $\mu$ = 4.21 at $T_{an}$ = 673 K | | K ($T_{an}$ = 673 K; max: 1.2 at 473 for (4 h annealing) | | | |
| SnSe | RT MS/fused silica; $T_{an}$ = 700 K | 650 nm | 0.4440 at 300 K; 12.19 at 675 K | 546 to 446 in 300 to 675 K range | 0.13 at 300; 2.4 at 675 K | - | expected: 0.28 at 675 K | 377 |
| *p*-SnSe/Co | thermal evaporation/ $Al_2O_3$ | 300–350 nm | $n$ = $10^{17}$ $\mu$ = 1.434 | | | | 0.32 at 573 K | 378 |
| (1 0 0)-orientated SnSe thin films | PLD/$SiO_2$/Si; $T_{an}$ = 673 K | 400 nm | 1.83 at 573 K | 350 at 300 K | 0.154 at 573 K | | | 278 |
| *a*-axis-oriented SnSe films | PLD | | | | ~4.72 at 600 K | | 1.2 at 600 K | 379 |
| **$CoSb_3$ films** | | | | | | | | |
| Co-excess and Sb-excess $CoSb_3$ | RF and DC magnetron co-sputtering/ flexible | 60-80 nm | 1000 for Sb-excess (82.7% Sb) | ~26 (82.7% Sb content) | Co-rich: 2.4 Sb-rich: 6.9; pristine: ~0.6 | | | 284 |
| Ag doped $CoSb_3$ | MS/glass | 170-187 nm | 65 at 623 K for 2.2 at% Ag film; $n$ = 1.33 × $10^{21}$ | 242 & 175 at 623 K for 0.3 & 1.6 at% Ag resp. | 2.97 for 0.3 at% Ag at 623 K | | | 285 |
| Polycrystalline *p*-$CoSb_3$ | RFMS/GaAs (100); $T_{an}$ = 750 °C | 71 nm | $n$ = 1.1 × $10^{18}$ | ~600 at 700 K | 200 at 700 K | | | 380 |
| Yb filled $CoSb_3$ skutterudite | RT DCMS/oxidized Si; 1020 K heat treatment | 130 nm | at RT: 61; $\mu$ = 80; $n$ = 4.8 × $10^{18}$ | at RT: -160; -270 at ~620 K | | $\kappa$ ~1.1 at 700 K | 0.48 at 700 K in $t_f$ = 130 nm | 287 |
| Indium filled $CoSb_3$ | co-sputtering/PI | 180 nm | 270 for 14% In filling | | | $\kappa_l$: 1.10 - 0.05 with rising In content | 0.05- unfilled; 0.56- 14% In filling | 288 |
| Co-Sb | RT RFMS/PI films; annealing (target: Co/Sb = 1:3 at. ratio) | | 350 for $T_{an}$ = 518 K | -75 at 473 K for $T_{an}$ = 518 K | 1.71 for $T_{an}$ = 518 K | | | 289 |
| $CoSb_3$ | DC magnetron co-sputtering/kapton type PI; annealing | | 390 at RT $\mu$ = 4.5 | 30.7 to 44.3 at RT ($T_{an}$s = 473 to 573 K) | 1.47 at 453 K | | | 290 |
| $CoSb_3$ | RF co-sputtering/ Si(100) at $T_{sub}$ ≈ 200 °C | 400-600 nm | | -50 at 900 K | 4.1 | | | 291 |
| *n*-type $CoSb_3$ | DC sputtering-annealing at 200 °C/Si(100) & $Al_2O_3$; | 190 nm | at RT, 7.87 & 5.9 for Si & $Al_2O_3$ | -250 at 550 K for both substrates | | $\kappa_l$ = 3 at RT | | 292 |
| | $T_{an}$ = 350 °C | | substrates resp. | | | | | |
| $CoSb_3$ | EChem/FTO | | $5.3 \times 10^3$ | 80 at RT | 31 | | | 294 |
| $CoSb_3$ | EChem/dep. potential of -0.97 V | 800 nm | $2.1 \times 10^3$ $\mu$ = 60; $n$ = $10^{20}$ | 58 at RT | 7.06 at RT | | | 295 |
| $CoSb_3$ | PLD and ion implantation (to substitute Fe on the Co site)/Si(100) | ~250 nm | $0.333 \times 10^4$ at 400 K | 254 at 420 K | 7 at 420 K | | | 307 |
| **other films** | | | | | | | | |
| Sn doped GeTe $t_f$ = 235-245 nm | DC magnetron co-sputtering/Si | | 26 at 425 K | | 27.8 at 300 K; 14 at 718 K | | | 250 |
| $n$-$Sr_{0.8}La_{0.2}TiO_3$ | MBE/$SrTiO_3$(001) | 20-700 nm | 3600 | -60 | 12.96 at 73 K | | | 381 |
| $n$-$SrTi_{0.8}Nb_{0.2}O_3$ | PLD/$LaAlO_3$ | 82 nm | 325; $n$ = ~$400 \times 10^{19}$ | -200 | 13 | 3.5 | 0.37 at 1000 K | 382 |
| $n$-$SrTi_{0.8}Nb_{0.2}O_{2.75}$ | PLD/ $LaAlO_3$ | 300 ± 10 nm | 614; $n$ = $132 \times 10^{19}$ | -125 | ~9.6 | | ~0.29 | 383 |
| $p$-type poly Si & $n$-type poly Si | PECVD/$SiO_2$/Si | 1.5 μm for both | $p$-type: 79 $n$-type: 1333 | 460.2; -87.7 | | 30.5; 12.6 (3ω method) | | 384 |
| $p$-Zn-Sb and $n$-Al-doped $Zn_{47}Al_2O_{51}$ | DCMS/kapton type PI | | ZnO: 560; $n$ = $10^{20}$; $\mu$ = 3.5; Zn-Sb :400; $n$ = 3.5 $10^{19}$; $\mu$ = 7.2 | -48; 153 (Zn-Sb) | | | | 385 |
| $n$-2D electron gas-type $SrTi_{0.8}Nb_{0.2}O_3$ | PLD/ $SrTiO_3$(001) and $LaAlO_3$(001) | 49 nm | 6250 at 300 K | -400 | ~$10^2$-$10^3$ | | 1.6 | 386 |
| $p$-CuI | reactive sputtering/ glass + PET | 300 nm | 143; $n = 11 \times 10^{19}$ | 162 | 3.7 | 0.55 | 0.21 at 300 K | 387 |
| Ag-doped ZnSb | DCMS/silica | 760 nm | | | 14.9 at 525 K (1% Ag) | | expected ~0.5 at 575 K | 388 |
| Single-Layer $MoS_2$ | CVD/$SiO_2$/Si | | field effect $\mu$s = 15 & 55 | $30\times 10^3$ at 280 K | | | | 389 |
| Graphene nanoribbon/$MoS_2$ heterostructure film | CVD; graphene on Cu and $MoS_2$ on sapphire; spin coating | | 7.03 at RT | -50.8 at 100 °C | 2.22 at 110 °C | | | 390 |

Table 2: Power generation performance of TEGs; ⦸ = cross section; PI= polyimide; ∥ = parallel

| *p*/*n* materials | Method | Substrate | Number of pairs | Δ*T* (K) | Output voltage (mV) | Output power (μW) | Power density μW/ $cm^2$ | Ref |
|---|---|---|---|---|---|---|---|---|
| Cu-Ni | photolithography /EChem | flexible polymer mold Si→Su8; TC leg length 80–150 μm | | 0.12 | | | 0.012 | 215 |
| $Bi_{0.5}Sb_{1.5}Te_3$/ $Bi_2Te_{2.7}Se_{0.3}$ | EChem | metallic | 160; 50 μm thick legs | 20 | 660 | 77 | 770 | 19 |
| $Bi_{2+x}Te_{3-x}$ | EChem, photolithography and etching | mold Si→Su8; ; *p* and *n* leg-length = 100 – 300 μm | | 40 | | | 278 | 214 |
| $Sb_2Te_3$/ $Bi_2Te_3$ $t_f$ = 500 nm | RF magnetron co-sputtering | kapton | 100 | 40 | 430 | 0.032 | | 391 |
| $Sb_2Te_3/Bi_2Te_3$ $t_f$ = 20 μm | EChem | Si | 242 | 22.3 | 294 | 5.9 | | 392 |
| Bi-Te alloys *p* & *n* column height = 150 μm | pulsed EChem with EG in | $SiO_2$/Si | 1 ⦸ = 300 $\mu m^2$ | 10 | 1.4 | 0.024 | | 211 |
| $Bi_2Te_3$-Cu, Cross-plane | Pulsed EChem | Si→Su8 mold | 71 pairs; 80–135 μm thick legs; ⦸ = 300 $\mu m^2$ | 38.64 | 215.5 | | 2434.4 | 393 |
| $Bi_{0.5}Sb_{1.5}Te_3$/ $Bi_2Se_{0.3}Se_{2.7}$ $t_f$ = 2 μm | magnetron co-sputtering | $SiO_2$ | 98 | 4 | 120.5 | 145.2; Δ*T* = 14.6 K for 160 mA | | 326 |
| $Sb_2Te_3/Bi_2Te_3$ $t_f$ = 10 μm | Pulsed EChem | Si | 127 | 52.5 | 405 | 2990 | 9200 | 394 |
| $Bi_2Te_3$-Cu and $Bi_2Te_3$-$Sb_2Te_3$ $t_f$ = 200 μm | EChem | $SiO_2$/Si | 24 | 2-4 | | | 1 & 4 resp. | 216 |
| $Bi_{0.68}Sb_{1.24}Te_{3.08}$ /$Bi_{1.92}Te_{3.08}$; RTG | EChem | PI | 32 | 57.8 | 170 | 0.247 | | 395 |
| $Sb_2Te_3/Bi_2Te_3$ $t_f$ = 200 μm | EChem | glass template/$SiO_2$/Si | 4 | 138 | 40.89 | 19.72 | | 396 |
| $Sb_2Te_3/Bi_2Te_3$ | lithography/ photoresist melting | Si | 127 | 123 | 18.5 | 3.14 | ~28.5 | 20 |
| $Sb_2Te_3/Bi_2Te_3$ | sputtering | $Al_2O_3$ | 8 | 333 | | 5 | | 397 |
| $Sb_2Te_3/Bi_2Te_3$ | coevaporation | CMOS | | 21 | | 0.7 | ~17.5 | 398 |
| $Bi_{0.4}Sb_{1.6}Te_3$ $Bi_2Te_{2.7}Se_{0.3}$ | flash evaporation | glass | 7 | 30 | ~84.0 | ~0.21 | ~0.07 | 399 |
| $Bi_{0.5}Sb_{1.5}Te_3$/ $Bi_2Te_{2.4}Se_{0.6}$ | flash evaporation | glass | 15 | 40 | | 45.0 | | 400 |
| Bi-Te Π-structure | EChem | $SiO_2$/(100 nm Au/10 nm Cr); leg height = 20 μm | 880; leg ⦸ = 50 × 50 μm$^2$ | | 17.6 | 0.96 | | 213 |
| CuI | reactive sputtering | PET | | 10.8 | 2.5 | 0.008 | | 387 |
| $Bi_{0.5}Sb_{1.5}Te_3$/ $Bi_{0.5}Te_{2.7}Se_{0.3}$ solar- TEG | MS | PI | 12 | | 150 under 30 mW/cm$^2$ illumination | | | 401 |
| $Bi_{0.5}Sb_{1.5}Te_3$/ $Bi_2Te_{2.7}Se_{0.3}$ solar- TEG | MS | PI | 12 | 100 | 220 | ~80 | | 402 |
| $Sb_2Te_3/Bi_2Te_3$ $t_f$ = 1 μm | RFMS | glass | 11 | 28 | 32 | 0.15 | | 333 |
| $Sb_2Te_3/Bi_2Te_3$ | sputtering | PI | 13 | 24 | 48.9 | 0.6935 | | 353 |
| $Sb_2Te_3/Bi_2Te_3$ | Pulsed laser ablation | AlN | 200 | 88 | 500 | | 1040 | 403 |
| Ni–Ag films $t_f$ = 120 nm | evaporation | flexible Si fiber | 7 | 6.6 | | 0.002 | | 404 |
| Zn-Sb/Al-doped ZnO | DC magnetron co-sputtering | PI | 10 | 180 | | 246.3 | | 385 |
| *p*-indium tin oxide (ITO)/*n*-indium oxide ($In_2O_3$); $t_f$ = 1.36 μm & 1.26 μm | RFMS | PI | 473 | | 4.61 and 11.38 | | | 405 |
| $Sb_2Te_3/Bi_2Te_3$ (20 layers) $t_f$ = 1.5 nm each layer | e-beam evaporation | Si | 128 × 256 elements | 0.5 K/μm | 51 | 0.021 | | 406 |
| $Bi_{0.5}Sb_{1.5}Te_3$/ $Bi_2Te_3$ $t_f$ = 300 nm each layer | RFMS; $T_{an}$ = 200 °C | Si | 4 | 50 | ~3.6 | 0.0011 | | 407 |
| $Sb_2Te_3/Bi_2Te_3$ $t_f$ = 20 μm | MS/EChem | $SiO_2$/Si | 3.8× 2.7× 0.8 mm | 25-100 | | 3-56 | 30-62 | 408 |
| $Sb_2Te_3/Bi_2Te_3$ $t_f$ = 100 μm | Pulsed EChem | $SiO_2$/Si | | 22 | 56 | | 3 | 351 |
| $Bi_{0.48}Sb_{1.52}Te_3$/- $Bi_2Se_{0.3}Te_{2.7}$ $t_f$ = 1 & 0.4 μm | evaporation | PI | 11 | 100 | 50 | | | 345 |
| *p*-type $Ag_{0.005}Bi_{0.5}Sb_{1.5}Te_3$ | MS | PI | 4 | 60 | 31.2 | | 1400 | 177 |
| $Sb_2Te_3/Bi_2Te_3$ | MS | PI | 25 | 20 | 78 | 7.9 | | 409 |
| $Ag_{1.8}Se$ | thermal evaporation | PI | 4 | 50 | | | 4680 | 410 |
| $MoS_2$ (*p*-type) and $WS_2$ (*n*-type) planner structure $t_f$ = 700 nm | RFMS | glass | 7.5 cm × 3.6 cm | 240 | 0.7 | | | 411 |

Table 3: Cooling performance of miniature TE devices

| *p*/*n* materials | Method | Configuration | Size ($mm^2$) | Pairs | Cooling performance, $\Delta T$ (K) | Cooling flux ($mW/cm^2$) | Ref |
|---|---|---|---|---|---|---|---|
| $Bi_2Te_3$ SL/$Bi_2Te_3$ SL | MOCVD | cross-plane | | 2 | 55.0 | 128 | 412 |
| $Sb_2Te_3/Bi_2Te_3$ | evaporation | cross-plane | 1& 2 stages | 1 & 3 | 13 & 19 | | 413 |
| poly-Si film/poly-Si film | E-beam evaporation/electroplating | in-plane | 100 | 62500 | 5.6 | | 384 |
| $Sb_2Te_3/Bi_2Te_3$ | EChem | cross-plane | 2.89 | 63 | 2.0 | | 414 |
| ($Bi_2Te_3/Sb_2Te_3$ SL)/ ($Bi_2Te_3/Bi_2Te_{2.83}Se_{0.17}$ SL) | MOCVD | cross-plane | 3.5 × 3.5 | 49 | 14.9 | $1.3 \times 10^5$ | 415 |
| $Si_{0.89}Ge_{0.10}C_{0.01}$/Si SL $t_f = 2\ \mu m$ | MBE | cross-plane | 50 × 50 $\mu m^2$ | | 2.8 at 25 K; 6.9 at 100 K | | 416 |
| $Sb_2Te_3/Bi_2Te_3$ clustered multipillar (MP) structure | EChem; MEMS | column-type; cross-plane | 6 × 6 | 100 ‖ ST; 100 ‖ BT MPs | 1.2 | | 384 |
| $Bi_{0.5}Sb_{1.5}Te_3$/ $Bi_2Te_{2.7}Se_{0.3}$ $t_f = 10\ \mu m$ | DC MS sputtering/Cu | cross-plane | 1 mm × 1 mm | 98 | 6 | $1.38 \times 10^5$ | 142 |
| $Sb_2Te_3/Bi_2Te_3$ $t_f$ = 700 nm for each layer | co-evaporation/ glass | | | 1 | 15 | | 335 |
| $Bi_2Te_3/Sb_2Te_3$ SL/ | MOCVD | cross-plane | | | 43.54 | 257600 | 175 |
| $\delta$-doped $Bi_2Te_{3-x}Se_x$ | | | | | | | |

## 3. Transverse thermoelectric effect

The transverse thermoelectric effect (TTE),[417] also called atomic layer thermopile (ALTP), originates from the anisotropy of the *S* of a material and requires some geometrical conditions such as a tilted angle $\beta$ between the orientation of the *c*-axis of sample, and the surface normal (Figure **22**b) i.e. when $\beta \neq 0^o$ or 90 $0^o$, the off-diagonal terms of the Seebeck tensor will be ≠ 0, and contribute to the TTE. In contrast to the conventional TE effect, where the electrical and thermal flows have the same direction under the temperature gradient ($\nabla T$), the electrical and thermal flows in TTE are perpendicular to each other. In other words, the introduction of a $\nabla T$ along the longitudinal direction can give rise to a transverse electric field in the TTE effect, which means the capability of managing a particular heat flux by adjusting the dimensions of the specimens.[418] Recently, TTE has also been observed in crystalline materials systems[419-421] evoking great interest in its potential use in the construction of power generators, detectors for laser radiation with sub-nanosecond temporal resolution,[422] and thermal sensors.[418, 423]

As shown in Figure **22**b, when a temperature difference $\Delta T_z$ is set up between the top surface and the bottom of the inclined sample along the *z*-axis, a voltage signal $V_x$ is generated along the *x*-axis and is expressed as:[421-422]

$$V_x = \frac{l}{2d} sin2\beta \cdot \Delta S \cdot \Delta T_z \qquad \text{Equation 12}$$

Where *l* is the diameter of the heat spot; *d* is the thickness of the sample; $\beta$ is the inclination/tilt angle of the principle crystal axis of the sample with respect to the surface normal. $\Delta S = |S_{\parallel} - S_{\perp}|$ is the difference of the *S*s along the in-plane ($S_{\parallel}$ or *S* along the *ab*-plane of the material) and out-of-plane ($S_{\perp}$ or *S* along the *c*-axis of the material) directions of the sample.

TTE effect has been observed in SnSe in *a*-axis inclined, oriented thin film samples (75 nm) which were fabricated via the PLD technique on *c*-axis miscut MgO single crystals (Figure **22**a).[424] Large open-circuit voltage signals along the film plane were detected when a very modest $\Delta T$ was applied across the film thickness via irradiation with thermal heaters or by beaming continuous wave (CW) laser beams ($\lambda$ = 532, 980, and 2200 nm). Besides, by adjusting the laser/heater power irradiating the film, the inclination angle of the film and the applied temperature differential across the film thickness, the amplitude of the transverse voltage signals was enhanced. At RT, the $\kappa$ and $\Delta S$ of the film were measured to be ~0.1 W/m-K and 30 μV/K respectively. **Figure 22**d illustrates the voltage response of the 10° tilted SnSe thin films to the 532 nm laser beaming at different $T_{sub}$. As the $T_{sub}$ rises, the amplitude of the laser-induced voltage signal diminishes owing to the reduction in $\Delta T_z$. Similarly, tilted BiCuSeO film coated with a very thin layer (4 nm) of gold nanoparticles (AuNPs) has shown enhanced light-induced transverse thermoelectric (LITT) effect.[425] At RT, the *ab*-plane $\rho$ and $S$ of the BiCuSeO film were ~11.5 mΩ-cm and 204 μV/K, resulting in a $PF$ of ~0.36 mW/m-K$^2$. A *c*-axis tilted BiCuSeO film (150 nm-thick) was deposited on a 20° miscut (001) $LaAlO_3$ substrate via a pulsed laser ablation of the BiCuSeO target and then sputter-coated with a 4 nm-thick Au nanoparticle layer. In both cases of continuous and pulsed light irradiation, the magnitude of the output voltage signal of the LITT effect increased owing to the improved photo-thermal conversion efficiency due to the presence of the AuNPs layer.

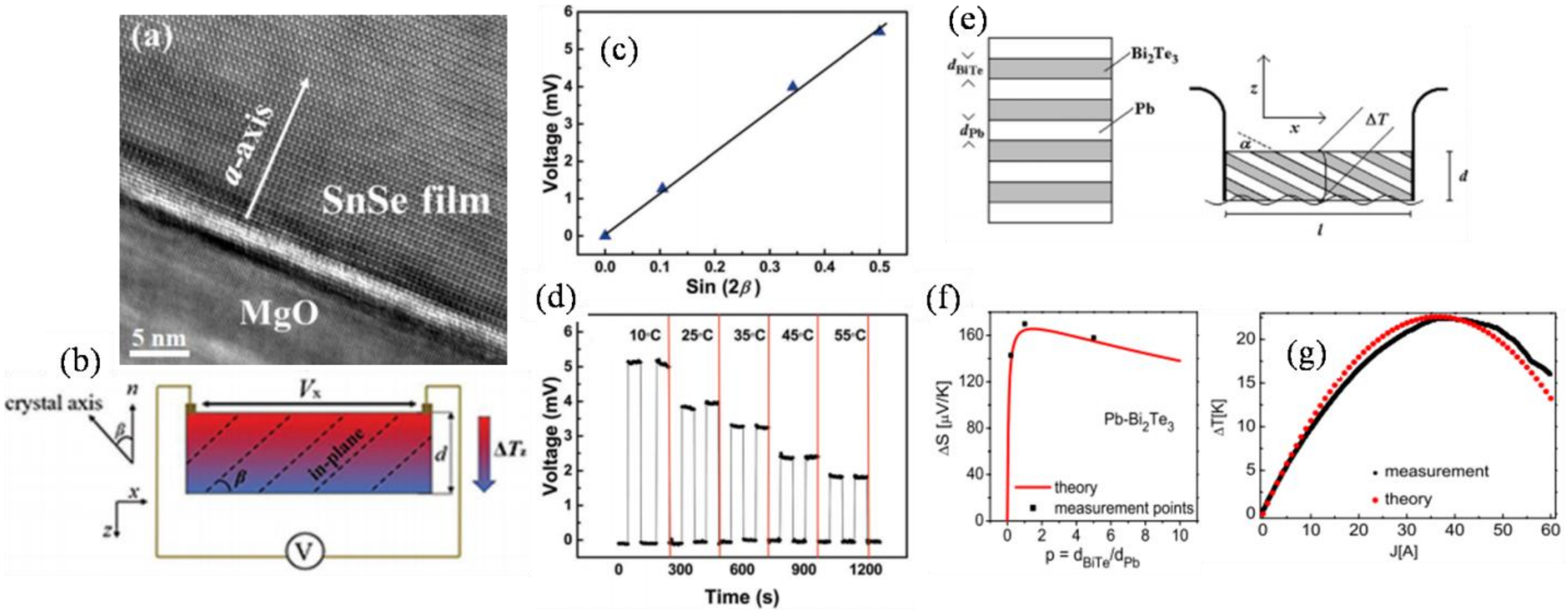

Figure 22: Caption: (a) HRTEM cross-sectional image of the SnSe thin film on MgO single crystal substrate; (b) illustration of the transverse effect (TTE); (c) the dependence of the magnitude of the 532 nm laser-induced voltage signal on the inclination angle of the film; (d) voltage responses of the inclined SnSe thin film to 532 nm laser beam illumination at different substrate temperatures. Adapted with permission from ref.[424] (e) Pb–$Bi_2Te_3$ multilayer stack and tilted sample with tilt angle $\alpha$, thickness ratio $p = d_{BiTe}/d_{Pb}$ prepared from stack, current connections, water cooling of sample bottom, measurement of $\Delta T$ between sample bottom and top side, sample dimensions, and coordinates. Adapted with permission from ref.[419] (f) anisotropic $\Delta S$ vs. $p$, for Pb–$Bi_2Te_3$ multilayer stacks; (g) measurement of $\Delta T$ for a Pb–$Bi_2Te_3$ sample with tilt angle $\alpha = 25^\circ$ and $p = 1$. Adapted with permission from ref.[419, 423]

An ultrafast transverse thermoelectric voltage of 7 ns rise time (i.e. TTE voltage response) has been observed in a highly conducting $La_{0.5}Sr_{0.5}CoO_3$ (LSCO) epitaxial film under the irradiation of a pulse laser ($\lambda$ = 248 nm) with a duration of 28 ns at RT.[418] LSCO thin film (thickness of 200 nm) was deposited on 5° vicinal cut $LaAlO_3$ (LAO) (100) substrates by PLD. This ultrafast response rate is owing to low resistivity of LSCO, which results in a shallow optical penetration depth and hence a short response time. Theoretical analysis, shows that in a tilted Pb–$Bi_2Te_3$ multilayer stacking structure with Pb ($\kappa$ = 35 W/m-K, $\sigma = 5 \times 10^6$/Ω-m, and $S$ = 0 µV/K) and $Bi_2Te_3$ ($\kappa$ = 2.3 W/m-K, $\sigma = 10^5$/Ω-m, and $S$ = 20 µV/K), the magnitude of $\Delta S = |S_{\parallel} - S_{\perp}|$ depends on the thickness ratio $p = d_{BiTe}/d_{Pb}$, where $d_{BiTe}$ and $d_{Pb}$ are the thicknesses of $Bi_2Te_3$ and Pb layers, respectively. **Figure 22**f shows the variation of $\Delta S = |S_{\parallel} - S_{\perp}|$ with $p$. The $\Delta S$ increases rapidly with increasing thickness of Pb layers, attaining a max. $\Delta S \approx$ 200 µV/K at $p$ = 1 and then starts to descend from beyond $p$ = 1. Since $S_{BiTe} \gg S_{Pb}$ and $\kappa_{BiTe} \ll \kappa_{Pb}$, $S_{\perp}$ is primarily determined by the $S$ of $Bi_2Te_3$. Further, because $\sigma_{BiTe} \ll \sigma_{Pb}$, $S_{\parallel}$ is miniscule in comparison to $S_{\perp}$, and therefore, $\Delta S \approx S_{BiTe}$. The same system was tested for its Peltier cooling performance. Stacks consisting of several alternating layers of Pb and $n$-type $Bi_2Te_3$ were prepared by a heating procedure. Tilted samples were obtained by cutting stacks obliquely to the stack axis. Tests were performed for a fixed $d_{BiTe}$ = 1 mm and varying $p$ = 0.2, 1, and 5 (Figure **22**e). Due to a large $\Delta S$ and large $\sigma$ but modest $\kappa$, samples had $\Delta T_{max}$ = 22 K transverse to the applied currents, resulting in a maximum cooling of the sample top side to about 280 K, which was 15 K lower than the ambient temperature (**Figure 22**g). A tubular shaped power generator fabricated using a composite comprising periodically laminated $Bi_{0.5}Sb_{1.5}Te_3$/Ni structure generated 1.3 W power (efficiency of 0.2%) under a $\Delta T$ = 83 K.[426]

## 4. Spin Seebeck effect (SSE) based devices

The conventional Seebeck effect observed in electric conductors relies on a temperature difference ($\Delta T$), which is converted into electric energy. The spin Seebeck effect (SSE) is the Seebeck effect's spin counterpart. SSE is the formation of a spin current, or flow of spin angular momentum, in a magnetic material as a result of a temperature gradient.[427] When a conductor is attached to a magnetic material, the SSE induces a spin-current injection into the conductor.[428] The SSE is applicable in the design of TE generators because the spin current generated by the SSE can be converted into a charge current via the spin-orbit interaction, or the inverse spin Hall effect (ISHE), in a conductive thin film (usually a paramagnetic metal film) adjoining the magnetic material.[428-429]

In conventional TEs the $zT$ of the energy conversion is strictly limited by the trade-off relationship between $S$ and $\sigma$, and $\kappa$. The *figure-of-merit* of SSE generating system is given by $z_{SSE}T = (S_{SSE}^2\sigma_{NM}/\kappa_{FM})T$ where $S_{SSE}$ and $\kappa_{FM}$ are the Seebeck coefficient and thermal conductivity of the ferromagnet (insulator or conductor), $\sigma_{NM}$ is the conductivity of the nonmagnetic layer and $T$ is the absolute temperature. Here, $\kappa_{FM}$ is decided by phonons in the FM layer and the magnitude of $S_{SSE}$ depends on magnons (elementary particles of a spin-wave) in the FM layer. Furthermore, $S_{SSE}$ and $\sigma_{NM}$ are heavily influenced by interface conditions at the FM/NM junction. Aside from $z_{SSE}T$, the TE efficiency of SSE device also depends on the thickness ratio of the FM and NM layers.[430] As a result, one viable approach to enhancing $z_{SSE}T$ would be to devise strategies for controlling phonons and magnons independently in the FM layer while simultaneously optimizing the FM/NM interface condition. The bilayer design of a spin-current-driven device has several advantages over conventional cuboid shaped (or parallelepiped) TE devices. The fundamental trade-off constraints (generation of Joule heat) that appear in traditional TE materials do not apply to the orthogonal energy conversion between heat and charge transport in distinct media in spin Seebeck systems. Furthermore, the spin-current-driven device may be easily scaled by simply expanding the area of a bilayer film, whereas scaling the traditional TE module requires series connections of alternating *p*–*n* pairs and other laborious steps.[3] Nonetheless, scaling the spin TE devices necessitates the development of facile magnetic insulator film processing, as well as property augmentation through a variety of variables. The SSE has been observed in ferromagnetic metals,[429, 431-433] semiconductors,[434-435] insulators,[436-440] and many other materials and therefore enables the design of insulator-based TE generators in conjunction with the inverse spin-Hall effect, which would have been impossible with conventional TE technology. Figure **23**a depicts the dynamics of spintronics, which is traditionally deals with transport studies of electron-carried spin current; the spin Seebeck effect (spin current generation via thermal energy); and the conventional Seebeck effect (charge current movement due to temperature difference). Figure **23**b and c show pure spin currents in non-magnetic metal and magnetic insulator respectively.

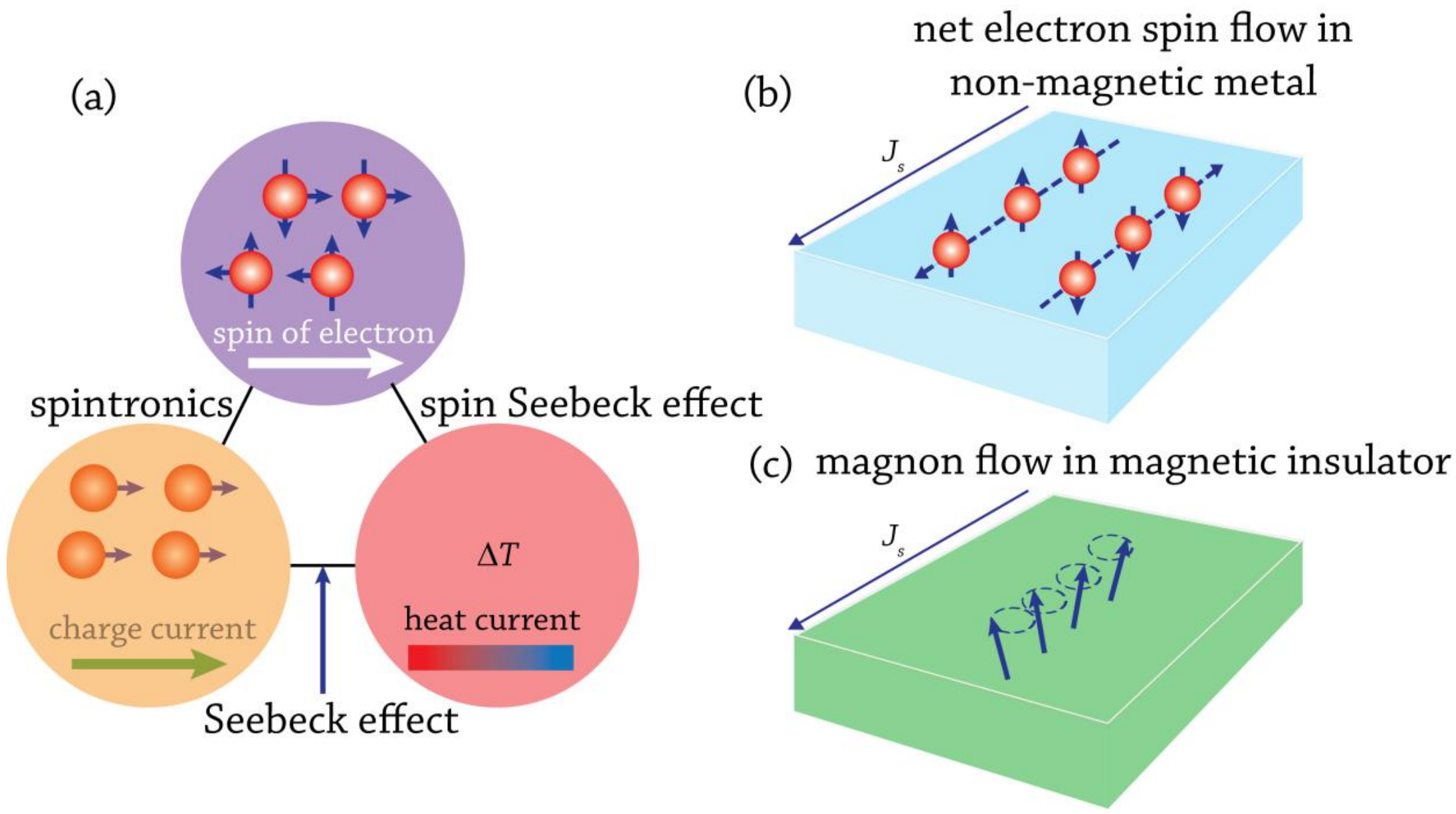

Figure 23: (a) Attributes of spintronics, SSE, and conventional Seebeck effect; (b) spin current in non-magnetic matel (NM) and magnetic insulator.

The intricacies of SSE and SEE voltage dynamics come from different physical mechanisms contributing to the SSE, including interface effects. Generally, the SSE is characterized using two universal device configurations: longitudinal spin Seebeck effect (LSSE) and transverse spin Seebeck effect (TSSE).[429] In LSSE, the spin current is parallel to the temperature gradient,[440-441] while in TSSE, the spin current is perpendicular to the temperature gradient.[442] The first report of a significant increase in transverse SSE (TSSE) in $LaY_2Fe_5O_{12}$ at low temperatures was published in 2010.[443] The TSSE results have proven contentious**,** however**.**[432-433, 444-448] As a result, most recent research is concentrated on the longitudinal configurations.

Figure 24c-d shows the longitudinal SSE (LSSE) device consisting of a ferromagnetic or ferrimagnetic insulator covered by a thin film of a strong spin-orbit coupling paramagnetic material, e.g. platinum (Pt). When a temperature gradient, $\nabla T$ is applied to the ferromagnetic layer perpendicular to the paramagnet/ferromagnet interface, because the interface spin-exchange interaction couples the localized magnetic moments in ferromagnet and conduction electrons in paramagnet, the SSE in the ferromagnet layer creates a spin current in the connected paramagnet layer. In the paramagentic layer, the ISHE converts this spin current into an electric field, $E_{\text{ISHE}}$ as a result of the spin–orbit interaction. The ISHE-induced $E_{\text{ISHE}}$ is generated following the relation

$$E_{ISHE} = (\mathrm{SHA} \cdot \rho)\, J_S \times \sigma \qquad \text{Equation 13}$$

Where SHA and $\rho$ denote the spin-Hall angle and electric resistivity of the paramagnetic layer, respectively, $J_S$ and σ are the spatial direction of the thermally generated spin current and the spin-polarization vector of electrons in the paramagnetic film, respectively. Here, the direction of the $E_{ISHE}$ induced by the ISHE in the conductive film is perpendicular to that of $\nabla T$ in the magnetic material (Figure 24c). $E_{ISHE}$ can be detected by measuring an electric voltage difference $V$ between the ends of the paramagnetic film. (Figure 24c inset). In a LSSE device configuration, the LSSE coefficient is defined as $S_{LSSE} = -E_{ISHE}/\nabla T$,[449] and $E_{ISHE} = V_{ISHE}/L$ where $V_{ISHE}$ is the measured voltage; $L$ is the distance between the electrical contacts on the ISHE film. The thermal gradient $\nabla T$ can be obtained as $\nabla T = \Delta T/t_z$ where $\Delta T$ is the temperature difference between the two surfaces of the sample and $t_z$ its thickness. The determination of a LSSE coefficient requires the simultaneous measurement of the $V_{ISHE}$ and the thermal gradient $\nabla T$.[449] Typically, the LSSE is characterized by applying an external magnetic field to align the magnetization of the soft magnetic insulator along the $x$ direction (Figure 24c). However, the external magnetic field is not required for the LSSE-based TEG, provided the device comprises a hard-magnetic material.[430, 450] There have been several studies on the impact of high-magnetic fields on LSSE in Pt/YIG systems with varying YIG thicknesses at varied temperatures that have yielded some results on the characteristic lengths of the LSSE.[451-453] TE conversion efficiency of the SSE-based device is low. Therefore its improvement is critical to be applicable for future spin-current-driven thermoelectric generator design.

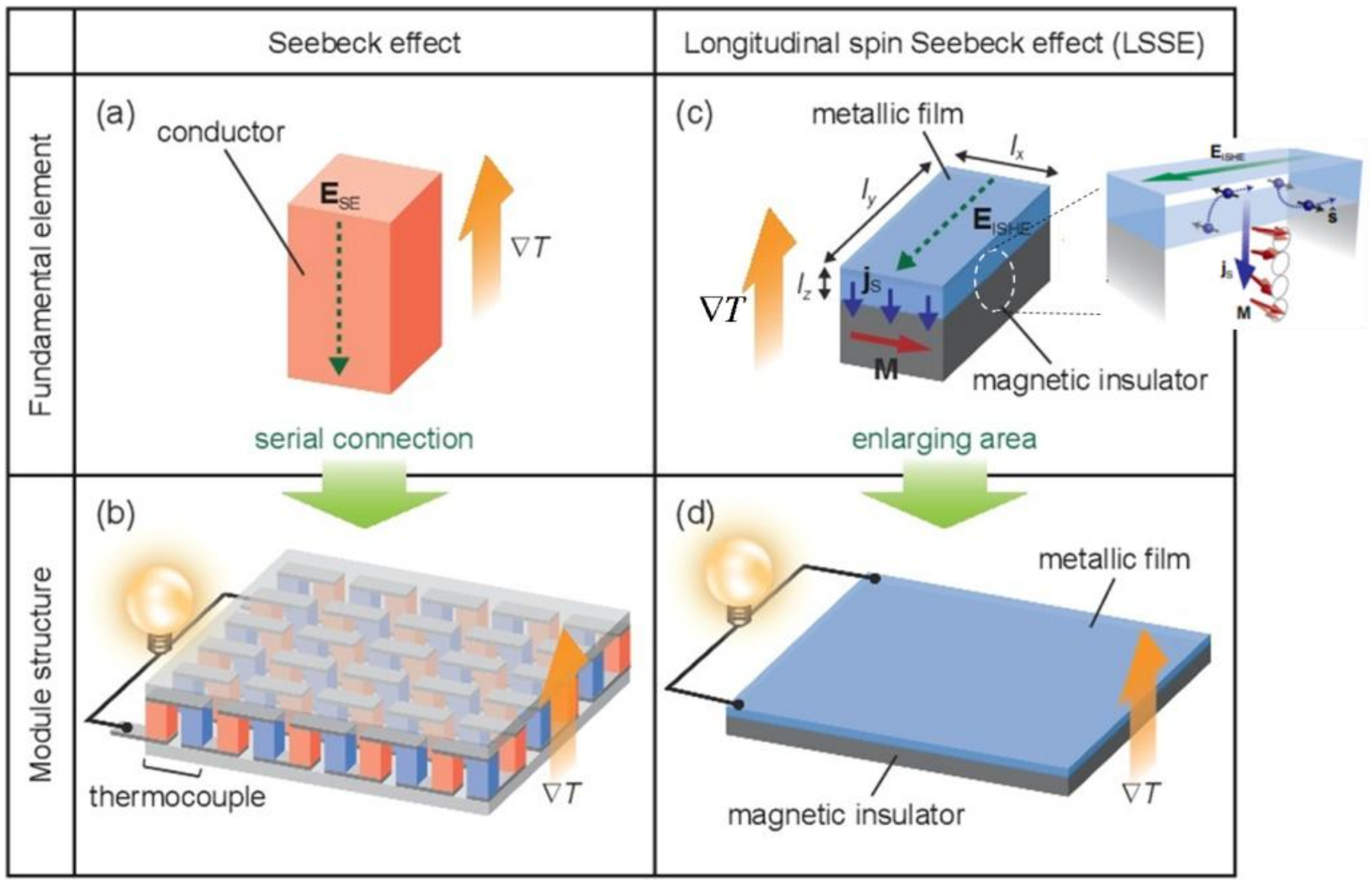

Figure 24: (a-b) Schematic illustrations of the fundamental element and module structure of the conventional TE device based on the Seebeck effect; (c-d) schematic illustrations of the fundamental element and module structure of the TE device based on LSSE. $\nabla T$, $E_{\mathrm{ISHE}}$, $M$, and $J_{\mathrm{S}}$ denote the temperature gradient, electric field generated by the Seebeck effect (due to ISHE), magnetization vector, and spatial direction of the thermally generated spin current, respectively. $l_x(y)$ is the dimension of the metallic film of the LSSE device along the $x(y)$ direction and $l_z$ is the thickness of the metallic film; Right side image in (c). A schematic illustration of the ISHE induced by the LSSE. $\hat{S}$ denotes the unit vector along the electrons pin polarization (\\$M$) in the metallic film. Modified and reproduced with permission from ref.[430]

SSE was first observed in $Ni_{81}Fe_{19}$/Pt films.[429] It was noticed that the spin voltage generated from a $\Delta T$ in a metallic magnet had a linear dependence on $\Delta T$ and only appeared when $Ni_{81}Fe_{19}$ and Pt layers are attached. The thermally induced spin voltage lingered even at distances far from the sample ends. Figure 25(a,1) and (a, 2) show $V$ at $H$ = 100 Oe as a function of $\Delta T$ when the Pt wire is on the lower and higher temperature ends of the $Ni_{81}Fe_{19}$ layer respectively. The magnitude of $V$ was proportional to $\Delta T$ at both ends of the layer. Figure 25(a, 3) shows that the electric voltage signal disappears in a plain $Ni_{81}Fe_{19}$ film when no Pt wire is attached.[429]

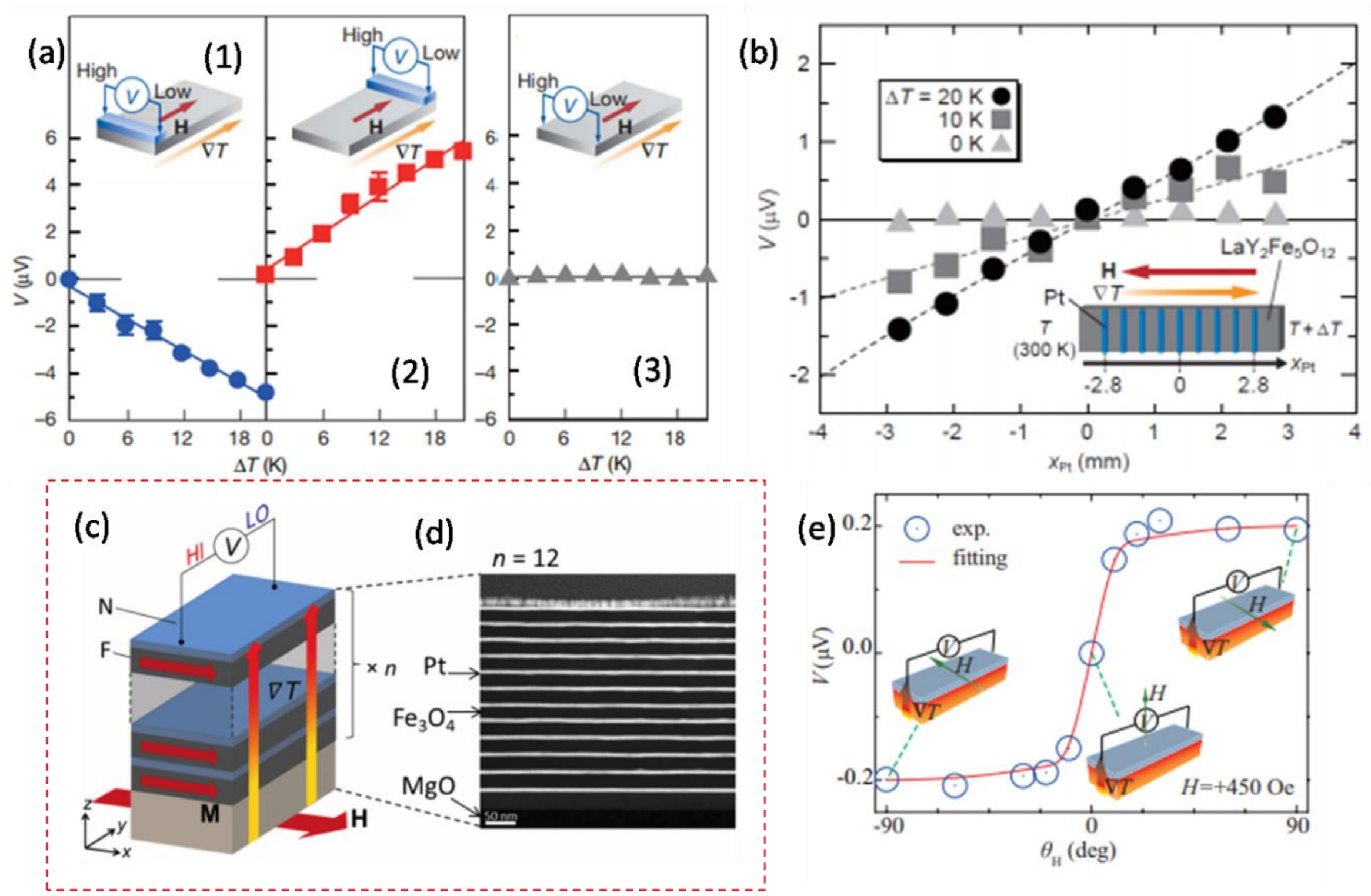

Figure 25: (a) $\Delta T$ dependence of the electric voltage difference $V$ between the ends of the Pt wire in the $Ni_{81}Fe_{19}$/Pt sample. Adapted with permission from ref.[429] (b) the dependence of $V$ on the position of the Pt wire from the centre of the $LaY_2Fe_5O_{12}$ layer along the $x$ direction for different values of $\Delta T$. Adapted with permission from ref.[427] Bottom left frame. $Fe_3O_4$/Pt bilayers (c) schematic illustration of the $[F/N]_n$ multilayers, $n$ denotes the number of $Fe_3O_4$/Pt bilayers. (d) the scanning transmission electron microscopy image of the cross section of a $[Fe_3O_4(23)/Pt(7)]_{12}$ sample. Adapted with permission from ref.[454] (e) $\theta_H$ dependence of voltage in the $IrO_2$/YIG sample at $H$ = +450 Oe. Reproduced with permission from ref.[455]

Uchida et al. reported electric-voltage generation from heat flowing in an insulator. Despite the absence of conduction electrons, a magnetic insulator $LaY_2Fe_5O_{12}$ found to convert a heat flow into spin voltage.[427] The single-crystal $LaY_2Fe_5O_{12}$ (111) film (3.9 μm-thick) was grown on a $Gd_3Ga_5O_{12}$ (111) substrate by liquid phase epitaxy (LPE). Sputtered Pt wires (15 nm) on film converted this spin voltage into electric voltage via the inverse spin Hall effect. Figure 25b shows the dependence of $V$ on $x_{Pt}$, the position of the Pt wire from the centre of the $LaY_2Fe_5O_{12}$ layer along the $x$ direction, in the $LaY_2Fe_5O_{12}$/Pt system for various values of $\Delta T$. To gauge the distribution, nine Pt wires were attached to the $LaY_2Fe_5O_{12}$ layer at intervals of 0.7 mm. The middle Pt wire was attached to the centre of the $LaY_2Fe_5O_{12}$ layer ($x_{Pt}$ = 0 mm). The spatial distribution of the thermally induced spin voltage was then characterized.

TE performance of spin Seebeck devices based on $Fe_3O_4$/Pt junction systems is reported.[454] The $Fe_3O_4$ (a ferromagnetic insulator) films were deposited on MgO(001), substrates by PLD using a KrF excimer laser, $\lambda$ = 248 nm, 10 Hz repetition rate, and $3 \times 10^9$ W/cm$^2$ irradiance. The Pt films were deposited by DC magnetron sputtering. The TE performance of two types of SSE devices were investigated: spin Hall thermopiles[456] and magnetic multilayers formed by repeated growth of *n* number of [F/N] bilayers (referred to as [F/N]*n*). The multilayer SSE devices yielded enhanced TE voltage and power compared to the conventional bilayer systems (Figure 25c-d, bottom left frames). The magnitude of the SSE thermopower of the single bilayer sample and spin Hall thermopile were 10 µV/K and 50 µV/K respectively.

The platinum (Pt) and yttrium iron garnet, $Y_3Fe_5O_{12}$ (YIG) bilayer system has drawn ample attention for studying the SSE [427, 438, 457] and for other spin dependent transport properties.[458-460] Pt is commonly used for spin current detection/spin-charge conversion due to its relatively large spin Hall angle.[461-463] The spin Hall angle (SHA) is a measurement of a material's ability to generate a transverse spin current from a charge current due to spin-orbit coupling and disorder in the spin Hall effect.[464] Because YIG is an insulator (band gap of 2.85 eV),[465] there is no direct injection of a spin-polarized electron current into the Pt layer. Therefore, spin pumping in YIG/Pt systems is only possible through exchange interactions between conduction electrons in the Pt layer and localized electrons in the YIG film. Since spin pumping is an interfacial effect, achieving a high spin-to-charge current conversion efficiency requires an optimal interface quality.[466] In addition to Pt, the other types of materials have also been used (including spin detecting layers): conductive films, simple/heavy metals (Au,[441, 467] Mo,[441] Ni,[430] Ta,[441, 468-469] Pd,[470] V,[471] Cr,[441] Nb,[456] Ir,[441] W,[441, 468, 472-473], and Ti[441, 474]), alloys (NiFe,[441, 475] IrMn,[476] FePt,[477] CuBi,[478] $Mn_2Au$,[479] and AuCu[480]), bilayer metals (Pt/Cu,[438, 472-473] Pt/FeCu,[481] Pt/Au,[482] CoFeB/Ti,[483] Pt/Ti,[474, 483] and Co/Cu,[484]), and oxides ($IrO_2$[455] and $SrRuO_3$,[485]).

Yttrium iron garnet (YIG) is an ideal ferrimagnetic insulator due to low intrinsic magnetic damping constant ($\alpha_o = 2.3 \times 10^{-4}$ for 20 nm thick film),[486] which permits long-distance magnon travel, high resistivity ~$10^{12}$ Ω-cm at room temperature (RT),[463] and a large band gap at RT. Its surface magnetic anisotropy plays an important role in the magnetization process and spin dynamics.[487], [438, 455, 470, 475, 480] The small $\alpha_o$ signifies the effectiveness of the delivery of the generated magnons (Bosons with a spin equal to one) in the ferrimagnetic insulator film to the interface with low loss, which is very important in spin thermoelectricity as well as in magnon spintronics. When a spin detection layer (for example, Pt) is deposited on top of the spin generating material, the system may have an effective damping constant ($\alpha_{eff}$) originating from both intrinsic and spin pumping at the interface ($\Delta\alpha$) between the spin generating layer and the spin detection layer due to effective spin-mixing conductance, $g_{eff}^{\uparrow\downarrow}$ (spin-pumping efficiency, a deterministic parameter in spin transport across the interface); therefore, $\alpha_{eff} = \alpha_o + \Delta\alpha$.[488-489] where, $\Delta\alpha = \frac{g\mu_B}{4\pi M_S}\ g_{eff}^{\uparrow\downarrow}\ \frac{1}{t}$. Here, $M_S$ and $t$ are the saturation magnetization and thickness of the magnetic material, $g$ is the $g$ factor, and $\mu_B$ is the Bohr magneton. The $g_{eff}^{\uparrow\downarrow}$ is determined by the thickness and saturation magnetization of the ferromagnetic layer, as well as the spin-diffusion length of the paramagnetic layer, spin-flip scattering associated with the interfacial structure (known as spin-memory loss),[490] and the spin-flip probability at the interface.[491] The $g_{eff}^{\uparrow\downarrow}$ and SHA are determined by using techniques: micro-wave-induced spin pumping,[463, 492-497] spin Hall magnetoresistance,[498-500] and nonlocal methods.[501-503] In these methods, microwaves or charge currents are applied (instead of temperature gradient) to paramagnet/ferromagnetic interfaces to characterize parameters.

The other candidates for magnetic insulators include garnet ferrites ($Y_{3-x}$(Ca,Nd,Gd,Bi)$_x$-$Fe_{5-y}$(Al, Mn, Ga,V, In, Zr)$_y O_{12}$,[441, 481, 504]; insulating paramagnets $Gd_3Ga_5O_{12}$(gadolinium gallium garnet) and $D_y$ScO3,[472, 505] and $Gd_3Fe_5O_{12}$[472]), spinel ferrites ((Mn, Zn)$Fe_2O_4$,[506] $NiFe_2O_4$,[507] $Fe_3O_4$,[508] $Co_xFe_{3-x}O_4$,[509] $CoCr_2O_4$,[510] $Ni_{0.2}Zn_{0.3}Fe_{2.5}O_4$[511]), hexagonal ferrites ($BaFe_{12}O_{19}$[450] and $Ba_{0.5}Sr_{1.5}Zn_2Fe_{12}O_{22}$[512]), perovskites ($La_2NiMnO_6$,[485] $La_{0.67}Sr_{0.33}MnO_3$,[485]), rutile ($MnF_2$)[473] corundum ($Cr_2O_3$),[513] and half metallic Heusler compounds.[514]

As stated earlier, the SSE signal is indirectly analyzed by measuring the ISHE. Therefore, not only are the bulk properties of magnetic materials important, but the pumping efficiency of the spin current across the interface is also very important. For metallic interfaces between metallic ferromagnet/nonmagnetic metal layer systems, interface effects can be critical in defining spin current effects and dominating observed signals.[515-516] The epitaxial growth of detecting layers with large SHA on oxides (such as YIG) is difficult, and earlier studies have demonstrated that the SSE signal may be improved by introducing an atomically smooth interface with a preferred orientation,[451] as discussed in the next paragraphs. At RT, different detection materials possess different SHAs and lead to different SSE signal amplitudes.[517] Further, the spin transmissivity probability can be spectrally dependent.[518] When magnon energy reaches comparable to the relevant energy scale of electrons, one can expect a significant shift in magnon transmissivity across different types of interfaces.[519] The spectrally dependent magnon transmissivity probability can change depending on the interface condition, resulting in interface-dependent LSSE signals as in the case of HM/YIG interfaces.[451] Further, spin pumping at the HM/YIG interfaces may be influenced by interfacial damping.[488] Therefore, understanding the process of spin transmission across the interface in regards to the atomistic interface structure/condition is very important.

Qiu et al. investigated the longitudinal spin-Seebeck effect (LSSE) effect in Pt/YIG bilayer systems.[428] The amount of the voltage induced by the LSSE is found to be sensitive to the Pt/YIG interface condition. Large LSSE voltage was generated in a Pt/YIG system with a better crystalline interface, while the voltage decayed sharply when an amorphous layer was introduced at the interface. In this study, the YIG films (~4.5 µm-thick) were grown on (111) gadolinium gallium garnet (GGG) substrates by using the LPE method. The YIG films were then annealed under oxygen at 1073 K for 2 hours to improve the crystal structure; then bombarded by ion beams to create amorphous layers. After the annealing or ion beam bombardment, 10 nm-thick Pt films were deposited on those YIG films using a PLD system and characterized. The same authors reported the observation of LSSEs in an all-oxide bilayer system comprising a conductive $IrO_2$ film and an ferrimagnetic insulator, YIG film.[455] $IrO_2$ (*n*-type transparent semiconductor ) was selected for detecting the SSE because of its relatively SHA.[496, 520] The single-crystalline YIG film (~4.5 µm) was deposited on a 0.5-mm-thick (111)-GGG substrate by using an LPE method. A 30-nm-thick $IrO_2$ film was then deposited on the YIG film by using an RF-sputtering method. When out-of-plane field angle, $\theta_H \neq 0^\circ$, a finite voltage signals appeared and their magnitude and sign consistently changed with $\theta_H$. However, no voltage signal was observed when the magnetic field is ⊥ to the film surface, i.e.$\theta_H = 0^\circ$ (Figure 25e bottom frames). Further, the LSSE voltage in the $IrO_2$/YIG sample was much smaller than that in the Pt/YIG sample, which may be due to the small spin-mixing conductance at the $IrO_2$/YIG interface.

Lee et al. investigated the effects of a surface coverage of large-area multilayer $MoS_2$ layer in Pt/$MoS_2$/YIG trilayers via LSSE measurement (Figure **26**a).[521] The magnitude of both electric voltage $V_{ISHE}$ in Pt/$MoS_2$/YIG trilayers and the spin current $J_S$ injected into the Pt layer were enhanced at a $MoS_2$ surface coverage of ~42%. No spin current was observed at 100% $MoS_2$ coverage. A combination of e-beam evaporator, CVD, wet-transfer technique, LPE, and sputtering techniques were used to prepare the trilayers. The thicknesses of Pt, $MoS_2$, and YIG layers were ~5 nm, ~20 nm, and 1 nm respectively. This improved spin-to-charge conversion efficiency in the trilayer system with $MoS_2$ surface coverage of 42% is ascribed to a large spin current and larger spin-mixing conductance at the trilayer interface compared to that of a Pt/YIG bilayer.

Bi-substituted yttrium iron garnet ($BiY_2Fe_5O_{12}$, a magnetic–insulator, abbreviated as Bi:YIG) and Pt films were coated on a (111)-oriented GGG ($Gd_3Ga_5O_{12}$) substrate using a spin-thermoelectric (STE) coating technique.[457] The fabrication steps are shown in Figure **26**(e, right side frame). Variation of SSE-induced TE voltage, $V$ with $\Delta T$ shows that $V$ increases in proportion to $\Delta T$ with a slope of $V/\Delta T$ = 0.82 μV/K. STE coating was also implemented on a 0.5-mm-thick glass (synthetic fused silica) substrate. Figure **26**f, right side frame) shows the measured voltage $V$ as a function of $H$ for $\Delta T$ = 3 K, showing a SSE-induced TE voltage, $V$ ~0.6 μV.

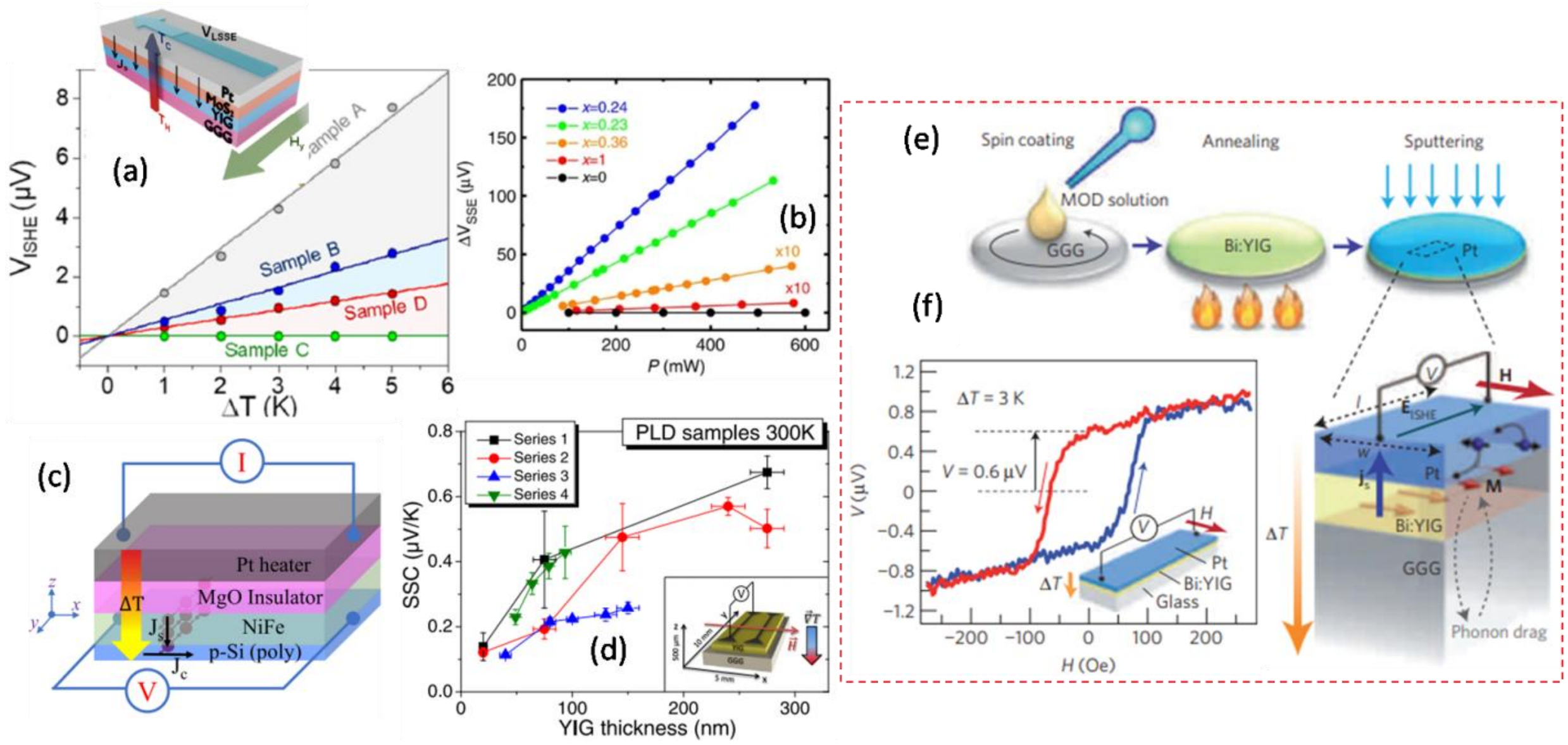

Figure 26: (a) LSSE characterization of Pt/$MoS_2$/YIG trilayers. $V_{ISHE}$ vs. $\Delta T$ for Samples A, B, C, and D at various $\Delta T$s ranging from 1 to 5 K; (inset) Illustration of LSSE measurement in Pt/$MoS_2$/YIG trilayers. Adapted with permission from ref.[521] (b) the heater power dependence of $V_{SSE}$ in five QL $(Bi_xSb_{1-x})_2Te_3$/YIG samples with various $x$ values. Adapted with permission from ref.[522] (c) The schematic showing the longitudinal spin-Seebeck effect measurement setup in NiFe (25 nm)/*p*-Si. Adapted with permission from ref.[523] (d) SSC as a function of YIG layer thickness for samples produced by PLD. The series highlights samples with identical interface. Adapted with permission from ref.[438] Right side frame (e) a schematic of the STE-coating process. The Bi:YIG film was formed by spin-coating the MOD solution, with subsequent annealing. Following this the Pt film was sputtered. Magnified image shows spin-current-driven

TE conversion in the STE coating. $\Delta T$ is the temperature difference applied including the substrate. In this set-up, the spin current $j_s$ is driven perpendicularly to the Pt/Bi:YIG interface and then converted into an electric field $E_{ISHE}$ as a result of the inverse spin hall effect (ISHE) in Pt; (f) TE voltage generation from the 5 × 2 $mm^2$ STE-coated glass for $\Delta T$ = 3 K. The *H*-field dependence of *V* shows SSE-induced TE conversion. Reproduced with permission from ref.[457]

Jiang et al. reported a spin-Seebeck voltage ($V_{SSE}$) of 175 µV in $(Bi_xSb_{1-x})_2Te_3$/YIG sample ($x$ = 0.24) topological specimen with an area of 900 µm × 100 µm (**Figure 26**b).[522] Lin et al. showed a similar $V_{SSE}$ with NiO spacer in between YIG and Pt (i.e. Pt/NiO/YIG). Also, they reported the highest SSC of 6 µV/K.[524]

It is reported that, at low temperatures, SSE in transverse Bi-YIG/Pt devices was elevated due to substrate-induced phonon drag effects influencing the movement of magnons.[443] Taking a cue from the previously known phonon-drag for the conventional Seebeck effect, a significant interaction between phonons and magnons was proposed, with the phonons pulling thermally excited magnons into the detection layer, therefore increasing the pumped spin current into the detection layer. The observed increase in the SSE signal at low temperatures has therefore been suggested to be due to phonon-magnon coupling.[525-528] However, the observed long-distance transmission of magnons in magnetic insulators favors the idea of a relatively weak interaction with phonons and impurities.[427] Recent temperature-dependent $\kappa$ measurements of YIG single crystals reveal that the phonon contribution to the $\kappa$ reaches its maximum around 25 K,14 which is 50 K lower than the known low-temperature SSE peak.[443, 473] Furthermore, the magnon contribution to the $\kappa$ has been determined to be < 5% at 8 K and rapidly declines with increasing temperature.[528] As a result, the viability of a phonon-drag SSE mechanism that relies on strong phonon-magnon interactions to explain YIG's SSE temperature dependency is debatable. The following section covers some instances of both scenarios.It is critical to understand the important interactions between magnons and phonons, as well as the length scales that control magnon transport.

There are some reports on the length scale of phonon and magnon transport in LSSE with YIG as the magnetic material. By measuring the magnetic field dependence of the $\kappa$ and specific heat of single-crystal YIG from 2 to 300 K. Boona et al. have estimated the bulk magnon thermal MFP within the framework of the kinetic gas theory and found this value is 100 microns at 2 K, and it decreases rapidly as the temperature is increased.[528] The result supports the idea that high-energy bulk magnons can not be solely or directly responsible for coherent spin transfer at the macro scale. Agrawal et al. estimated the diffusion length for thermal magnons as ~500 nm in the YIG/Pt system.[529] According to their experiment, a specific volume of the YIG film near the interface contributes to the LSSE, and the temporal dynamics of the LSSE voltage are dependent on the magnon transport in the YIG volume. Kehlberger et al. investigated the origin of the SSE in insulating ferrimagnetic, YIG of different film thicknesses (20 nm to 50 μm) at RT and 50 K.[438] $Y_3Fe_5O_{12}$ samples covered with Pt (i.e. YIG/Pt) were employed in the investigation. Even within a group of samples with identical magnetic properties, thickness-dependent changes of the longitudinal spin Seebeck signal were observed. With increasing YIG film thickness, an increase and saturation of the spin Seebeck coefficient (SSC) was observed (Figure **26**d). This characteristic behavior is attributed to a finite propagation length of thermally excited magnons, created in the bulk of the ferromagnetic material. For PLD samples, at RT, the mean propagation length of thermally excited magnons was ~100 nm. For LPE grown samples, at RT, the SSC was constant for thicknesses above 1.5 μm and the calculated maximum magnon propagation length was ~1 μm. Measurements at 50 K indicate the magnon propagation length is coupled to the absolute system temperature and the magnon propagation length was calculated to be 7 μm.

Huang et al. revealed that, in the YIG/Pt structure, the Pt film itself turns strong ferromagnetic because of its close proximity with the magnetic properties of the YIG layer.[530] If this result is true, then the observed voltages attributed to SSE are not merely due to the spin-Seebeck effect, but also have a contribution resulting from the anomalous Nernst effect in Pt.[531]. Siegel et al., on the other hand, demonstrated that magnon transport theory is credible in explaining observed SSE results.[439] They also found that substrate-induced phonon-drag effects typically observed in the transverse device setup do not appear in the longitudinal device setup. Besides, based on their magnetoresistance measurements, they have refuted the notion of magnetic proximity effects are causing Pt films to become magnetic. In $Ga_{0.84}Mn_{0.16}As$/GaAs system, the amplitude of the SSE changes in proportion to the $\kappa_l$ of the GaAs substrate and the phonon-drag contribution to the thermoelectric power of the GaMnAs, indicating that phonons drive the spin redistribution. The spin-Seebeck coefficient reaches a maximum of ~5 μV/K when the $\kappa_l$ and phonon–electron drag (PED) are maximal.[532] Jaworski and coworkers have observed a large spin Seebeck voltage (~mV/K) in non-magnetic semiconductor (*n*-InSb).[435] This large spin Seebeck voltage is reported to have originated from the simultaneous strong phonon–electron drag and spin–orbit coupling in InSb. In $\gamma-Fe_2O_3$/Pt system, Jimenez-Cavero et al. have experimentally quantified the bulk magnon accumulation and interfacial contributions to the LSSE.[533] The bulk magnon accumulation dominated the SSE in films of different thicknesses. However, the interfacial contribution was not small. It represented from ~33% to ~12% of the total voltage. In addition, bulk LSSE coefficient values were found to be 0.68 V/m-K for the 14.5 nm thick sample and 0.46 V/m-K for the 77 nm thickn sample. The interfacial LSSE coefficient was estimated to range from ~0.1-1 V/m-K. The length of a thermal magnon accumulation in $Fe_2O_3$ is reported to be 29 nm.

Studies have shown that SSE is very sensitive to the paramagnetic metal (PM)/ferromagnetic materials (FM) interface conditions: surface magnetic anisotropy, surface crystallinity, and surface roughness. Aqeel et al. investigated the conditions of the interface quality on the SSE of the YIG/Pt bilayer system.[460] The magnitude and shape of the SSE were found to be influenced by mechanical treatment of the YIG single crystal surface. The saturation magnetic field for the SSE signal increased from 55.3 mT to 72.8 mT with a mechanical treatment and this change in the magnitude of saturation magnetic field can be associated with the presence of a perpendicular magnetic anisotropy due to the treatment generated surface strain or shape anisotropy in the Pt/YIG system. It is reported that the magnetic-field dependence of the LSSE in a Pt/YIG system deviates from the bulk magnetization curve of the YIG slab in a low field regime.[487] This deviation is ascribed to the presence of intrinsic perpendicular magnetic anisotropy (PMA) near the YIG surface, which reduces the magnitude of the in-plane magnetization component, thus lowering the spin pumping efficiency across the Pt/YIG interface. The effect of the near-surface magnetic anisotropy on the SSE persisted even when the surface roughness of YIG was very small.

Saiga et al. investigated the effect of annealing on the SSE signal in the polycrystalline YIG/Pt system.[534] It was observed that the annealing YIG in air at 1073 K before Pt deposition increases the spin Seebeck voltage by 2.8 times that of a non-annealed sample. This increase is alluded to the improved interface crystallinity. Measurements of $\kappa$ and spin Seebeck thermopower in nanostructured YIG show a reduction in both (when compared to a YIG single crystal), with the latter suffering a significant reduction, implying that YIG nanoparticles suppressed the propagation of long-wavelength magnons.[535] In this experiment, YIG samples with different nano sizes obtained from ball milling were sintered using SPS.

Empirical evidence suggests that in $Ni_{81}Fe_{19}$/Pt system, the $\alpha_{eff}$ increase due to spin pumping is inversely proportional to the thickness of the $Ni_{81}Fe_{19}$,[489, 536] attributed to the interfacial character of the spin-pumping effect. Similarly, the $\alpha_{eff}$ of the YIG/Pt samples (with YIG thicknesses of 20, 70, 130, 200, and 300 nm) is shown to be enhanced for smaller YIG film thicknesses, ascribed to the interface condition on spin-pumping effect ( i.e. increased surface/volume for thinner samples).[488] With increasing YIG film thickness, ISHE-voltage increased and then started to saturate above thicknesses near 200-300 nm.[488] The temperature-dependent LSSE in heavy metal (HM)/YIG systems was investigated as a function of YIG film thickness, magnetic field strength, and different HM layers.[451] The LSSE signal exhibited a significant increase with decreasing temperature, resulting in a prominent peak at low temperatures. For thick YIG films, $\xi$ rises with decreasing temperature, allowing more magnons to reach the detecting layer and enhancing the LSSE signal (Figure 27a-b). Because at higher temperatures $\xi$ is restricted in thinner YIG films, the peak temperature shifts towards higher temperatures as film thickness decreases, and the total amplitude of the LSSE signal falls as well. Where $\xi$ is the LSSE's characteristic length, defined as the effective propagation length of all thermally excited magnons arriving at the neighboring detection layer. Figure 27c shows the $\sigma_{LSSE}$ vs. film thickness at 300, 200, and 120 K. The LSSE progressively rises with increasing thickness and saturates at a certain thickness, indicating that only magnons excited at distances smaller than ξ from the interface may contribute to the observed ISHE signals. Also, it was observed that in YIG, $\xi$ monotonically increases with decreasing temperature ($\xi \sim 1/T$) quite similar to $\xi \sim 1/\alpha$,[537-538] showing a longer propagation length correlates well with a lower magnetic damping (Figure 27d). The $\sigma_{LSSE}$ of the W/YIG hybrid was negative LSSE signal due to the negative value of the spin Hall angle of W (SHA = -0.14).[492] The $\sigma_{LSSE}$ of Pd/YIG << $\sigma_{LSSE}$ of Pt/YIG due to the smaller spin Hall angle (SHA= 0.01) and resistivity ($\rho_{\mathrm{Pd}}$ = 0.25 μΩ-m) of Pd compared to that of Pt (SHA= 0.068, $\rho_{\mathrm{Pt}}$ = 0.42 μΩ-m).[539-540] It was also shown that the interface has a significant role in the temperature dependency of the LSSE signal (Figure 27e-f). The LSSE peak location shifted significantly when the capping layers were changed. The interfacial structure of YIG/metal films altered depending on the metal layer on top, indicating that the temperature dependency of the LSSE is not only driven by the magnetic material, but may also be impacted by the interface between the metal capping layer and ferromagnets. It has been shown experimentally that the spin-pumping efficiency (spin-mixing conductance) in YIG/Pt structures is independent of the YIG thickness indicating the exchange interaction at the YIG/Pt interface determines the effectiveness of spin pumping.[541] The experimental investigation of the ISHE induced by the spin pumping in $Ni_{81}Fe_{19}$ (10 nm thick)/Pt system (with Pt thicknesses of 10, 30, 50, and 75 nm) revealed that the ISHE voltage ($V_{ISHE}$) is inversely proportional to the thickness of paramagnetic Pt films and was reproduced by using an equivalent circuit model, which is stated to be valid when Pt film thickness >> spin diffusion length in Pt (~7 nm)[540].[542]

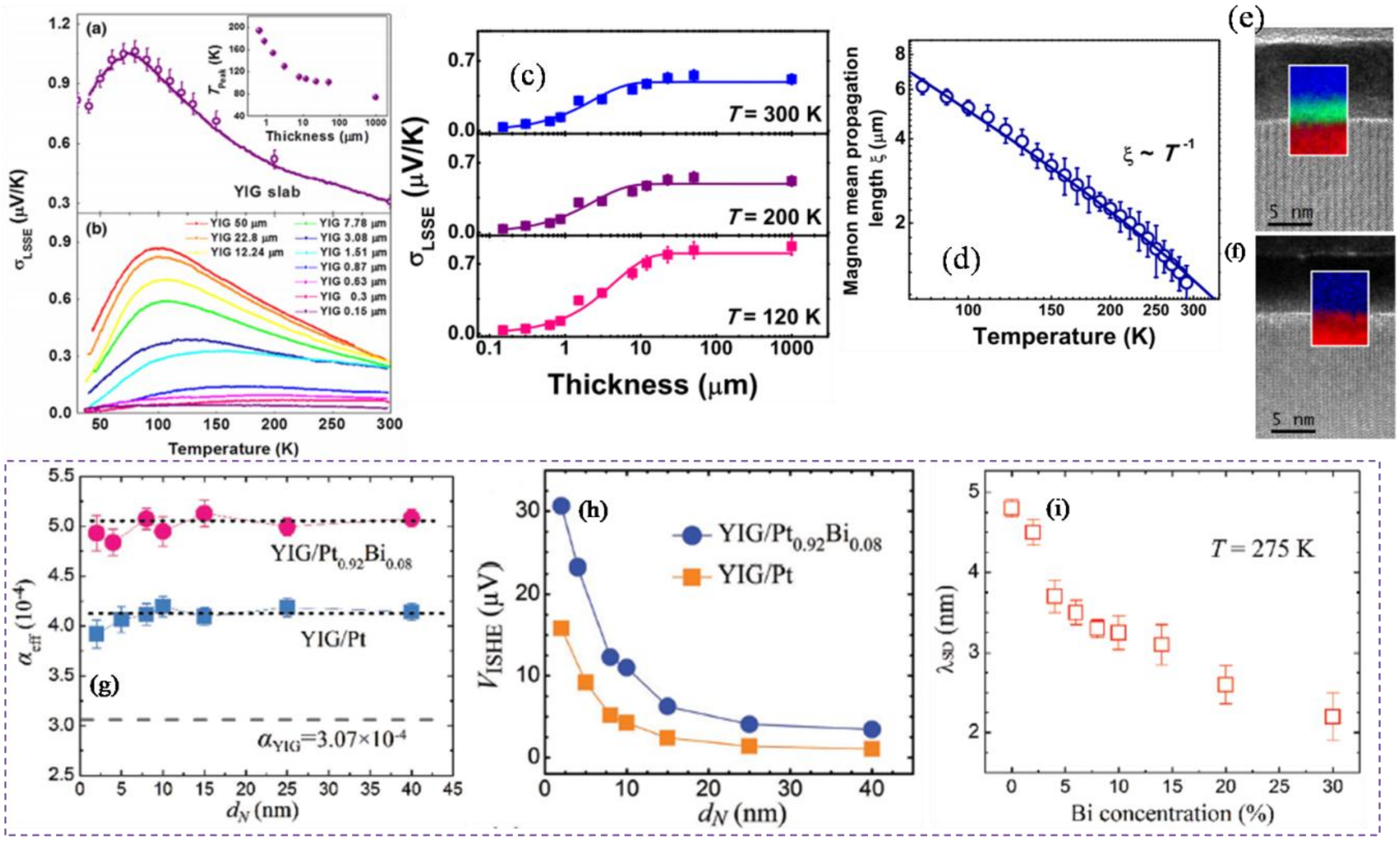

Figure 27: Temperature-dependent LSSE coefficients ($\sigma_{LSSE}$) of (a) YIG slab and (b) YIG thin films of different thicknesses. The open symbols in (a) represent the results recorded by field-sweep mode measurements. The inset of (a) shows the peak position of the $\sigma_{LSSE}$ vs.$T$ curves as a function of film thickness; (c) Thickness-dependent $\sigma_{LSSE}$ measured at a fixed temperatures of 300 K, 200 K, and 120 K. The solid lines are the fitting results according to the atomistic spin model; (d) Temperature-dependent effective magnon propagation length $\xi$. The solid line is the fit to the power law of $\xi \sim 1/T^n$, where the fitted exponent $n \approx 1$; (e-f) HRTEM images of YIG/Cu/Pt and YIG/Pt layer stacks, respectively. Inset in each image is a colored panel that indicates the elemental (Fe, Cu, Pt) distribution within the interfaces of the two samples. Fe (red), Cu (green), and Pt (blue). Adapted with permission from ref.[451] Bottom frame: $YIG/Pt_{0.92}Bi_{0.08}$ system (g) $Pt_{0.92}Bi_{0.08}$ thickness ($d$N) dependence of the $\alpha_{eff}$. The gray dashed line is for bare YIG. The blue square and pink circle represent YIG/Pt and $YIG/Pt_{0.92}Bi_{0.08}$, respectively; (h) thickness ($d$N) vs.ISHE voltage; (i) spin diffusion length of PtBi alloy at RT as a function of Bi concentration. Adapted with permission from ref.[543]

Pt is an earlier detection material for spin-Hall-effect at RT. However, the SHA of Pt is insufficient to satisfy the needs for future spintronic devices; Pt is expensive. In $Mn_2Au/Co_{40}Fe_{40}B_{20}$ bilayer system, a $\boldsymbol{g_{eff}^{\uparrow\downarrow}}$ of 3.27 ± 0.02 × $10^{18}/m^2$ and a record high SHA of 0.22 for $Mn_2Au$ are reported.[479] RT sputter-grown 3*d* light transition metal vanadium film exhibited SHA of -0.071 ± 0.003 (comparable to 0.076 ± 0.007 of Pt).[471] In CoFeB/$C_{60}$ bilayer system, the spin diffusion length of $C_{60}$ was determined to be 2.5 nm; the SHA of the sample is stated to be 0.055. With an increase in $C_{60}$ layer thickness, the $\boldsymbol{g_{eff}^{\uparrow\downarrow}}$ is increasing due to a reduction in surface roughness. The spin pumping voltage rises as the thickness of $C_{60}$ increases. It was explained that owing to π-σ hybridization, the curvature of the $C_{60}$ molecules may improve the SOC (spin-orbit coupling) strength.[544] A significant SHA ≈ -0.24 is recorded in $Cu_{1-x}Bi_x$ alloy ($x$ = 0.5% Bi) at 10 K,[478] but it is unsuitable for practical applications because of the extremely low temperature. $Pt_{0.92}Bi_{0.08}$ alloy exhibited better spin pumping and high spin–charge conversion in YIG ($La_{0.03}Y_{2.97}Fe_5O_{12}$)/$Pt_{0.92}Bi_{0.08}$ bilayer system at RT (Figure 27g-i).[543] The spin diffusion length and spin Hall angle (SHA) of the $Pt_{0.92}Bi_{0.08}$ alloy were 3.3 ± 0.1 nm and 0.106 ± 0.005 (10.06%), respectively. The spin–charge conversion of this alloy was reported to be twice that of bare Pt. The increase in SHA was explained based on the skew scattering from Bi impurity states through the spin–orbit interaction. For the system, YIG (300 nm)/$Pt_{0.92}Bi_{0.08}$ (10 nm) layer, the reported values of $V_{ISHE}$, $\sigma$, $\alpha_{eff}$, and $\boldsymbol{g_{eff}^{\uparrow\downarrow}}$ were 11.01 μV, 3.36 × $10^6$ S/m, 5.04 × $10^{-4}$, and 5.67 × $10^{18}/m^2$ respectively.[543]

Kim et al. investigated the effect of interface conditions on the LSSE in a Pt/polycrystalline $NiFe_2O_4$ (NFO) system.[545] Different interface conditions were generated by treating the surface of NFO slabs with different polishing forces/post-annealing temperatures before the Pt deposition. It was found that mesoscale surface defects (pores, gouges, and cracks) and the surface roughness play a significant role in determining the magnitude of LSSE signals. One sample exhibited a SSC of 0.58 μV/K (larger than those previously reported for slab type samples of Mn–Zn ferrite ≈ 0.071 μV/K,[546] Ni–Zn ferrite ≈ 0.091 μV/K,[547] and Zn ferrite ≈ 0.084 μV/K[548], and twice that of the YIG sample ≈ 0.28 μV/K[535]).

In a polycrystalline nickel-ferrite NFO/Pt bilayer system, Kim et al. achieved concurrent optimization of the magnon and phonon transport in bulk NFO and NFO/Pt interface quality following a heat treatment scheme.[549] The phase separation of NFO via heat treatment yielded a hierarchical microstructure of nano-sized NFO embedded in micro-sized NiO precipitates (Figure 28a-b). This structure aided in selectively scattering the phonons while not disturbing magnons, leading to reduced $\kappa$ without changing spin Seebeck coefficient. LSSE in a YIG/Pt system was enhanced by adding an intermediate layer of high spin diffusion length $C_{60}$ (i.e. making YIG/$C_{60}$/Pt system).[550] The presence of the $C_{60}$ layer reduced the conductivity mismatch between YIG and Pt and the surface magnetic anisotropy of YIG, giving rise to the enhanced spin-mixing conductance and hence the elevated LSSE. This enhancement was dependent on temperature and the thickness of the $C_{60}$ layer.

The charge potential mediated by structural inversion asymmetry causes the proximity induced Rashba effect, which has been observed in Si thin films and can be controlled by varying the thickness of the sandwitched layer. Bhardwaj et al. reported a large increase in SSE in NiFe (25 nm)/*p*-Si (polycrystalline, thicknesses of 5 nm, 25 nm, and 100 nm) bilayer specimens (Figure **26**c).[523] The spin-Seebeck voltage showed a three-fold increase in the case of 5 nm *p*-Si specimen as compared to the 25 nm and 100 nm *p*-Si specimens. The inverse spin-Hall effect was proposed to occur due to proximity induced Rashba spin orbit coupling (SOC) at the NiFe/*p*-Si interface, which increases significantly with reduction in *p*-Si layer due to structure inversion asymmetry. This design eliminated the requirement of Pt for spin to charge conversion. The measured voltage of 100.3 μV is one of the largest reported spin Seebeck voltages (for the area of ~160 × 10 $\mu m^2$).

Under bending stress, the performance of conventional TE cuboids-shaped devices degrades. For a wide range of practical applications, TE devices must be flexible and geometrically accommodative. To prepare a LSSE-based flexible TE sheet, a 500 nm thick $Ni_{0.2}Zn_{0.3}Fe_{2.5}O_4$ ferrite film was spray-coated on a 25-μm-thick PI substrate; a Pt film with a thickness of 5 nm was deposited on the $Ni_{0.2}Zn_{0.3}Fe_{2.5}O_4$ film by means of magnetron sputtering to obtain a bendable Pt/$Ni_{0.2}Zn_{0.3}Fe_{2.5}O_4$ film.[511] The variation of TE voltage with heat-flux was linear. Tests on the dependence of the heat-flow sensitivity ($V/q$) on the curvature ($1/r$) show that the bending stresses applied to the Pt/$Ni_{0.2}Zn_{0.3}Fe_{2.5}O_4$ films did not alter the TE conversion process.

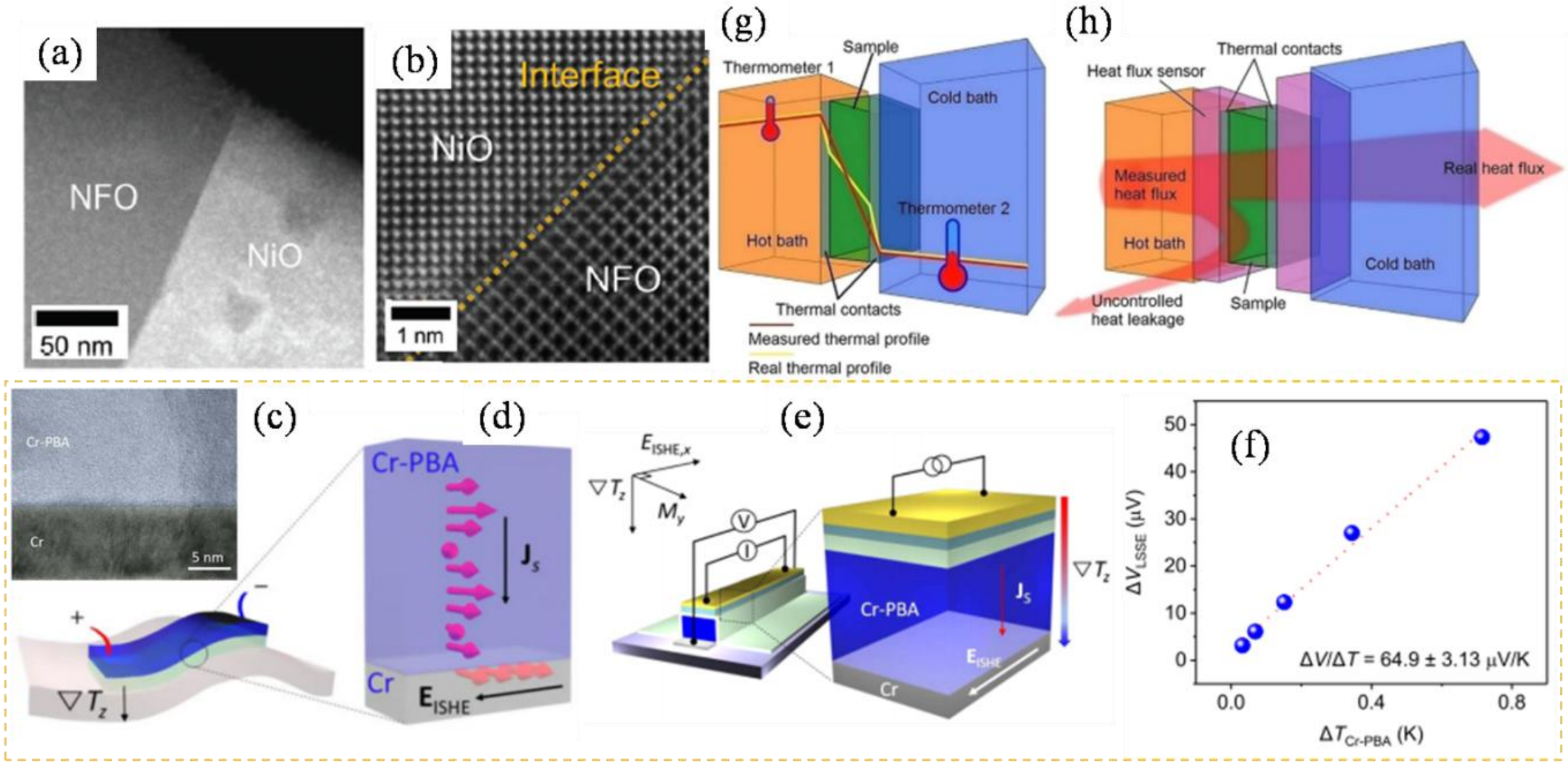

Figure 28: Microstructure and composition analysis of NFO matrix/Fe-doped NiO precipitates. (a) Cross-sectional HAADF-STEM image; (b) high resolution HAADF-STEM image of the micro-NiO/nano-NFO interface. Reproduced with permission from ref.[549] (g) schematic representation of the direct temperature measurement; thermometers are placed on the hot and cold baths each one at thermal equilibrium. The red line is the assumed linear temperature profile between the two heat baths. The yellow line represents the real temperature profile taking into account thermal resistances at the interfaces; (h) schematic representation of the heat flux measurement; the approximation of the measured heat flux with the real quantity which flows into the sample depends on the amount of lost heat (small arrow). Reproduced with permission from ref.[449] Lower frame. (c) Electrodeposited Cr-PBA thin film (inset) and illustrations of spin TE energy conversion under $\nabla T_z$ (bottom image) and the mechanism of LSSE associated with thermally generated magnons and their conversion into a charge current via the ISHE (right side image); (e) illustration of the LSSE characterization of the Cr-PBA/Cr bilayer; (f) $\Delta V_{LSSE}$ vs. $\Delta T$ in a Cr-PBA film with slope ≈ $\Delta V_{LSSE}/\Delta T$ ≈ 64.9 ± 3.13 μV/K at 100 K. Adapted with permission from ref.[551]

Organic-based magnets may be viable magnetic insulator options due to lower spin–orbit coupling and reduced electron-lattice scattering, which may significantly reduce the damping constant. Furthermore, organic-based magnetic films can be prepared at lower temperatures, and their versatile synthesis pathways may enable scalable deposition processes. Vanadium tetracyanoethylene ($V(TCNE)_x$, $x \approx 2$), for example, has demonstrated spin-polarized carrier injection,[552-553] coherent magnon production, and spin pumping.[554]

$Cr^{II}[Cr^{III}(CN)_6]_x \cdot nH_2O$ (Cr-PBA) is known to have high magnetic-ordering temperature($T_c$) ~240 K[555-557] and belongs to the family of molecule based magnets called Prussian blue analog (PBA). Cr-PBA has been tested for spin TE applications because of its versatile synthetic pathways, poor spin-lattice interaction coupled with low $\kappa$, which might lead to efficient propagation of thermally excited magnons while preserving a larger temperature gradient.[551] In a electrochemical deposition process, a thin film (10 nm) of Cr was employed as a working electrode to grow PBA film to obtain Cr-PBA/Cr bilayer, which was directly used as an spin TE device with a smooth interface for effective spin pumping (Figure 28c-f). The schematic depiction for the LSSE characterization of the Cr-PBA/Cr STE device is shown in Figure 28e. The $\nabla T_z$ was established by Joule heating of a top Au line. The vertical flow of magnons pumps a pure spin current into the adjacent Cr layer. The induced spin flow is then transformed into a longitudinal charge current, resulting in $E_{ISHE}$. In the LSSE configuration, the Cr-PBA/Cr bilayer (with 5 mm length, 100 μm width, and 1.4 μm thickness of Cr-PBA) has generated a spin Seebeck voltage, $\Delta V_{LSSE}/\Delta T$ ~64.9 ± 3.13 μV/K at 100 K (Figure 28f) in addition to displaying a damping constant of ~$7.5 \times 10^{-4}$. This damping value is larger than the intrinsic damping constant of Cr-PBA film $(2.4 \pm 0.67) \times 10^{-4}$ due to an additional damping constant caused by spin pumping at the Cr-PBA/Cr heterojunction.

As discussed previously, the determination of a LSSE coefficient requires simultaneous measurement of the $V_{\mathrm{ISHE}}$ and the thermal gradient $\nabla T$. While the measurement of $V_{\mathrm{ISHE}}$ can be done very accurately, $\nabla T$ is not easily available by direct measurements.[449] For the determination of $\nabla T$, $\Delta T$ method (direct temperature measurement, Figure 28g)[517] has generally used method ever since the first observation of LSSE. This method uses thermocouples to measure the temperature of the thermal baths. Recently, the heat flux method was proposed, where the contribution of the thermal resistance of the contacts is not a factor, Figure 28h.[440] Sola et al. tested the reproducibility of the LSSE measurements on YAG (Yttrium Aluminum Garnet)/YIG/Pt system using these two experimental methods and suggested that an approach based on the heat flux method with a low uncertainty is the appropriate one.

## 5. Concluding remarks and outlook

Novel methodologies for improving the TE performance of various types of TE materials have made significant advances over the last two decades. Despite that, their use in real-world TE applications is not keeping pace with that of other energy-harvesting technologies. In this scenario, it is critical to address the issues of improving the *PF* of TE materials with intrinsically low $\kappa$ via band structure and microstructure manipulation. 2D materials exhibit a high anisotropy along the perpendicular and in-plane directions. Further, issues related to cost-effective synthesis of 2D materials must be sorted out as the current methods are ineffective. In low-dimensional TE thin films, it is much easier to tailor electrical transport and thermal transport compared to TE bulks. Thin film technology bears great advantages for fabricating high-performing TE thin-films and devices. Considerable efforts have been made to optimize the electrical and thermal properties in films, including tuning the carrier concentration, defect generation, and engineering the band structure. However, the synergetic control of electrical and thermal properties is still a challenge. The implementation of quantum confinement is not being realized widely. In thin films, embedding well controlled nanostructures to create unhindered paths of charge carriers (e.g. with nanowires/2D nanostructures) but blocking the transmission of phonons will be an effective method in improving the TE performance. Controlling crystalline nanoparticle phase composition, their volume fractions, size, and distribution (distance between nanoparticles) to improve electron and phonon transport in a film of a certain thickness by regulating fabrication conditions and subsequent heat treatment is a complicated and material-specific process. Some results may be obtained from studies such as how the heat treatment of a system with chemically inert nanoprecipitates spread between two layers affects the TE characteristics of the system. Computational research might provide guiding criteria for developing and fabricating TE films.

High temperature film deposition methods are not energy-efficient. Stringent ultra-high vacuum requirements, equipment/maintenance costs, small area deposition capability of many methods are other unfavorable factors. However, the tradeoff between energy expenditure and the salient features associated with TE device miniaturization and their reliability, on the other hand, has the potential to overwhelm the various constraints of the film fabrication methods. The current high-performing TE materials are incompatible with industrial CMOS technology making it difficult to integrate these materials into Si ICs. The noteworthy property of Si-based heterostrucures, on the other hand, is their basic compatibility with Si technologies, which play a prominent role in the field of solid-state electronics. Therefore, development of Si based TE materials is critical.

Scientific literature on the electrochemical deposition of *n* and *p*-type $Bi_2Te_3$-based compounds is rife with deposition mechanisms, the type of electrolyte solutions, and deposition protocols. The electrodeposition of bismuth telluride-based films is cost effective and well suited for designing μTEGs. However, the poor crystalline quality of the films is a negative factor prompting moderate transport properties such as low charge $\mu$. Adapting new synthesis protocols: pulsed electrochemical methods, additives and post annealing treatment under a controlled atmosphere are pivotal. The electrodeposition of thick and efficient *p*-type (Bi, Sb)$_2$Te$_3$ compounds is still a challenge. At present, the performance of TE devices is restricted and heavily reliant on internal resistance, which must be reduced. Many studies have reported TE thin film deposition on flexible substrate such as PI. However, the PI substrate does not withstand very high temperatures. Some studies have reported the use of CNT films/scaffolds containing randomly distributed CNT bundles as substrates. But the degree of similarity in CNT dispersion in two separate CNT films may not be the same. The electrical and thermal interface resistance at the interface between the TE layer and the CNT film may affect the nature of electrical and thermal properties. Depending on the distribution of CNT in the film, TE properties may vary from one sample to another. Solution-based CNT film preparation techniques may be required to disperse CNT homogeneously without leaving any residue (e.g. surfactant impurities) and this remains a challenge.

The cosolvent route to obtain TE films is a viable and cost-effective option. However, synthesis of soluble precursor solutions for a wide range of TE semiconductors using less toxic solutions is still a challenge. Further, controlling the microstructures in films obtained via cosolvent method is not explored yet.

Many studies we reviewed do not mention the thermal stability of TE films, which is an essential feature to consider in addition to TE performance. Recurrent heating cycles and rapid temperature fluctuations near operational temperatures can induce strain, thermal fatigue, and fractures, as well as interdiffusion of the components or even breakdown, degrading *zT* values. After multiple heating cycles, the structural resilience of the nanostructures may cease to exist, bringing morphological changes that can lower $\sigma$ (due to discontinuity in charge carrier path). Therefore, data on compositional changes via in-situ XRD and information on morphological changes via TEM as a function of temperature are needed to test/ascertain the root cause of thermally induced breakdown in *zT*. It should be noted that, although acceptable film thicknesses appear to be possible even at greater lattice mismatches in a stacked film structure, the metastable layers may partially relax after further heat treatments.

At present, 3ω and time-domain thermoreflectance (TDTR) are the two most commonly used methods to characterize the thermal properties of thin films. But they are not necessarily popular considering the complicated measurement steps. Designing experiments to characterize the film properties with low uncertainty in a short time is important.

The development of SSE based systems is in a nascent state. As already discussed, SSE has been observed in different materials, including insulators. Therefore, SSE material does not have to be primarily TE. This begs the question of the relevance of current TE materials considering the well established fruitful strategies of enhancing their TE performance (including the materials with $\kappa$ ~theoretical limit), which can not be ignored. The spin degree of freedom augments functionality and may improve the performance of conventional TE device. In SSE-inducing materials, balancing the phonon and charge carrier transport/spin current of either magnons or spin-polarized electrons is one of the main topics of future research on SSE materials. The separation of phonon and magnon currents could be envisaged by promoting selective scattering mechanisms in the SSE-material so that the heat carrying phonons are effectively scattered while the transport of magnons is not hindered (e.g. disordered SL alloys with nanoscale features). It is critical to develop new magnetic materials for magnonics. Materials with small damping, high saturation magnetization, and high Curie temperature are required.

Fabrication of ultra-thin high quality films (e.g. YIG) of high-dynamical quality and required thickness is very essential, as spin–orbit torque is interfacial. Similar to TE periodic structures, the fabrication of layers of SSE based systems is labor intensive and expensive, which do not portend well for the scope of a large-scale application. Due to scaling issues, inorganic magnetic insulator films are not suitable for practical spin TE applications since it is difficult to obtain large-area films. Further, epitaxially grown single crystals at high temperatures are required. The search for low-cost methods, or the development of new techniques, is critical for translating prototype design to cost-effective large-scale fabrication. Despite the investigation of a wide range of systems, the factors influencing spin transfer across FM/NM interfaces remain elusive. Pt is inappropriate for wider TE applications because of its high cost and its spin-charge conversion efficiency is insufficient to provide appropriate TE performance. Further, there is no clear consensus on the spin diffusion length of Pt. The reported values range from 0.5-14 nm. Such a vast spectrum of values does not auger well for designing and predicting device functionality. As a result, low-cost Pt substitutes are an essential requirement. Materials for SSE signal generation and spin-to-charge conversion must be abundant on Earth. Experimental inverstigations are needed to determine how to attain extended spin lifetimes and spin diffusion lengths in graphene and other 2D materials (via functionalization). The design of eco-friendly and inexpensive spin voltage generating organic composites is tangible from the practical point of view. Si has established a well-recognized niche in the microelectronics industry. It remains to be seen whether the research efforts of using Si-based materials (or any other earthly abundant material) as the raw material for TE and spin voltage based devices can be materialized.

## Notes

The author declares no competing financial interests. No funding.